\documentclass[fleqn,usenatbib]{mnrasfix}

\usepackage[T1]{fontenc}

\DeclareRobustCommand{\VAN}[3]{#2}
\let\VANthebibliography\thebibliography
\def\thebibliography{\DeclareRobustCommand{\VAN}[3]{##3}\VANthebibliography}

\usepackage{graphicx}	
\usepackage{amsmath}	
\usepackage{amssymb}	

\usepackage{newtxtext,newtxmath}

\title[A model selection approach to dynamical modelling]{A novel approach to optimize the regularization and evaluation of dynamical models using a model selection framework}

\author[M. Lipka et al.]{
Mathias Lipka $^{1,2}$\thanks{E-mail: mlipka@mpe.mpg.de}, Jens Thomas$^{1,2}$
\\
$^{1}$Max-Planck-Institut für extraterrestrische Physik, Giessenbachstrasse, D-85748 Garching\\
$^{2}$Universitäts-Sternwarte München, Scheinerstrasse 1, D-81679 München, Germany
}

\date{Accepted XXX. Received YYY; in original form ZZZ}

\pubyear{2021}

\begin{document}
\label{firstpage}
\pagerange{\pageref{firstpage}--\pageref{lastpage}}
\maketitle

\begin{abstract} 
Orbit superposition models are a non-parametric dynamical modelling technique to determine the mass of a galaxy's central supermassive black hole (SMBH), its stars, or its dark-matter halo. One of the main problems is how to decide which model out of a large pool of trial models based on different assumed mass distributions represents the true structure of an observed galaxy best. We show that the traditional approach to judge models solely by their goodness-of-fit can lead to substantial biases in estimated galaxy properties caused by varying model flexibilities. We demonstrate how the flexibility of the models can be estimated using bootstrap iterations and present a model selection framework that removes these biases by taking the variable flexibility into account in the model evaluation. We extend the model selection approach to optimize the degree of regularisation directly from the data. Altogether, this leads to a significant improvement of the constraining power of the modeling technique. We show with simulations that one can reconstruct the mass, anisotropy and viewing angle of an axisymmetric galaxy with a few percent accuracy from realistic observational data with fully resolved line-of-sight velocity distributions (LOSVDs). In a first application, we reproduce a photometric estimate of the inclination of the disk galaxy NGC 3368 to within 5 degree accuracy from kinematic data that cover only a few sphere-of-influence radii around the galaxy's SMBH. This demonstrates the constraining power that can be achieved with orbit models based on fully resolved LOSVDs and a model selection framework.

\end{abstract}


\section{Introduction}
Revealing the internal structure of external galaxies is a challenging, yet essential task for a broader understanding of galaxy evolution. Since observations of external galaxies are limited to the projected kinematics and the surface brightness of the luminous galaxy components the common approach is to establish dynamical models of an observed galaxy and to compare their projections to the corresponding observations. A collisionless galaxy component, such as the ensemble of its stars, follows the collisonless Boltzmann equation and can be fully characterized by its phase-space distribution function. Therefore a dynamical model can, at least in principal, describe such a system of stars if it can sufficiently emulate the real distribution function. If a galaxy could not be assumed to be in equilibrium, dealing with all the freedom in the distribution function during the modelling process would be hopeless. However, in equilibrium the Jeans Theorem can be invoked \citep[e.g.][]{Binney2008} and the distribution function has a relatively simple structure which can be expressed as a function of the integrals of motion meaning it takes the form of an orbit superposition. 

Even though stellar systems with a high degree of symmetry can sometimes be modelled well using analytic distribution functions, a more universal approach to dynamical modelling is generally required to deal with the full range of compatible distribution functions. Schwarzschild modelling is an efficient numerical approach based on the superposition of stellar orbits to construct such dynamical models. In its initial version, established by \citet{1979ApJ...232..236S}, such orbit models were designed to replicate a given triaxial density distribution in a self-consistent manner. In the subsequent decades the Schwarzschild technique was extended to include the fitting of kinematic observations \citep[e.g.][]{1985ApJ...295..349L, 1997ApJ...488..702R, 1999ApJS..124..383C, 2004MNRAS.353..391T,2008MNRAS.385..647V}, making it possible to constrain the distribution function of an observed galaxy more tightly and enabling the estimation of intrinsic properties of stellar systems such as black hole mass \citep[e.g.][]{1998ApJ...493..613V,2000AJ....119.1157G,2002ApJ...578..787C,2003ApJ...583...92G,2013AJ....146...45R}, mass-to-light ratios \citep[e.g.][]{2005MNRAS.360.1355T,2006MNRAS.366.1126C,2011MNRAS.415..545T}, dark matter halos \citep[e.g.][]{2005MNRAS.360.1355T,2007MNRAS.382..657T,2009ApJ...691..770T,2013AJ....146...45R,2018MNRAS.477..254L}, the velocity dispersion anisotropy and orbital structure \citep[e.g.][]{2008MNRAS.385..614V,2009MNRAS.393..641T,2014ApJ...782...39T,2017MNRAS.470.3959K}, and the intrinsic shape of the stellar distribution \citep[e.g.][]{2020MNRAS.491.1690J}.

Several different implementations of the Schwarzschild Method with varying degrees of symmetry have been described \citep[e.g.][]{1997ApJ...488..702R,1999ApJS..124..383C,2000MNRAS.319...43S,2000MNRAS.314..433H,2000AJ....119.1157G,2003ApJ...583...92G,2004ApJ...602...66V,2004MNRAS.353..391T,2008MNRAS.385..647V,2015MNRAS.450.2842V,2020ApJ...889...39V,2020MNRAS.500.1437N}, but the main steps of the method are rather general. Very briefly: If an assumed gravitational potential is given, one needs to compute tens of thousands of representative orbits in the available phase space. These orbits are combined to a galaxy model, where each orbit can carry an adjustable number of stars. These so-called orbital weights (or orbital occupation numbers) together with each orbits intrinsic and projected properties then determine the dynamical model. Through optimization of the orbital weights, the model can be adapted to the observations. In this sense, the orbital weights are the variables of the model. While different algorithms are used for this weight optimization, they are all tied to a $\chi^2$-minimisation framework.

With the rise of integral-field spectrographs in the last decade, the amount of kinematic data that can be used to constrain these models has increased significantly. While many early studies had to rely on line-of-sight velocity distributions (LOSVDs) that were parameterized by Gauss-Hermite series expansions \citep[][]{1993MNRAS.265..213G,1993ApJ...407..525V} up to fourth order (often along a 1d slit \citep[][]{1994MNRAS.269..785B}, it is now often possible to reliably obtain higher Gauss-Hermite moments \citep[e.g.][]{2009MNRAS.399.1839K,2020ApJ...891....4L} or measure the fully resolved, non-parametric LOSVDs over the 2d sky area occupied by the galaxy \citep[][]{2019ApJ...887..195M}. The number of measured data points per galaxy has thus increased from a few dozen to several thousands. Still, one of the characteristics of the Schwarzschild Method is that one is usually faced with a situation where the number of free model parameters is significantly larger than the number of observational data points, mainly because the number of orbits is typically very large and each orbit is associated with a free orbital weight parameter. For simple, linear models this would imply a negative number of degrees of freedom of the resulting $\chi^2$ distribution \citep[][]{1992nrfa.book.....P} and the solution of the $\chi^2$-minimization is non-unique \citep[cf.][]{2004ApJ...602...66V,2006MNRAS.373..425M,2020MNRAS.500.1437N}. Under the assumption that a model is generally detailed enough to deal with all the structure in the data, one would consequently expect that every data set is perfectly fitted due to the models comparatively huge number of free parameters, resulting in a $\chi^2 \rightarrow 0$. In practice, this does not happen for two reasons. Firstly, data with a realistic amount of noise can not be fitted perfectly when the orbital weights are constrained to be non-negative (see \citealt[][]{2006MNRAS.373..425M} for a discussion). Secondly, a perfect fit is suppressed when a regularisation term is applied in the weight optimisation, e.g. via the commonly used maximum penalized likelihood or maximum entropy approaches. The smoothing induced by such a regularization term allows to discard physically implausible solutions and prevents overfitting by effectively reducing the flexibility of the dynamical model. Despite the differences in the individual implementations of regularization, the result is always the same: The $\chi^2$ of the regularized model is significantly larger than zero even though the model's parameters typically outnumber the data points. This suggests that the \textit{effective} number of free parameters is smaller than the nominal number of variables in the model, implying a reduction in the model's flexibility (i.e. its ability to fit noisy data). Beyond the regularization another core aspect in reducing the flexibility of Schwarzschild models is the prior constraint on the orbital weights to be non-negative to avoid negative phase-spaces densities in the model. 

The reduction of model flexibility is a generic property of regularized models, of models with prior constraints imposed on their free parameters, or in general of non-linear models \citep[cf.][]{2010arXiv1012.3754A}. As a consequence, because the \textit{effective} number of parameters is an unknown, the interpretation of the absolute value of $\chi^2$ becomes less obvious. This is particularly important, because in most Schwarzschild applications, the primary interest is not in the distribution of the orbital weights. Instead, one aims to determine the mass composition of a galaxy, which requires the comparison of a number of Schwarzschild models that were obtained with different trial gravitational potentials. Traditionally this comparison is done by evaluating the $\chi^2$ values inherited from the orbital weight optimization of each trial model. Given the fact that these $\chi^2$ values depend on the model flexibility (or its \textit{effective} number of free parameters) it may be important to take this into account in the evaluation process.

The goal of this paper is threefold. Firstly, we provide a method to estimate the flexibility of Schwarzschild Models, or the effective number of free parameters respectively, that can robustly deal with all the smoothing constraints, positivity constraints and non-linearities of the models. Secondly, we demonstrate that varying model flexibilities lead to biased $\chi^2$ surfaces when evaluating orbit models obtained using different mass distributions, viewing angles or assumed regularisation powers. Thirdly, we provide a model selection framework that allows to overcome these biases that can arise in the simple $\chi^2$ comparison framework. The new framework not only improves the constraining power of Schwarzschild models significantly, but also enables a new data-driven approach to optimize the amount of regularization. All these considerations can probably be generalised to other non-linear, non-parametric, non-dynamical models that involve regularisation or prior constraints.

As the main application example, we investigate the reconstruction of the viewing angle (or equivalently the intrinsic flattening) under which an axisymmetric galaxy is observed. The question whether it is feasible to constrain the inclination/intrinsic shape via dynamical modelling is as of yet a point of contention. In early works \citet{2002MNRAS.335..517V} argue that they constrained the inclination of M32 to $70\degr \pm5\degr$ using axisymmetric, three-integral, Schwarzschild models. However, follow-up results of \citet{2005MNRAS.357.1113K}, who tried to recover the inclination of semi-analytic galaxy models with axisymmetric models, suggest that different inclinations are degenerate even under ideal conditions. \citet{2006MNRAS.366.1126C} confirmed these results for a large sample of early-type galaxies in so far as they state that Schwarzschild models with a wide range of inclinations are able to fit the observed kinematics well within the errors. \citet{2007ApJ...670..105O} modelled the bulge of the Seyfert 1 galaxy NGC 4151 edge-on and with an inclination $i=23\degr$ obtained from the observed ellipticity of the large-scale disk, with the result that the edge-on model provides better fits to the kinematic constraints than the $23\degr$ model, suggesting either that the bulge is not aligned with the disk or that the inclination recovery using dynamical modelling is biased. Similarly, \citet{2007MNRAS.382..657T} found that a surprising amount of the 17 early-type galaxies they dynamically modelled with axisymmetric three-integral models are fitted best when the orbit model is viewed edge-on and they argue there may be a small inclination bias favoring edge-on Schwarzschild models. In this paper we will demonstrate that the assumed inclination strongly affects the flexibilty of orbit models which could explain the observed edge-on preference of previous studies. We will show that the inclination of axisymmetric galaxies can be well constrained from typical observational data with the new model selection framework as it considers the model flexibility in the evaluation of the fitted models. 

The paper is organized as follows: In section~\ref{sec:overview+problem} we outline our motivation for using a model selection approach in the evaluation of Schwarzschild models in detail. In section~\ref{sec:ansaetze} we lay out two bootstrap methods which facilitate the quantification of a model's intrinsic flexibility. A discussion of an evaluation approach based on model selection techniques follows in section~\ref{sec:Model_selection}. We then test the model selection with respect to the inclination reconstruction on a number of simulated galaxies in section~\ref{sec:simulated}. In section~\ref{sec:mass_parameters} we extend the analysis to mass parameters. Section~\ref{sec:regularization} further refines our approach by including the regularization power. Equipped with all these insights we model the disk galaxy NGC 3368 in section~\ref{sec:NGC3368} with our refined model selection framework and set it into contrast to the prevalent $\chi^2$-minimization. Sec.~\ref{sec:discussion} discusses general aspects related to modelling degeneracies and the inclination recovery. The paper concludes with a summary in section~\ref{sec:conclusions}.

\section{Motivating a model selection approach to the Schwarzschild technique}
\label{sec:overview+problem}

As mentioned in the Introduction above, we want to demonstrate the power of the model selection framework using the example of Schwarzschild orbit superposition models. The main reasons for picking up Schwarzschild models are: (i) This technique is very general and allows to model any kind of galaxy without apriori assumptions upon the orbital structure; (ii) The method is observationally not restricted to moments of the velocity distribution of the stars but instead can deal with the full information contained in the line-of-sight velocity distributions \citep[e.g.][]{2004MNRAS.353..391T, 2019ApJ...887..195M, 2020ApJ...889...39V, 2020MNRAS.500.1437N}. The specific implementation of the Schwarzschild Method that we consider in this paper assumes axially symmetric galaxies and is described in detail in \citet{2004MNRAS.353..391T}. 

In Schwarzschild models, one first specifies a trial mass distribution and viewing angle via parameters like the stellar mass-to-light ratio $\Upsilon$, black-hole mass $M_\bullet$, dark-halo parameters and inclination $i$. Then one calculates an orbit library with thousands of orbits whose projected properties form the building blocks of the superposition model. By computing the best-fitting orbital weights one can assess how well the trial mass distribution allows to explain the observational data. A systematic search through the space of trial mass models then leads to the identification of a best-fit mass distribution for each galaxy. This is only a very brief sketch of the method (we provide a more detailed description in App.~\ref{appendix:modelling} and Sec.~\ref{subsec:selection_parameters}) but it already demonstrates one important aspect: the fact that one {\it first} needs to specify a set of mass parameters (and the inclination) \textit{before} the orbit distribution can be determined will turn out to be significant. As a consequence it is helpful to conceptually distinguish two distinct parameter layers that together define the properties of every Schwarzschild model: Minimization parameters (orbital weights $\mathbfit{w}_i$) and selection parameters (all other parameters, i.e. $M_\bullet, \Upsilon, i$...). It is the latter parameters that require an evaluation using a \textit{model selection} framework. 

We present our reasoning for this distinction of parameter layers in Secs.~\ref{subsec:selection_parameters} and \ref{subsec:parameter_estimation}. We also recapitulate our regularization approach to finding the optimum minimization parameters (i.e. orbital weights) in \ref{subsec:parameter_estimation}. The theoretically less interested reader may skip ahead to Sec.~\ref{subsec:spherical} where we show that a simple $\chi^2$ evaluation where all parameters are treated equally can lead to a bias, using the inclination as an example. As we will show throughout this paper, other selection parameters such as the stellar mass can also suffer from such a bias.

\subsection{Selection vs. minimization parameters}
\label{subsec:selection_parameters}
It is tempting to consider the question of identifying the best-fit Schwarzschild model in terms of a huge $\chi^2(M_\bullet,\Upsilon,i,\ldots,\mathbfit{w}_i)$ minimisation, where all the parameters are treated equally as free parameters of a \textit{single 'global' Schwarzschild model}. This concerns mass parameters like $M_\bullet$ and $\Upsilon$, viewing angles like the inclination $i$ and the orbital weights $\mathbfit{w}_i$ (cf. App.~\ref{appendix:modelling}). It is therefore common practice to assume that the best choice for $M_\bullet,\Upsilon,i,\ldots$ is given by the trial set that resulted in the smallest $\chi^2$ without much concern of the role of orbital weights. 

Using a sophisticated Bayesian framework \citet{2006MNRAS.373..425M} acknowledges that there often exist many possible combinations of orbital weights (or equivalently distribution functions) that are consistent with given observational data and an assumed trial potential. \citet{2006MNRAS.373..425M} argues that a more appropriate approach sums over all \textit{possible} distribution functions for a given potential. This is achieved by marginalizing over the orbital weights and weighing the corresponding likelihoods by a suitable prior.  \citet{2006MNRAS.373..425M} argues that this approach allows a more accurate calculation of the likelihood for a given potential, and that the odds of one set of trial parameters $(M_\bullet,\Upsilon,i,\ldots)$ over another can be evaluated by the ratio of the likelihoods of the two potentials they generate.

The Bayesian framework of  \citet{2006MNRAS.373..425M} does not specifically address the problem that the orbital weights are not independent parameters of a \textit{single, global} Schwarzschild model. Instead, the orbital weights function as (linear) coefficients for the fundamental building blocks of the model: the projected properties of the orbits which are different for each trial potential and, thus, need to be recalculated from scratch for each trial potential (cf. Appendix~\ref{appendix:modelling}). 
In other words, the subjects to which the orbital weights $\mathbfit{w}_i$ refer to, \textit{depend} on the choice of the parameters $M_\bullet,\Upsilon,i,\ldots$ which have to be specified to generate the orbits. This means there is no straight-forward way to define a \textit{single} model with a common parameter space $(M_\bullet,\Upsilon,i,\ldots,\mathbfit{w}_i)$ as the basis for the comparison of $\chi^2$ values obtained for different sets of trial parameters like $M_\bullet,\Upsilon$ or $i$. Instead one actually compares the goodness-of-fit of \textit{distinct} statistical models, each with its own individual space of orbital weights $\mathbfit{w}_i$.

On a more fundamental level, the Schwarzschild technique is an {\it exact} method to find phase-space distribution functions $f$ that obey the Collisionless Boltzmann equation:
\begin{equation}
    \mathbfit{v} \cdot \frac{\partial f}{\partial \mathbfit{x}}-\frac{\partial \phi}{\partial \mathbfit{x}} \cdot \frac{\partial f}{\partial \mathbfit{v}}=0
	\label{eq:CBE}
\end{equation}
(where $\phi$ is the gravitational potential and $f$ is the phase-space distribution function of the system under study). We can formulate the Schwarzschild technique by considering a partition of phase space into a huge number of small cells. The distribution function $f$ is represented by the large number of phase-space densities $f_i$ in each of these small local cells. The observables of the model, such as the binned LOSVDs $\mathbfit{l}_\mathrm{mod}$ of the model or its binned 3d luminosity density $\mathbfit{d}_\mathrm{mod}$,  can be derived from the $f_i$ by simple phase-space integrations, e.g. the amount of light in bin $k$ reads
\begin{equation}
\mathbfit{d}_{\mathrm{mod},k} = \int_{S_k} f \, \mathrm{d}^3r  \mathrm{d}^3v, 
\end{equation}
where the integral goes over the whole velocity space and over the subset $S_k$ of the configuration space that represents bin $k$ (cf. App.~\ref{appendix:modelling} for the exact definitions and vector notation of $\mathbfit{l}_\mathrm{mod}$ and $\mathbfit{d}_\mathrm{mod}$). One could argue that these $f_i$ could be used as the fundamental parameters in a {\it single, global} Schwarzschild model. 

In fact, when we do a $\chi^2$ minimization, we assume a statistical model for the data, specifically that each measurement $\mathbfit{l}_{\mathrm{obs},k}$ was drawn from a Gaussian distribution. The width of this distribution is commonly assumed to be known and approximated by the observational error. Only the mean, which is given by $\mathbfit{l}_{\mathrm{mod},k}$  has to be determined in the modelling process. Hence, our {\it statistical model} is completely determined when the $\mathbfit{l}_\mathrm{mod}$ are known, which is the case when the $f_i$ are specified. For this, we do not even need to assume a gravitational potential or mass model. And up to this point one could indeed formulate a {\it single, global} dynamical phase-space model with the $f_i$ being {\it independent} free parameters. 

However, one issue with such a phase-space model is that arbitrary distributions of the $f_i$ are of little interest, because most of them will be physically unrealistic or even unphysical. Besides the constraint $f_i > 0$ (positive phase-space density) another crucial requirement is given by the fact that we are only interested in equilibrium solutions, or more specifically, in solutions that represent a stationary state. The Jeans Theorem then implies that two phase-space cells $i$  and $j$  that happen to be located along the same orbit necessarily have to carry the same $f_i = f_j$. It is this constraint -- that the phase-space density is constant along orbits -- which gives equilibrium solutions of Eq.~(\ref{eq:CBE}) the form of orbit superpositions and allows us to use orbits as building blocks for the models. In practice, thus, the Jeans Theorem introduces many, many nonlinear equality constraint equations for our statistical model if we formulate it on the fundamental level of the $f_i$. These equations require the assumption of a gravitational potential and will change as soon as the parameters of the assumed mass distribution (like $M_\bullet$ etc) will change: the equilibrium distributions depend on the mass structure. Hence, the set of feasible points, i.e. the set of $f_i$ that fulfill the constraint equations will change as well. This shows that the selection parameters are not free parameters in the same sense as the $f_i$ because they are required to define the constraint equations that regulate the domain of the model. In fact, when these constraint equations change, one effectively deals with a different statistical model with a potentially different model complexity.

The above chain of arguments can be applied to any parameter that affects the gravitational potential and, thus, the stellar orbits, e.g. the stellar $\Upsilon$ or the DM halo parameters. It also holds for the inclination $i$, or more generally, for the viewing angles. One reason is that assuming different viewing angles changes the deprojection and with it the stellar contribution to the gravitational potential. Even if the latter is kept fixed, however, the (self-consistency) constraints $\mathbfit{d}_\mathrm{data}=\mathbfit{d}_\mathrm{mod}$ imposed on the model's stellar density do change (cf. App.~\ref{appendix:modelling}). This, in turn, means that the set of feasible points changes with the same consequences as discussed above.

Taking all this together it is clear that parameters like $M_\bullet, \Upsilon, i$ and the DM halo parameters on the one side and the orbital weights $\mathbfit{w}_i$ on the other, do not "operate" on the same level. It is conceptionally helpful to categorize them into two different parameter classes: selection parameters and minimization parameters. The selection parameters define a {\it family of models} $\mathcal{F}(M_\bullet,\Upsilon,i,\ldots)$. Each specific set of selection parameters defines a member of this family, i.e. a single statistical model for the measured LOSVDs. The {\it free parameters} of such a member model, i.e. the parameters to be minimised, are the orbital weights $w_i$ associated with the specific set of selection parameters. From one model to another the  $w_i$ are assigned to different sets of orbits and they cannot be easily interchanged with each other across different models.

As illustrated above treating the problem of finding the optimum selection parameters as a selection of multiple distinct candidate models is a straight-forward approach, whereas a parameter estimation approach using $\chi^2(M_\bullet,\Upsilon,i,\ldots,\mathbfit{w}_i)$ of a single global Schwarzschild model is rather ambiguous due to the dependence of the orbital weights on the selection parameters. Furthermore the model selection framework is nothing but a generalization of the parameter estimation via $\chi^2$ which would simplify to the latter if the multiple (sub-)models would stem from a single, statistical model (cf. Sec.~\ref{sec:Model_selection}). Similarly, the generalization of the Bayesian framework introduced by \citet[]{2006MNRAS.373..425M} to a (Bayesian) model selection framework is straight forward and model selection criteria such as the Akaike criterion can be derived from it by choosing a suitable prior which penalizes flexibility \citep[cf.][]{burnham_anderson_burnham_2002}. 

\subsection{Identifying the optimum minimization parameters}
\label{subsec:parameter_estimation}

Finding the optimal set of minimization parameters (i.e. orbital weights $\mathbfit{w}_i$) for a given set of selection parameters is in principle a direct parameter estimation problem. The goal is to minimise
\begin{equation}    
    \chi^2=\sum_{j}^{N_\mathrm{kin} \times N_\mathrm{vel}}\left(\frac{\mathbfit{l}_{\mathrm{obs},j}-\mathbfit{l}_{\mathrm{mod},j}}{\Delta \mathbfit{l}_{\mathrm{obs},j}}\right)^{2}.
	\label{eq:fitquality}
\end{equation}
where $N_{kin}$ is the number of spatially resolved bins and $N_{vel}$ the number of velocity bins in a single spatial bin. However, in Schwarzschild models one is often faced with the problem that the free parameters $\mathbfit{w}_i$ (i.e. the number of orbits calculated for a given trial mass distribution) outnumber the observational constraints $\mathbfit{l}_{\mathrm{obs},k}$. The related optimisation problem is therefore underconstrained. In order to be able to decide on a set of unique $\mathbfit{w}_{i}$ and simultaneously prevent overfitting it is common practice to introduce regularization in the fitting process. We do this by maximizing the entropy-like term proposed by \citet{1988ApJ...327...82R}:
\begin{equation}    
    \hat{S}=S-\alpha \chi^2
	\label{eq:regularization}
\end{equation}
where $\alpha$ is a regularization parameter and S is the Boltzmann entropy defined by:
\begin{equation}    
    S=-\sum_{i}^{M}w_{i}\cdot \ln{\left(\frac{w_{i}}{V_{i}}\right)}.
	\label{eq:entropy}
\end{equation}
Here, $V_{i}$ is the phase-space volume of orbit $i$ such that $w_i/V_i$ is the phase-space density along the orbit. The entropy term in eq.(~\ref{eq:regularization}) guarantees that the optimisation problem has a unique solution for the orbital weights $\mathbfit{w}$ (cf. the extensive discussion in \citealt{2020MNRAS.500.1437N}). The choice of equation~\ref{eq:entropy} as the smoothing function is somewhat arbitrary though. The motivation behind maximising eq.~(\ref{eq:entropy}) is that it yields $w_i \sim V_i$ and, hence, $f_i \approx \mathrm{const}$ (in the absence of other constraints). In other words, it tends to smooth the corresponding physical phase-space distribution function rather than the distribution of the $w_i$, which are only parameters. \citet{2020MNRAS.500.1437N} discuss how the more general form 
\begin{equation}    
    S=-\sum_{i}^{M}w_{i}\cdot \ln{\left(\frac{w_{i}}{\omega_{i}}\right)}
	\label{eq:general_entropy}
\end{equation}
can be used to explore the full range of all (possibly degenerate) solutions for a given set of kinematic observations. 

The regularization parameter $\alpha$ determines the smoothness of the model distribution function, meaning a small $\alpha$ implies the entropy term in equation~\ref{eq:regularization} dominates and the phase-space density is smooth, while a larger $\alpha$ leads to a better fit but increases the possibility of overfitting. This raises the question which amount of regularization is optimal to represent the phase-space density of the observed galaxy. Using Monte-Carlo simulations of isotropic rotator models \citet{2005MNRAS.360.1355T} describes an approach to estimate the optimum $\alpha$ for NGC 4807. However, such simulations would be required for every newly modelled galaxy as the optimum $\alpha$ depends on the obtained data and the underlying galaxy structure. In section~\ref{sec:regularization} we will come back to this issue and provide a method to determine a more optimal amount of smoothing by treating $\alpha$ as a selection parameter. In general, all the intrinsic parameters that control the behaviour of the models can be treated as selection parameters. Although we will not discuss it in full depth in this paper, another important selection parameter besides $\alpha$ is the number of orbits used in the models.

In conclusion, the modelling process can be described using two fundamentally different parameter layers that together define the properties of an orbit model. For the primary parameter layer the modeller has to choose a set of selection parameters, including mass and library parameters, which define the fundamental building blocks of a Schwarzschild model. In contrast, the minimization parameters that form the second layer, i.e. the orbital weights, are estimated using a form of parameter estimation for the given observational constraints. In our case this parameter estimation is performed by maximizing the entropy-like quantity in equation~\ref{eq:regularization}. This estimation process of the minimization parameters is only concerned with finding a good set of orbital weights. In fact, it is only possible if all selection parameters required for the calculation of the orbits are fixed and established, i.e. if a specific model out of the family of all possible models has been selected. Consequently one has to construct multiple trial models, if the goal is to optimize the selection parameters. To this end one has to sample the selection parameter space, estimate the best set of $w_{i}$ for each of the resulting orbits (using eq.~\ref{eq:regularization}), and compare the properties of each model afterwards. Since these models consist of different building blocks they are fundamentally distinct and one has to establish a framework to evaluate which one of them results in the best approximation of the galaxy's underlying structure.

As already stated above, the common approach to evaluate different models $\mathcal{F}_\mathcal{A}$ and $\mathcal{F}_\mathcal{B}$ with different selection parameters $M_\bullet, \Upsilon, \ldots$ based on  $\Delta \chi^2=\chi^2_{\mathcal{A}}-\chi^2_{\mathcal{B}}$ (cf. equation \ref{eq:fitquality}) is not optimal. We will show that judging models solely in terms of their goodness-of-fit can induce biases in the recovered selection parameters. The next section demonstrates such a case for a selection parameter that is very heavily impacted by this: The inclination of axisymmetric models.

\subsection{Biases in selection parameters: The inclination as an example}
\label{subsec:spherical}
We tested the Schwarzschild technique described in the previous section by applying it to a simulated spherical galaxy. For this purpose we created a number of mock LOSVDs obtained from an N-body model of an isotropic spherical galaxy \citep{1990ApJ...356..359H}. We sampled the sphere with $N=10^9$ particles and added Monte-Carlo noise to the corresponding LOSVDs. The underlying Hernquist density we used had a total mass $M=50\cdot10^{10}M_{\sun}$, a mass-to-light ratio $\Upsilon=1$ and an effective radius of $10$kpc. The stellar kinematics of the N-body was projected to a grid of 80 angular and radial bins reaching out to approximately 2.3 effective radii. In total we simulated 10 independent observations of this spherical galaxy by adding random uncorrelated Gaussian noise with a standard deviation of two percent of the maximum LOSVD value in the respective spatial bin. Quantifying this noise in terms of the Gauss-Hermite coefficients \citep[cf.][]{1993ApJ...407..525V} this translates to an average error of $\Delta \sigma / \sigma \approx 2\%$ and $\Delta h_{4}\approx0.02$. The resulting Gauss-Hermite coefficients of one of these mocks are shown in Fig.~\ref{fig:spherical_kinmap} and \ref{fig:spherical_central_kinmap} of the appendix. Without the Gaussian noise the N-body models follow the semi-analytic Gauss-Hermite profiles of \citep[cf.][]{2005A&A...432..411B} very closely, which means that the odd Gauss-Hermite coefficients are zero everywhere, the velocity dispersion $\sigma$ has a maximum at intermediate radii and $h_4$ is slightly negative but increases significantly for $r \to 0$.     

\begin{figure}
	\includegraphics[width=\columnwidth]{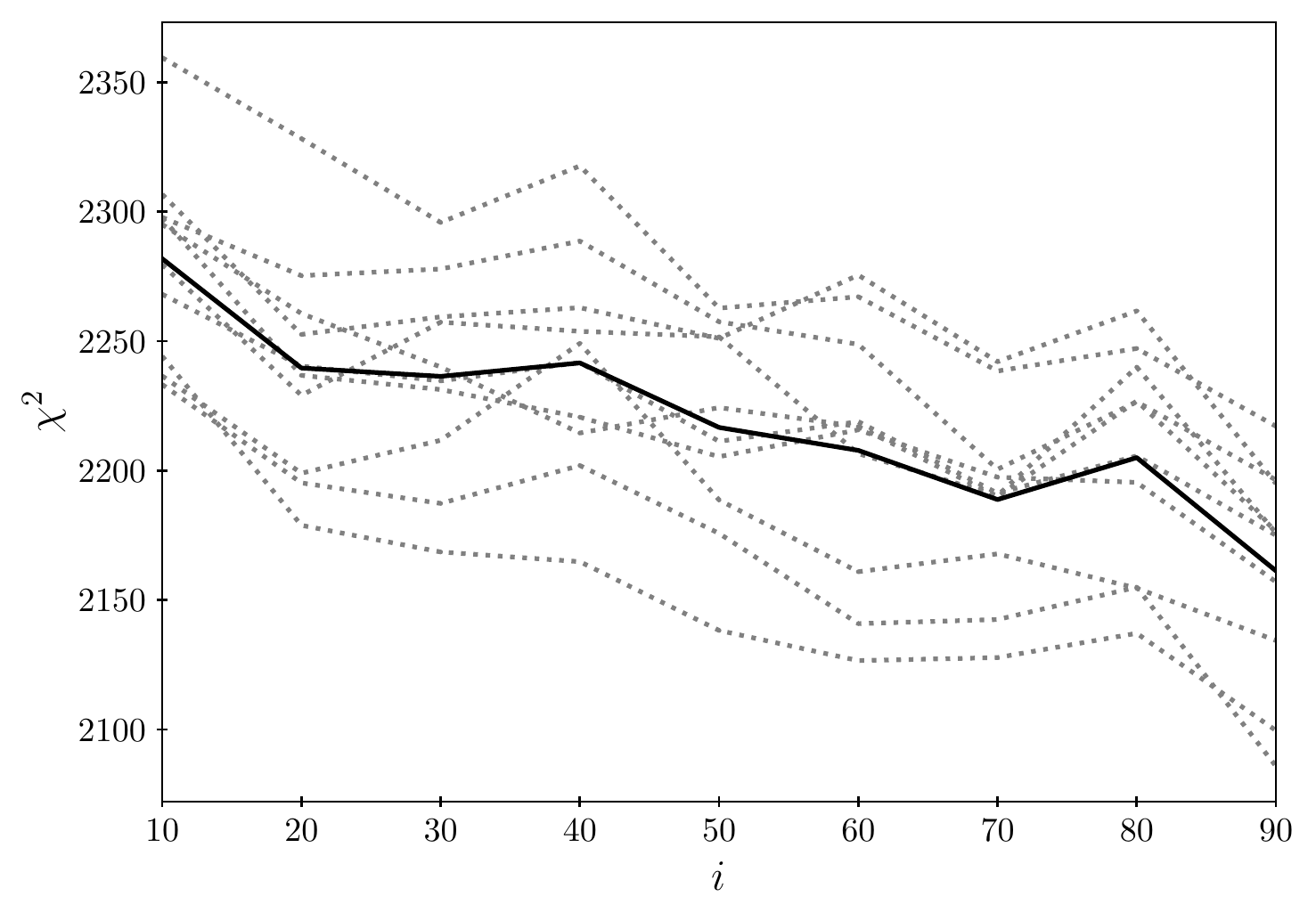}
    \caption{Axisymmetric Schwarzschild fits to a spherical N-body simulation. \textit{Dashed, grey lines} represent fits to different mock data sets obtained by adding random noise to the same original spherical model kinematic. The \textit{solid line} is the corresponding arithmetic mean. The number of kinematic data points is $N_{\mathrm{data}}=2640$. Even though the input model is spherical, the $\chi^2$ varies as a function of the inclination assumed in the fit.}
    \label{fig:spherical_chi2}
\end{figure}

We modelled the spherical mocks with differently inclined, axisymmetric Schwarzschild models using the same spherical Hernquist density that was used to create the N-body in the first place. Thus, effectively we used the same orbit library for all models, yet the orbits were projected under different viewing angles. The resulting $\chi^2$ values as function of the inclination are shown in Fig.~ \ref{fig:spherical_chi2}.

All 10 mocks exhibit a significant bias towards edge-on models, implying that the ability of an orbit model to fit data depends on the angle it is observed at. The reason for this behaviour is the axisymmetry of our Schwarzschild models: Every orbit in our model exists in a prograde and a retrograde version. In the case of an edge-on model the LOSVD of the prograde and retrograde version have opposite signs and are each a unique contribution to the model's LOSVDs in equation~\ref{eq:model LOSVD}. This is contrasted by a face-on model where the prograde and retrograde version are identical in projection, thus effectively reducing the number of unique base functions in equation~\ref{eq:model LOSVD}. This weakens the flexibility of our model LOSVDs to fit the observations and consequently leads to a higher $\chi^2$ of face-on models when compared to their edge-on counterparts. Since this effect is intrinsic to the modelling technique, the edge-on bias will also be present when non-spherical galaxies are modelled, thus diminishing the possibility of successfully constraining the actual inclination of a galaxy. Therefore we should aim to quantify the variable model flexibility and use that to correct our results for the inclination constraints.

\section{Model flexibility}
\label{sec:ansaetze}
In section~\ref{sec:overview+problem} we showed that evaluating Schwarzschild models based on their $\chi^2$ can suffer from a bias if models with varying flexibility are compared. Therefore we    introduce two methods in section~\ref{subsec:bootstrapping} with the goal to estimate this flexibility, followed by an outline of three model selection approaches in section~\ref{sec:Model_selection} which exploit this information to improve the constraining power of our models.  

\subsection{Estimating the model flexibility in Schwarzschild models}
\label{subsec:bootstrapping}
In statistical modelling a model's general ability to fit data is quantified by its number of free parameters m, because every free parameter reduces the degrees of freedom of the $\chi^2$ distribution. This means if we assume uncorrelated Gaussian noise and quantify the deviations of N observed data points to a statistical model with m free parameters then the expectation value of the respective $\chi^2$ will be $E(\chi^2)=N-m$ with variance $Var(\chi^2)=2(N-m)$. Therefore it is only appropriate to compare models using $\chi^2$ if they have the {\it same number of free parameters}.    

For linear models without prior constraints the number of free parameters is a well defined property that can be quantified using a linear algebra framework \citep[e.g.][]{2013}. However, our Schwarzschild models are non-linear due to the entropy term introduced in the estimation process of the orbital weights (equation \ref{eq:regularization}). But even if they were linear the estimation of the number of free parameters would be a non-trivial task because we demand the orbital weights $w_i$ to be non-negative. This requirement is a prior constraint that limits the accessible parameter-space and thus reduces the flexibility of the models in an unpredictable fashion \citep[e.g.][]{2010arXiv1012.3754A}. This is because the amount by which priors reduce the flexibility of a model depends on the data that are getting fitted, i.e. in this case on the structure of the galaxy under study and on the noise pattern of the data. Therefore we need a more generalized approach to quantify the actual flexibility of our Schwarzschild models, like the definition introduced by \citet{doi:10.1080/01621459.1998.10474094} which extends to non-linear models (see details below). 

We call the quantity describing this flexibility the number of effective free parameters $m_{\mathrm{eff}}$. Both, the non-linearity and the non-negativity prior affect the flexibility of the Schwarzschild models in a complicated fashion. This prevents a direct calculation of the model flexibility and we rely on bootstrap iterations to estimate $m_{\mathrm{eff}}$. After having fitted a Schwarzschild model to the observed LOSVDs we start a number of $K$ bootstrap iterations. In each iteration we add random Gaussian noise to the original fit $\mathbfit{l}_\mathrm{fit}$. The standard deviation of this artificially added noise is based on the observational error $\Delta \mathbfit{l}$. When $\mathbfit{l}_\mathrm{fit}$ is a statistically "good" fit, i.e. if it can quantitatively explain the actual data in line with the estimated observational errors, then the new bootstrap data $\mathbfit{l}_\mathrm{bootdata}$ should mimick the observed data: each bootstrap data point is redrawn from the (assumed) same distribution as the corresponding real data point. If the model is not a good fit (e.g. if the mass parameters are completely wrong) this assumption is not valid, fortunately such models can be discarded easily and are of no further interest. Given this bootstrap assumption (i.e. the bootstrap samples are representative of the observed data), we can estimate the flexibility of the original fit model by fitting each of the $K$ bootstrap data sets with the same selection parameters that were used for the original fit. We denote bootstrap fits as $\mathbfit{l}_\mathrm{bootfit}$. We tested two independent methods to estimate $m_{\mathrm{eff}}$ based on these bootstrap fits. The first one measures the reduction of $\chi^2$ directly by calculating the $\chi^2$ before the bootstrap fit 
\begin{equation}    
    \chi^2_{\mathrm{prior}}=\sum_{j=1}^{N}\left(\frac{\mathbfit{l}_\mathrm{bootdata,i}-\mathbfit{l}_\mathrm{fit,i}}{\Delta \mathbfit{l}_i}\right)^{2}
	\label{eq:chi2_prior}
\end{equation}
and after it:
\begin{equation}    
    \chi^2_{\mathrm{posterior}}=\sum_{j=1}^{N}\left(\frac{\mathbfit{l}_\mathrm{bootdata,i}-\mathbfit{l}_\mathrm{bootfit,i}}{\Delta \mathbfit{l}_i}\right)^{2}
	\label{eq:chi2_posterior}
\end{equation}
We can then exploit the expectation values $E(\chi^2_{\mathrm{prior}})=N$ and $E(\chi^2_{\mathrm{posterior}})=N-m_{\mathrm{eff}}$ to estimate $m_{\mathrm{eff}}$ by averaging over all bootstrap iterations $K$:
\begin{equation}
m_{\mathrm{eff}}=\frac{1}{K}\sum_{k=1}^{K}m_k=\frac{1}{K}\sum_{k=1}^{K}(\chi^2_{\mathrm{prior},k}-\chi^2_{\mathrm{posterior},k})
	\label{eq:delta_chi2}
\end{equation}
An alternative approach to estimating $m_{\mathrm{eff}}$ is to calculate:
\begin{equation}
m_{\mathrm{eff}}=\frac{1}{K}\sum_{k=1}^{K}\sum_{i=1}^N\frac{1}{\Delta \mathbfit{l}_i^2}\left(\mathbfit{l}_\mathrm{bootfit,k,i}-\mathbfit{l}_\mathrm{fit,i}\right)\left(\mathbfit{l}_\mathrm{bootdata,k,i}-\mathbfit{l}_\mathrm{fit,i}\right)
	\label{eq:covariance}
\end{equation}
If we assume that the expectation values of the bootstrap fit and data are given by $E\left(\mathbfit{l}_\mathrm{bootfit,i}\right)=E\left(\mathbfit{l}_\mathrm{bootdata,k,i}\right)=\mathbfit{l}_\mathrm{fit,i}$ equation~\ref{eq:covariance} equals the sum of normalized covariances:  
\begin{equation}
m_{\mathrm{eff}}= \frac{1}{K}\sum_{k=1}^{K}\sum_{i=1}^N \frac{cov\left(\mathbfit{l}_\mathrm{bootfit,k,i},\mathbfit{l}_\mathrm{bootdata,k,i}\right)}{\Delta \mathbfit{l}_i^2}
	\label{eq:covariance_2}
\end{equation}
In this form $m_{\mathrm{eff}}$ is equivalent to the concept of generalized degrees of freedom for non-linear models developed by \citet{doi:10.1080/01621459.1998.10474094}. The above bootstrap approach is versatile and can be adopted to estimate the flexibility of very complex statistical models such as orbit superposition models. However, it comes at the cost of computational performance as it requires several additional model fits, i.e. identifying the optimum orbital weights (cf. Sec.~\ref{subsec:parameter_estimation}) for each of the constructed bootstrap data sets. Fortunately, though, the bootstrap iterations do not require additional orbit integrations as one can reuse the orbit library of the original model fit. 

\begin{figure}
	\includegraphics[width=\columnwidth]{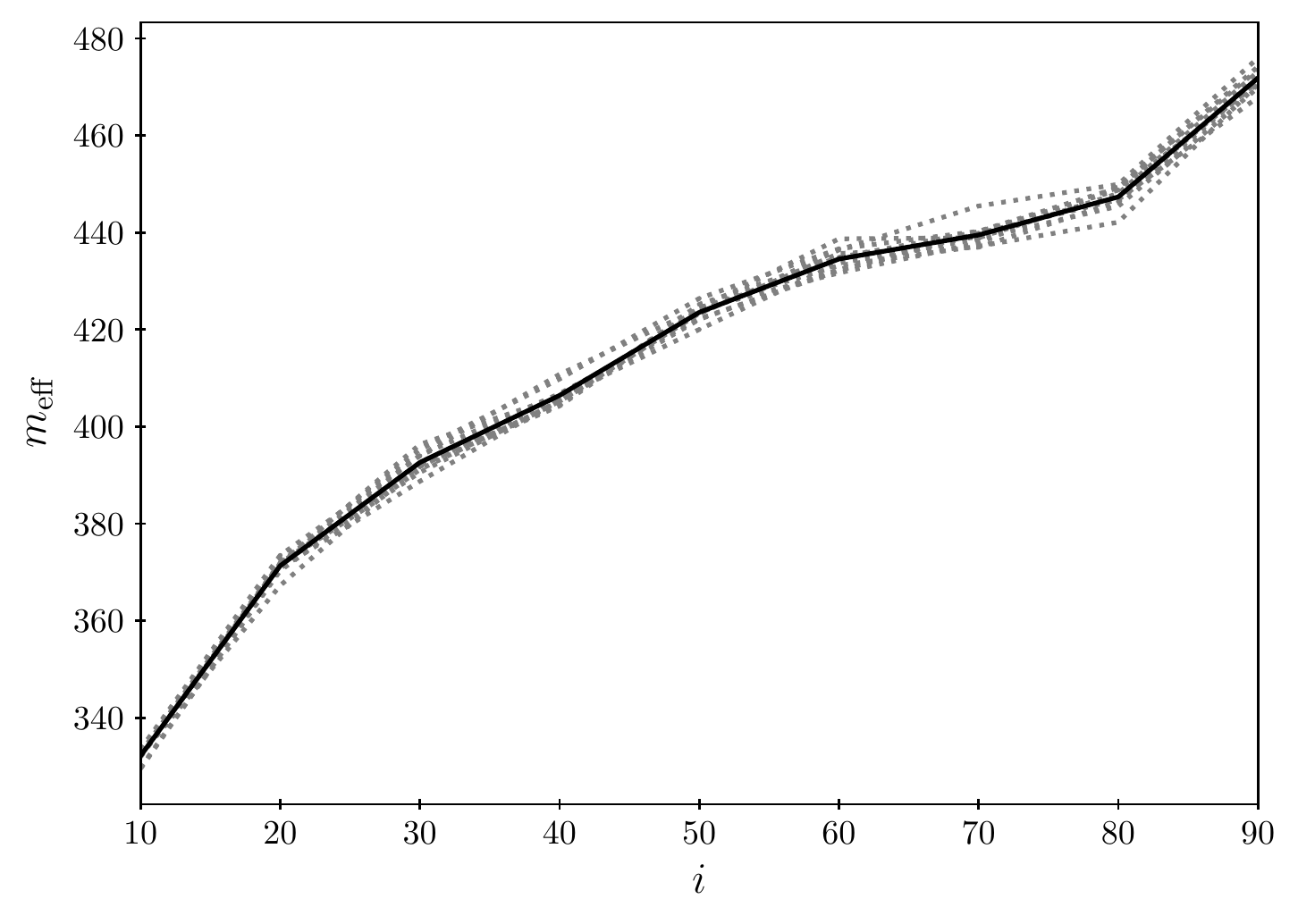}
    \caption{The number of effective parameters $m_{\mathrm{eff}}$ as estimated by Eq.~\ref{eq:covariance} of the models shown in Fig~\ref{fig:spherical_chi2}. The unexpected $\chi^2$ variation in Fig.\ref{fig:spherical_chi2} can be explained by the varying model flexibility.}
    \label{fig:spherical_m}
\end{figure}

\begin{figure}
	\includegraphics[width=\columnwidth]{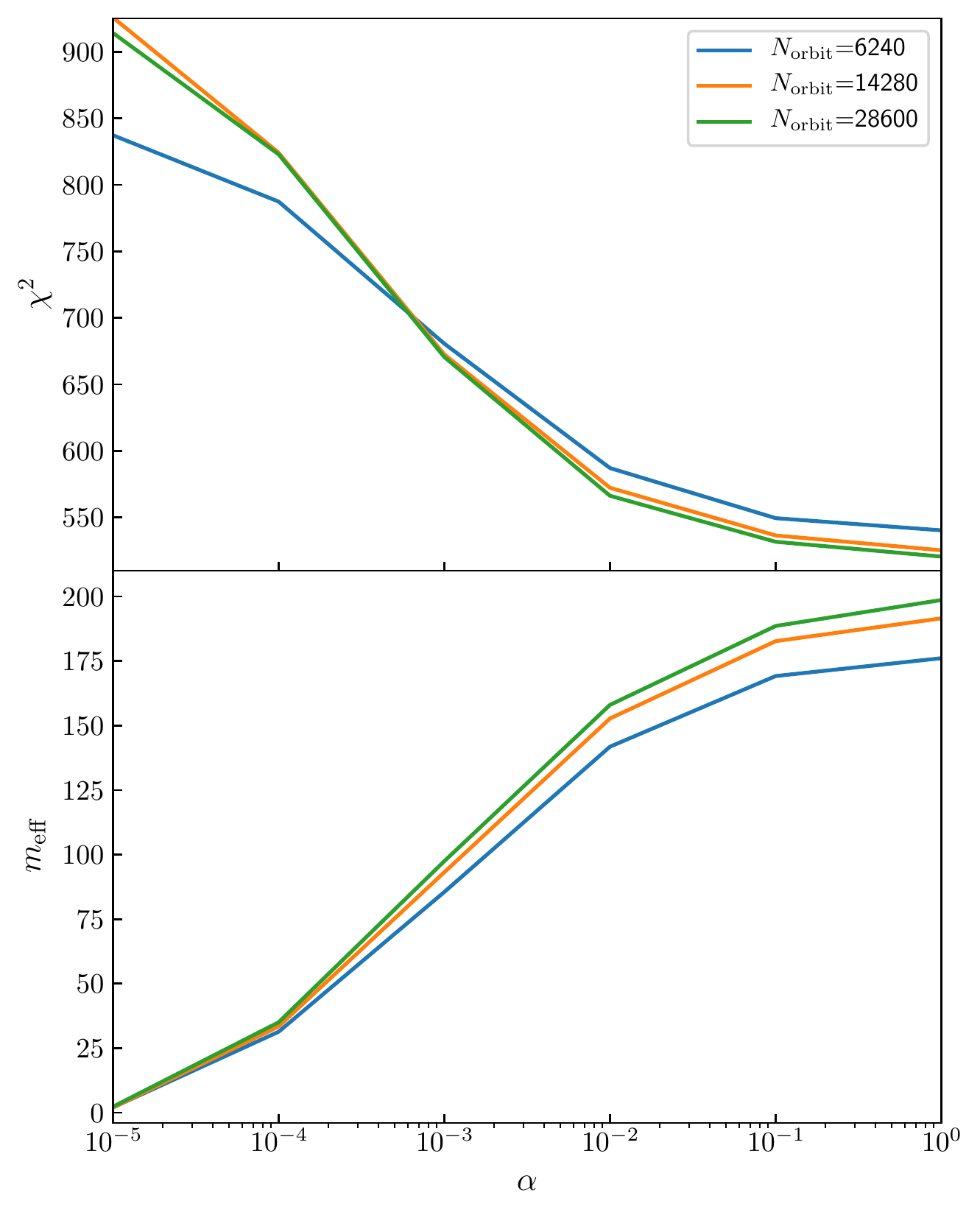}
    \caption{\textit{Top panel}: $\chi^2$ as a function of the regularization parameter $\alpha$ for three orbit libraries with different number of orbits $N_{\mathrm{orbit}}$. A set of 782 kinematic data points of a \textit{single} mock of an edge-on toy galaxy is being fitted by the orbit models. All orbit models are viewed edge-on and employ the correct density and gravitational potential that was used to create the mock data in the first place. \textit{Bottom panel}: The same as the top panel but for the number of effective parameters $m_{\mathrm{eff}}$. More orbits and less regularization lead to an increase in the intrinsic model flexibility, resulting in correspondingly better fits.}  
    \label{fig:regularization}
\end{figure}

As a first proof-of-concept we applied our bootstrap methods to simple polynomial function models where the number of free parameters (the number of polynomial coefficients to be recovered) can be counted. We found that equations~\ref{eq:delta_chi2} and~\ref{eq:covariance} are equivalent within the adopted numerical precision: both can be used to calculate the number of free parameters correctly. For the more complex Schwarzschild models we found a slight offset, which is likely because the condition $E(f_{i})=\mathbfit{l}_\mathrm{fit,i}$ is not fulfilled for all data points when the orbital weights are forced to be non-negative but the bootstrap noise is assumed to follow a Gaussian distribution (which sometimes implies negative LOSVD values). Fortunately, this offset is approximately constant thus our results do not depend on the chosen estimation approach. For the rest of the paper, we use the covariance approach (eq.~\ref{eq:covariance}). 
We applied both estimates to the modelling of the spherical toy galaxy introduced in section~\ref{subsec:spherical} using $K=50$ bootstrap iterations. The resulting $m_{\mathrm{eff}}$ as a function of the inclination are depicted in Fig.~\ref{fig:spherical_m}. We found that the behaviour of $\chi^2$ is mirrored by the number of effective parameters $m_{\mathrm{eff}}$: where the $\chi^2$ gets lower, $m_{\mathrm{eff}}$ increases. In fact, the $\chi^2$ variation in Fig.~\ref{fig:spherical_chi2} can be entirely explained by $m_{\mathrm{eff}}$. This supports our hypothesis that the apparent edge-on bias is caused by the varying flexibility of the axisymmetric models with the inclination. 
Using the above bootstrap estimation methods we also identified a multitude of other factors influencing the general ability of a Schwarzschild model to fit a given set of data. The most dominant other selection parameter is the regularization value $\alpha$. In Fig.~\ref{fig:regularization} $m_{\mathrm{eff}}$ is shown as a function of $\alpha$ for three edge-on orbit libraries with different number of orbits $N_{\mathrm{orbit}}$. All three libraries are constructed with the correct gravitational potential that was used to create the mock observation they attempt to fit. A decrease in $\alpha$ suppresses the freedom of the orbital weights and, thus, restricts the model's flexibility. In the limit of $\alpha \to 0$ the model becomes completely rigid resulting in $\chi^2\approx N$. For very large values of $\alpha$ both, $\chi^2$ and $m_{\mathrm{eff}}$, appear to converge to a constant value. We will come back to the regularization in Sec.~\ref{sec:regularization}. An increase in orbits typically leads to a (non-linear) increase in model flexibility, resulting in correspondingly smaller $\chi^2$ values. However, for very small $\alpha$ the orbit library with only 6240 orbits achieves a better $\chi^2$ even though its intrinsic flexibility $m_{\mathrm{eff}}$ is consistently lower than that of the other libraries. This is because all orbit models are very rigid for such small values of $\alpha$ and the kinematic data of the toy galaxy shown in Fig.~\ref{fig:regularization} was generated using a maximum entropy Schwarzschild model with exactly $N_{\mathrm{orbit}}=6240$. We will describe the construction of such mocks in more detail in Section~\ref{subsec:hernquist}.

\section{Selection parameter optimisation via model selection}
\label{sec:Model_selection}
The effective parameters quantify the model flexibility that is synonymous with the size of the parameter subspace accessible to the orbital weights. In principle any selection parameter can influence this freedom, meaning models with different sets of selection parameters should not be be compared without taking into account their actual flexibility. While we showed how we can estimate this flexibility $m_{\mathrm{eff}}$ in the last Sec.~\ref{sec:ansaetze} the question remaining is how we can use this information to choose the model with the best set of selection parameters. Since this means we need to choose the "best" model out of a pool of models with different flexibilities the only option is to work within a model selection framework. We tested three different model selection approaches. The first and what we call intuitive approach is based on the following assumption: if we want to treat all models a priori equivalent then we should demand that the expectation value of our evaluation statistic should be identical for all models. Therefore we should minimize: 
\begin{equation}
\chi^2+m_{\mathrm{eff}}
	\label{eq:intuitve}
\end{equation}
because $E(\chi^2+m)=N_{\mathrm{data}}$ holds for all models that are able to emulate the observed data. The second, information based approach is to maximize the Akaike information criterion (AIC) or equivalently to minimize:
\begin{equation}
\chi^2+2m_{\mathrm{eff}}
\label{eqn: Akaike}
\end{equation}
This means that the AIC approach penalizes flexible models more than the intuitive approach in equation~\ref{eq:intuitve} does. Since both of the above approaches differ only by the relative importance of the model flexibility $m_{\mathrm{eff}}$, our third, more generalized selection approach is to minimize:  
\begin{equation}
\chi^2+w_m m_{\mathrm{eff}}
\label{eqn: calibrated}
\end{equation}
where the factor $w_m$ is a free parameter calibrated using a set of simulations described in the following section \ref{sec:simulated}. Eq.~\ref{eqn: calibrated} includes the intuitive approach ($w_m=1$) and the AIC approach ($w_m=2$) as special cases. 

\section{Application: Inclination recovery in simulated galaxies}
\label{sec:simulated}
We created a number of toy galaxies with different inclinations to test the model selection framework proposed in the previous section. The goal was to recover their true inclination. Analytical distribution functions for non-spherical models are mostly not very realistic in terms of the orbital anisotropy they imply.  Consequently we decided to create the noise-free LOSVDs of our toy galaxies on the basis of Schwarzschild models in realistically flattened 3d mass distributions. This approach is very flexible as modifications to the entropy term can be used to generate almost any desired orbital anisotropy (cf. \citealt{2020MNRAS.500.1437N}). We cross-checked the validity of this toy model generation on the Hernquist sphere: the modelling results obtained using either LOSVDs from the $N$-body-particle sampled analytic distribution function (Sec.~\ref{subsec:spherical}) or from suitable Schwarzschild models in the Hernquist potential did not show any significant differences. Using this Schwarzschild model construction we simulated the kinematics of both, late-type and early-type galaxies. The former with maximum-entropy models of a realistic mass distribution obtained by deprojecting the surface brigthness profile of a real disk galaxy and the latter with models of a flattened Hernquist distribution. Table~\ref{tab:toy_gal_table} is a comprehensive list of all toy galaxies we investigated and their most relevant characteristics. To create realistic mock observations of these toy galaxies the stellar kinematics of the resulting orbit model must be projected to the sky. To this end we projected the kinematics to spatial grids borrowed from real SINFONI and MUSE observations \citep[e.g.][]{2019ApJ...887..195M} and convolved them with the corresponding seeing. For the early-type galaxies we used a Voronoi grid that covers a large FoV and extends well beyond the toy galaxy's effective radius. To project the late-type galaxies we used regular grids with a smaller FoV but higher spatial resoultion. The FoV is still large enough to cover the sphere of influence of a hypothetical black hole with a mass typical for the respective galaxy. We simulated 10 independent observations of each toy galaxy by adding random uncorrelated gaussian noise with a standard deviation of two percent of the maximum LOSVD value in the respective spatial bin (cf. Sec.~\ref{subsec:spherical}). Since we generated rotation in some of the toy galaxies (see below) not all of them intrinsically represent a maximum-entropy state. 
\begin{table*}
	\centering
	\caption{Table of tested toy galaxies with different intrinsic axis ratios $q$, angular momentum biases $\lambda$ and inclinations $i$. Beyond that, we varied the velocity resolutions $\Delta v_{\mathrm{los}}$ of the LOSVDs for some models. We used three different spatial grids for the projected kinematic of these galaxies that are typical for modern observations with IFU spectrographs. The wide field FoV covers at least the kinematics within the galaxy's effective radius. The other two grids have smaller FoVs, however even the smallest field comfortably covers a hypothetical sphere of influence of a black hole typical for the size of such a toy galaxy. Furthermore all three spatial grids have increased resolution in the centre where such a sphere influence would be located.}
	\label{tab:toy_gal_table}
	\begin{tabular}{lcccccc} 
		\hline
		Name  & Density & Axis ratio q & Angular momentum bias $\lambda$ & Inclination & $\Delta v_{\mathrm{los}}[km/s]$ & Spatial grid\\
		\hline
     Galaxy A & Early-type &  0.6 & 0.0 & $60^{\circ}$ & 76 & wide field\\ 
     Galaxy B & Early-type  &  0.4 & 0.0 & $70^{\circ}$ & 76 & wide field\\ 
     Galaxy C & Early-type  &  0.6 & 0.0 & $45^{\circ}$ & 76 & wide field\\ 
     Galaxy D & Early-type  &  0.6 & 0.5 & $60^{\circ}$ & 76 & wide field\\ 
     Galaxy E & Early-type  &  0.6 & 1.0 & $60^{\circ}$ & 76 & wide field\\ 
     Galaxy F & Early-type  &  0.6 & 0.0 & $60^{\circ}$ & 65 & wide field\\ 
     Galaxy G & Early-type  &  0.6 & 0.0 & $60^{\circ}$ & 54 & wide field\\ 
     Galaxy H & Late-type & $\sim$0.2 & 1.0 & $55^{\circ}$ & 30 & intermediate field \\ 
     Galaxy I & Late-type & $\sim$0.2 & 0.0 & $55^{\circ}$ & 30 & intermediate field\\  
     Galaxy J & Late-type & $\sim$0.2 & 1.0 & $55^{\circ}$ & 30 & small field\\ 
     Galaxy K & Late-type & $\sim$0.2 & 0.0 & $55^{\circ}$ & 30 & small field\\ 
		
	\end{tabular}
\end{table*}

After the addition of noise, all resulting mock observations were modelled with Schwarzschild models using $6240$ orbits and a regularization parameter of $\alpha=1.67$. For our toy galaxies this regularization value was large enough to ensure that $\chi^2(\alpha)$ has roughly converged to a constant value, in other words the flexibility of the respective models has plateaued and a further increase in $\alpha$ does not improve the goodness-of-fit significantly. To recover the mock galaxies' inclinations, we had to model them under different assumed viewing angles. To do so, we projected the true intrinsic density of each toy galaxy on the sky and deprojected the resulting mock images assuming the various inclinations probed by the dynamical models. For these deprojections we used the Metropolis-Algorithm of \citet{1999MNRAS.302..530M}. It is usually not required to probe every viewing angle as not all inclinations are necessarily compatible with a given photometry. In axisymmetric systems in particular the observed flattening on the sky determines a minimum possible inclination that corresponds to an infinitely thin distribution. But even for triaxial galaxies the range of plausible viewing angles can be narrowed down  \citep[cf.][]{2020MNRAS.496.3076D}. 

\subsection{Simulated galaxies}
\label{subsec:hernquist}
The early-type toy galaxies have a mass-to-light ratio $\Upsilon=1$ and are based on the flattened density profile:
\begin{equation}
\rho\left(a\right)=\frac{M}{2\pi} \frac{a_{\mathrm{scale}}}{q} \frac{1}{a(a+a_{\mathrm{scale}})^3}
\label{eqn: Hernquist}
\end{equation}
which is similar to the Hernquist sphere \citep[cf.][]{1990ApJ...356..359H}.
The variable \textit{a} parametrizes the spheroidal isodensity surfaces. While not all early-type toy galaxies have the same intrinsic axis ratio q, their total mass $M=5\cdot 10^{11}M_{\sun}$, effective radius $r_{\mathrm{eff}}=10$kpc and distance $d=141.8$Mpc are always the same. 
The density for the late-type toy galaxies was obtained by deprojecting the photometry of a real late-type galaxy (NGC 3489, cf. \citealt{2010MNRAS.403..646N}). Beyond that we also borrowed the assumed inclination ($i=55\degr$), distance ($d=12.1$Mpc) and seeing from said galaxy for the setup of the late-type galaxies. 

After establishing the density distribution for each galaxy we generated its kinematics by constructing maximum-entropy Schwarzschild models based on this density. The kinematics of the models were then projected onto the sky and convolved with realistic spatial point spread functions (PSF). The (non-parametric) LOSVDs for the early-type galaxies were projected onto a set of 34 Voronoi bins using the method of \citet{2003MNRAS.342..345C}. To make the mock data as realistic as possible, this observational setup and the respective PSFs were borrowed from real SINFONI and MUSE observations of a massive elliptical galaxy (ESO 325-G004). The mock SINFONI observations cover the inner $1\arcsec$ with high enough spatial resolution to resolve a central supermassive black hole while the corresponding MUSE observations encompass the larger-scale kinematics of the galaxy out to about 2 half-light radii. The typical size of these Voronoi bins varies with radius from about $\sim 0.1 \, \mathrm{kpc}$ (or $\sim 0.17 \, \arcsec$) in the centre to $\sim 3 \, \mathrm{kpc}$ (or $\sim 4 \, \arcsec$) in the outer bins. An example Gauss-Hermite map for the MUSE grid of one of the resulting mock observations is shown in Fig.~\ref{fig:GalaxyD_muse} in App.~\ref{appendix:kinmaps}. For the projection of the Schwarzschild kinematics of late-type galaxies we tried two regular grids with differently sized FoVs (cf. Tab.~\ref{tab:toy_gal_table}) resulting in 30 and 70 spatial bins respectively. For the PSF convolution we borrowed the observed PSFs of the same late-type galaxy NGC 3489 that we used to create the intrinsic density distribution.

All maximum entropy galaxies display characteristic features of non-rotating flattened systems such as smaller observed velocity dispersion $\sigma$ and $h_4$ on the minor axis, but overall they behave similar to the spherical toy galaxy of Sec.~\ref{subsec:spherical} in that their LOSVDs are symmetric and have no net rotation. Since this is not representative of most real galaxies \citep[cf.][]{2011MNRAS.414..888E}, we also modelled rotating toy galaxies with the goal to investigate how this affects the inclination recovery. We introduced rotation by exploiting the symmetry of our axisymmetric orbit libraries: each orbit of our libraries exists in a prograde and a retrograde version. In a maximum-entropy model, the weights of the two orbits in a prograde-retrograde pair, denoted by $w_i^+$ and $w_i^-$, are identical and consequently the orbits cancel each other's rotation signal. We can simply manipulate this prograde-retrograde balance to create a toy galaxy with net rotation by reassigning new orbital weights:
\begin{equation}
\begin{split}
    \hat{w}^{+}_{i}&=\frac{1+\lambda}{2}\cdot (w^{+}_{i}+w^{-}_{i}),\\
    \hat{w}^{-}_{i}&=\frac{1-\lambda}{2}\cdot (w^{+}_{i}+w^{-}_{i})
\end{split}
\label{eqn: rotation}
\end{equation}
Where the free parameter $\lambda \in [-1,1]$ biases the total angular momentum $L_z$ along the library's symmetry axis. For $\lambda=1$ only prograde orbits contribute to the observed kinematics, resulting in a maximum positive rotation signal, while $\lambda=-1$ implies a maximum negative rotation signal as only retrograde orbits are populated. We can also recover our original non-rotating, maximum entropy model by choosing $\lambda=0$. Using this weight manipulation we created two rotating mock galaxies (see Table~\ref{tab:toy_gal_table}) with an angular momentum bias of $\lambda=1$ (Galaxy E) and $\lambda=0.5$ (Galaxy D). In addition to the intrinsic axis ratio q, the inclination i and the angular momentum bias of our toy galaxies we also experimented with a change of the velocity binning $\Delta v_{\mathrm{los}}$ by altering the maximum velocity $v_{\mathrm{max}}$ of the LOSVDs while keeping the number of velocity bins $n_{\mathrm{vel}}$ constant. 

Qualitatively all simulated galaxies showed similar behaviour when modelled under different assumed inclinations. Therefore we first aim to outline this common behaviour using Galaxy D as a generic case. This Galaxy has an axis ratio $q=0.6$ and was originally projected at an angle $i=60\degr$, resulting in an apparent axis ratio $q'\approx0.72$. In axial symmetry then, inclinations $i\lesssim44\degr$ are not compatible with the projected surface brightness and do not have to be considered. When modelling the galaxy, we sampled the inclination parameter linearly in the interval $[50^{\circ},90^{\circ}]$ with a step size of $\Delta i=10^{\circ}$.  
\begin{figure}
	\includegraphics[width=\columnwidth]{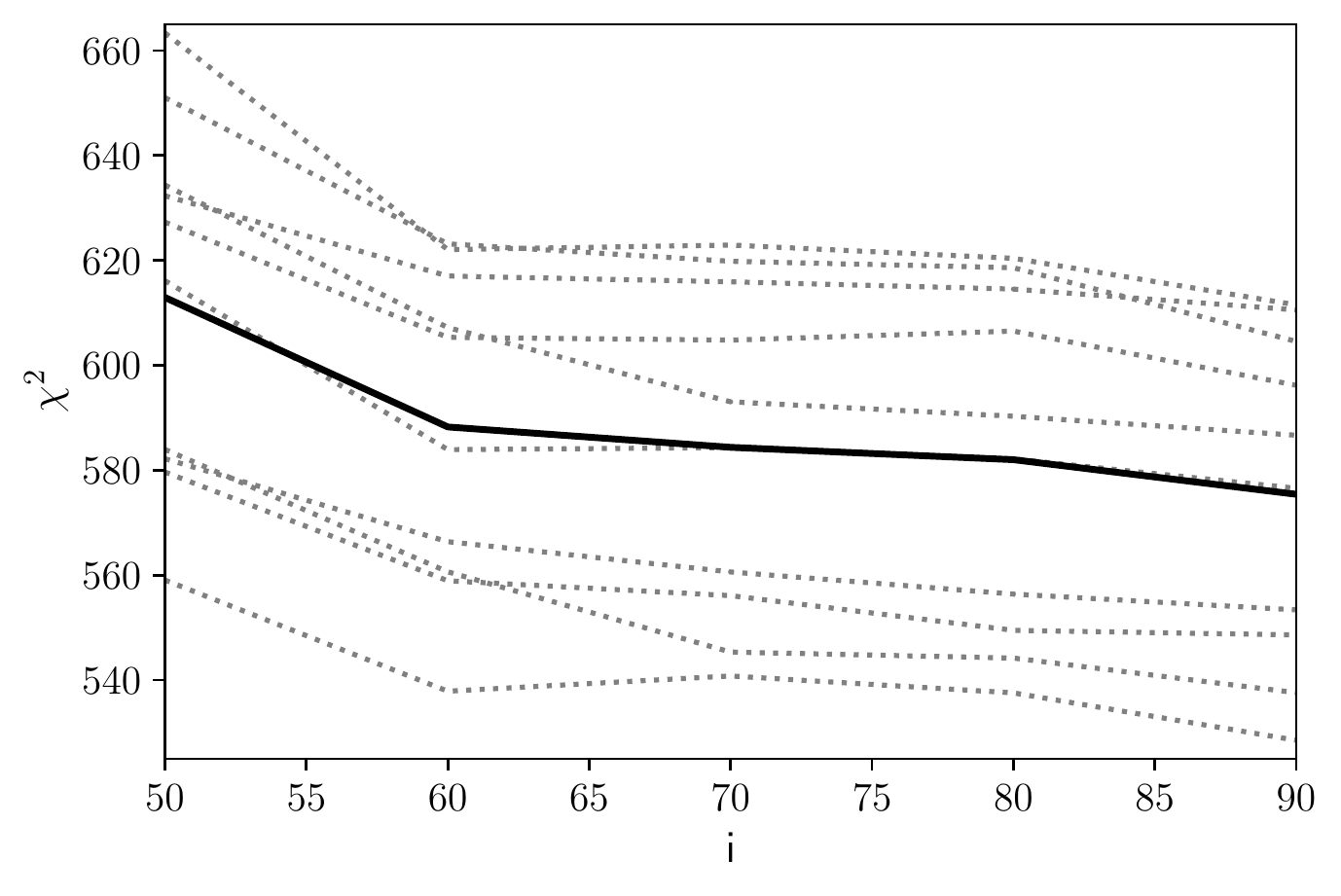}
    \caption{$\chi^2$ as a function of the assumed library inclination when modelling toy galaxy D. \textit{Dotted lines:} $\chi^2$ for the individual mocks. \textit{Solid line:} The arithmetic mean $\chi^2$ of all 10 mocks. The true inclination of the galaxy is $i=60 \degr$. All models had $6240$ orbits and a fixed regularization parameter $\alpha=1.67$.}
    \label{fig:GalaxyD_chi2}
\end{figure}

\begin{figure}
	\includegraphics[width=\columnwidth]{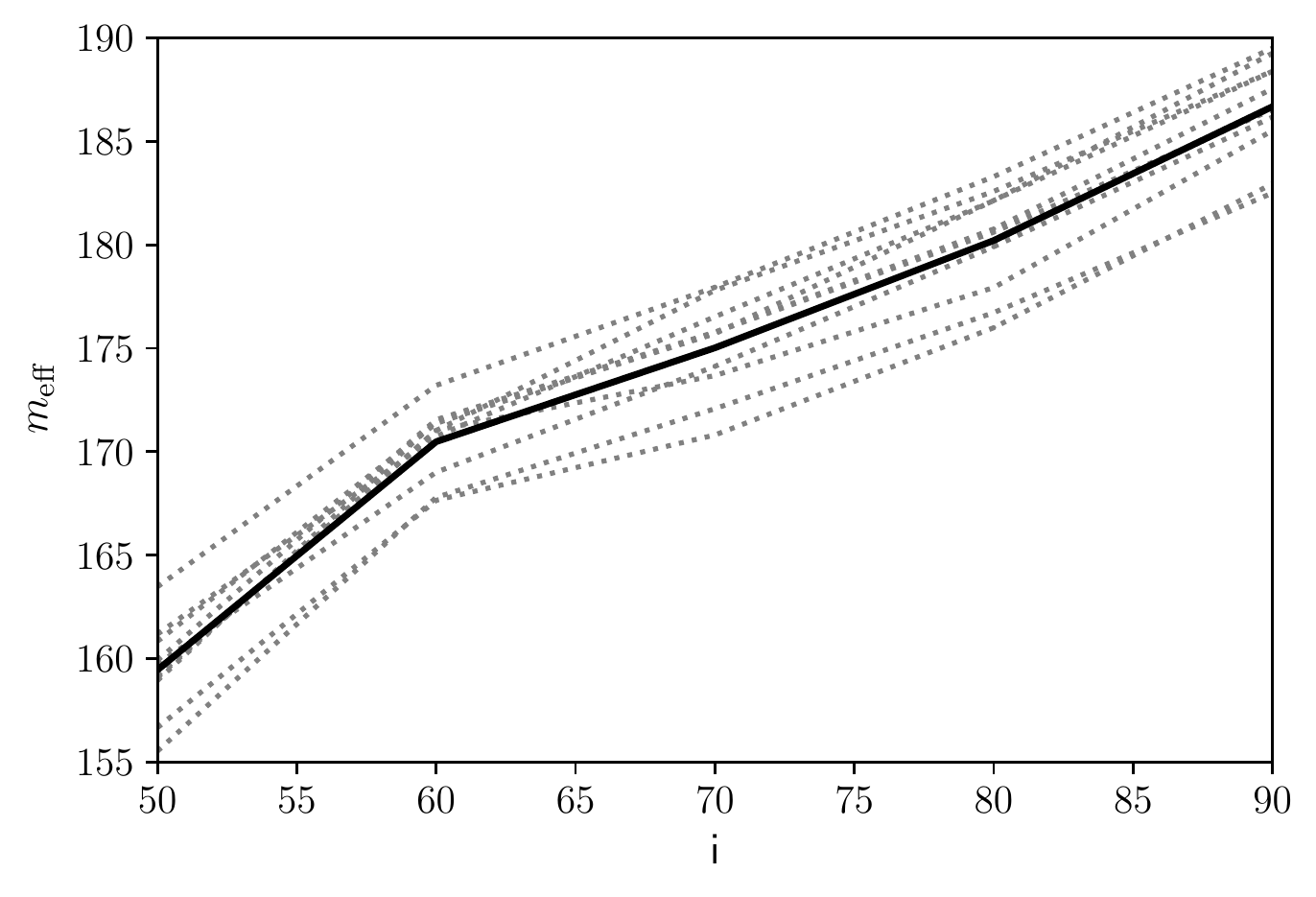}
    \caption{As Fig.~\ref{fig:GalaxyD_chi2} but for the number of effective parameters $m_{\mathrm{eff}}$ of the model fits to the mocks of Galaxy D that are shown in Fig.~\ref{fig:GalaxyD_chi2}.}
    \label{fig:GalaxyD_parameters}
\end{figure}

\begin{figure}
	\includegraphics[width=\columnwidth]{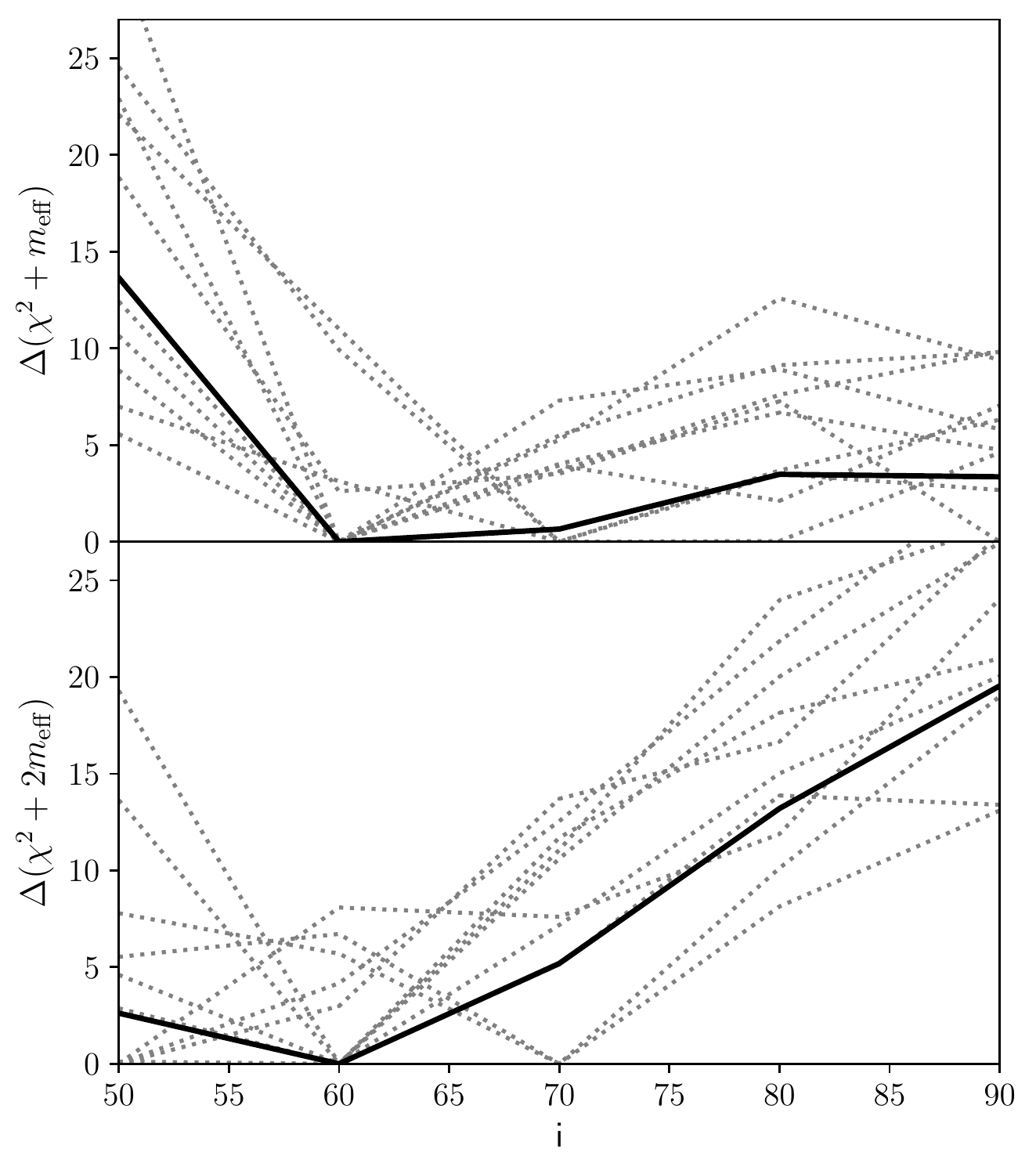}
    \caption{Similar to Fig.~\ref{fig:GalaxyD_chi2} but for a model selection framework. The absolute minimum of $\chi^2+w_m m_{\mathrm{eff}}$ has been subtracted for each mock, such that only the relativ differences $\Delta$ to this minimium is plotted. \textit{Top panel:} Inclination recovery using the "intuitive" approach $\chi^2+m_{\mathrm{eff}} \rightarrow \mathrm{min}$ (equation~\ref{eq:intuitve}). \textit{Bottom panel:} Inclination recovery using the AIC approach  $\chi^2+2m \rightarrow \mathrm{min}$ (equation~\ref{eqn: Akaike}). For both approaches we find that the apparent inclination bias that results from a simple $\chi^2$ minimisation (Fig.~\ref{fig:GalaxyD_chi2}) disappears when the changing model flexibility (Fig.~\ref{fig:GalaxyD_parameters}) is taken into account. The true inclination of Galaxy D is $i=60\degr$. }
    \label{fig:GalaxyD_modelselection}
\end{figure}

\begin{figure}
	\includegraphics[width=\columnwidth]{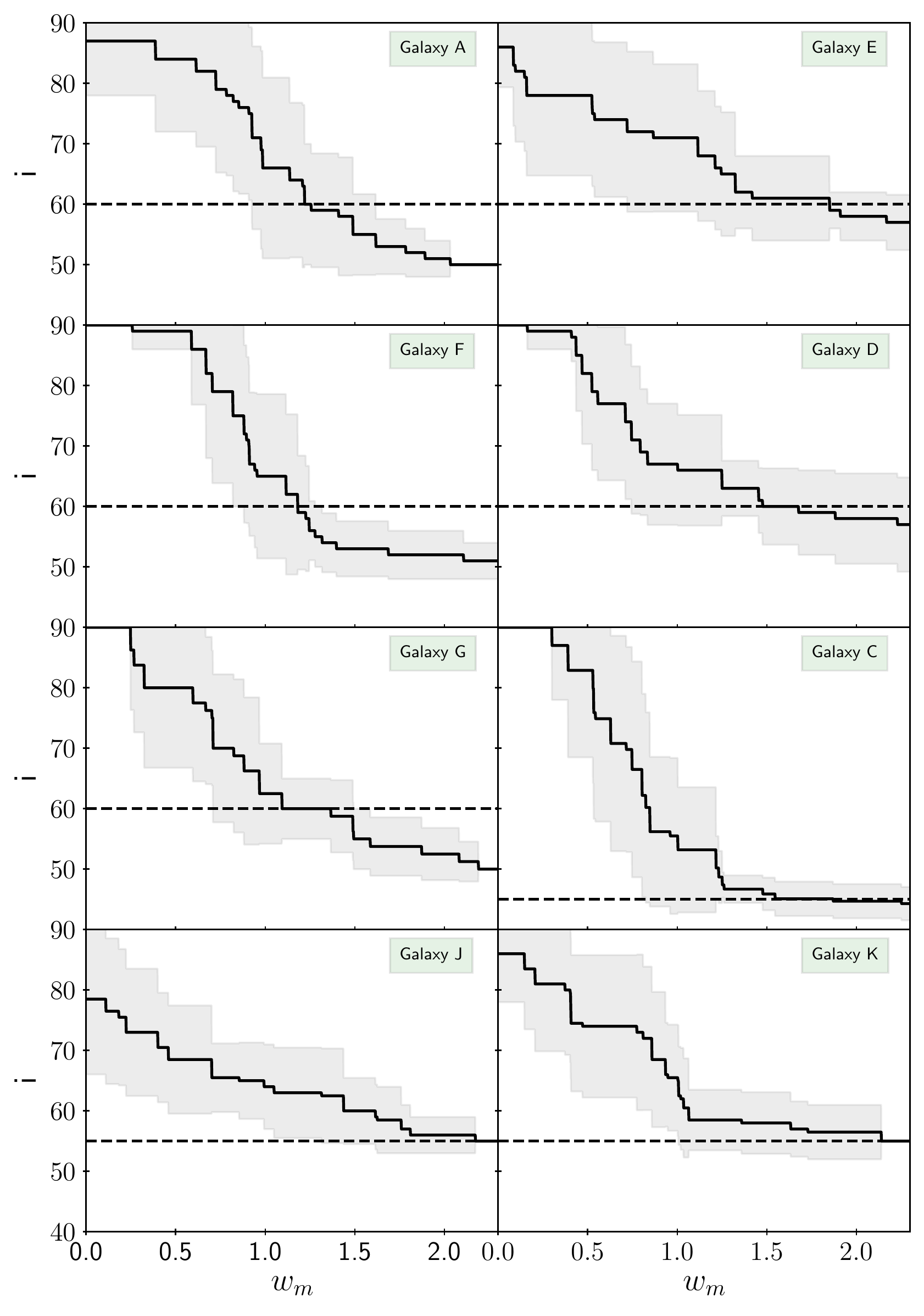}
    \caption{Inclination recovery as a function of the weight parameter $w_m$ for some of the toy galaxies of table~\ref{tab:toy_gal_table}. \textit{Solid Line:} The recovered inclination averaged over fits to all 10 mock observations of each simulated galaxy. \textit{Grey area:} The corresponding 1$\sigma$ level estimated from the mock sample. \textit{Dashed Line:} The true inclination of each toy galaxy. All simulated galaxies suffer from a dominant edge-on bias when evaluated with $\chi^2$ ($w_{m}=0$). For the early-type galaxies we find that a model selection with an intermediate $w_{m}\in[1.2,1.8]$ achieves an optimum inclination recovery. For the late-type galaxies a slightly higher $w_{m}\rightarrow2$ appears to be optimal. However, all these results depend on the regularization $\alpha$ as we will demonstrate in Sec.~\ref{sec:regularization}. }
    \label{fig:calibration_toy_galaxies}
\end{figure}

Despite the fact that the galaxy is fairly inclined with a true inclination $i=60\degr$, the $\chi^2$ of the Schwarzschild models suggest that the edge-on model is the best fit on average (Fig.~\ref{fig:GalaxyD_chi2}). This undesired behaviour mirrors the situation in the Hernquist sphere discussed in Sec.~\ref{subsec:spherical}. It is is caused by a drastic change in the model's flexibility with inclination, as can be seen in Fig~\ref{fig:GalaxyD_parameters}. The Figure shows that the number of effective parameters increases rapidly as a function of inclination.
Thus, the lower $\chi^2$ at higher inclination are not necessarily an indication that the fits to the data are intrinsically better at larger $i$. Rather, the $\chi^2$ gets lower because the model is more flexible. In other words, the measure for which $\chi^2$ represents a "good" fit has shifted towards lower values. In fact, if we take the varying model flexibility into account and compare the models within our model selection framework, then the inclination bias disappears (Fig~\ref{fig:GalaxyD_modelselection}). For both, the intuitive and the AIC approach, we can recover the galaxy's true inclination on average very well. If any, then the fits to the individual mock realisations seem to scatter slightly above the true inclination for the intuitive approach and slightly below the true inclination for the AIC approach, but on average any related bias is very small, $\Delta i \sim 5\degr $. Still, the distribution of the individual fits suggests that even better results might be achieved with some optimized $w_m \in [1,2]$.
In fact, we repeated this analysis for all toy galaxies of table~\ref{tab:toy_gal_table}. For a representative subset of these simulated galaxies, Fig.~\ref{fig:calibration_toy_galaxies} shows the recovered inclination again averaged over fits to ten mock data realisations (solid black lines) as a function of $w_m$. The Figure shows that the behaviour of the fits to Galaxy D described above is actually generic. For $w_m = 0$ (simple $\chi^2$ minimization) we find that the "best-fit" inclination is almost always $i=90\degr$ or close to it. With increasing $w_m$, the influence of the varying model flexibility on the recovered inclination becomes stronger and the "best-fit" inclination moves away from $i=90\degr$. 

For most early-type galaxies, the true inclination is very well recovered with $w_m \approx 1.5$, while larger $w_m$ lead to a slight bias towards too small inclinations. Similarly, the inclination recovery for the late-type galaxies  is improved when selecting models with $w_{m}>1$, however, optimum results are achieved in the Akaike regime ($w_{m}=2$). A bias towards too low inclinations is strongly suppressed for the late-type toy galaxies as inclination angles significantly smaller than the true $i=55\degr$ are excluded on the basis of the apparent flattening in the photometric data.

Given the modelling results of all toy galaxies in Table~\ref{tab:toy_gal_table} we conclude that a simple $\chi^2$ minimisation for the {\it selection parameter} $i$ leads to a very significant bias. This is caused by the fact that edge-on models are much more flexible than models with a lower inclination and, hence, yield much smaller minimum $\chi^2$ values. However, when taking the differences in model flexibility into account by penalising each model with an additive term proportional to the effective number of free parameters, the inclination recovery actually works very well. With the new model selection framework, we could recover the inclination of all toy galaxies to within $\Delta i = 5\degr$ on average. 

The best results were obtained when weighting the influence of the model flexibility with a proportionality factor of $w_m \approx 1.5$, which is a value right between the "intuitive" approach and the AIC. We will come back to this in Sec.~\ref{sec:regularization} where we also include the regularisation into the analysis.

\subsection{Flexibility and underlying galaxy structure}
\label{subsec:underlying structure}
A unique feature of the flexibility of non-linear statistical models is that the number of effective parameters can generally depend on the underlying data generating process \citep[cf.][]{doi:10.1080/01621459.1998.10474094}. In the case of our dynamical models this means that the flexibility is not a universal property of the model alone, i.e. a property of the orbit library, but instead can additionally depend on the underlying galaxy structure. 

\begin{figure}
	\includegraphics[width=\columnwidth]{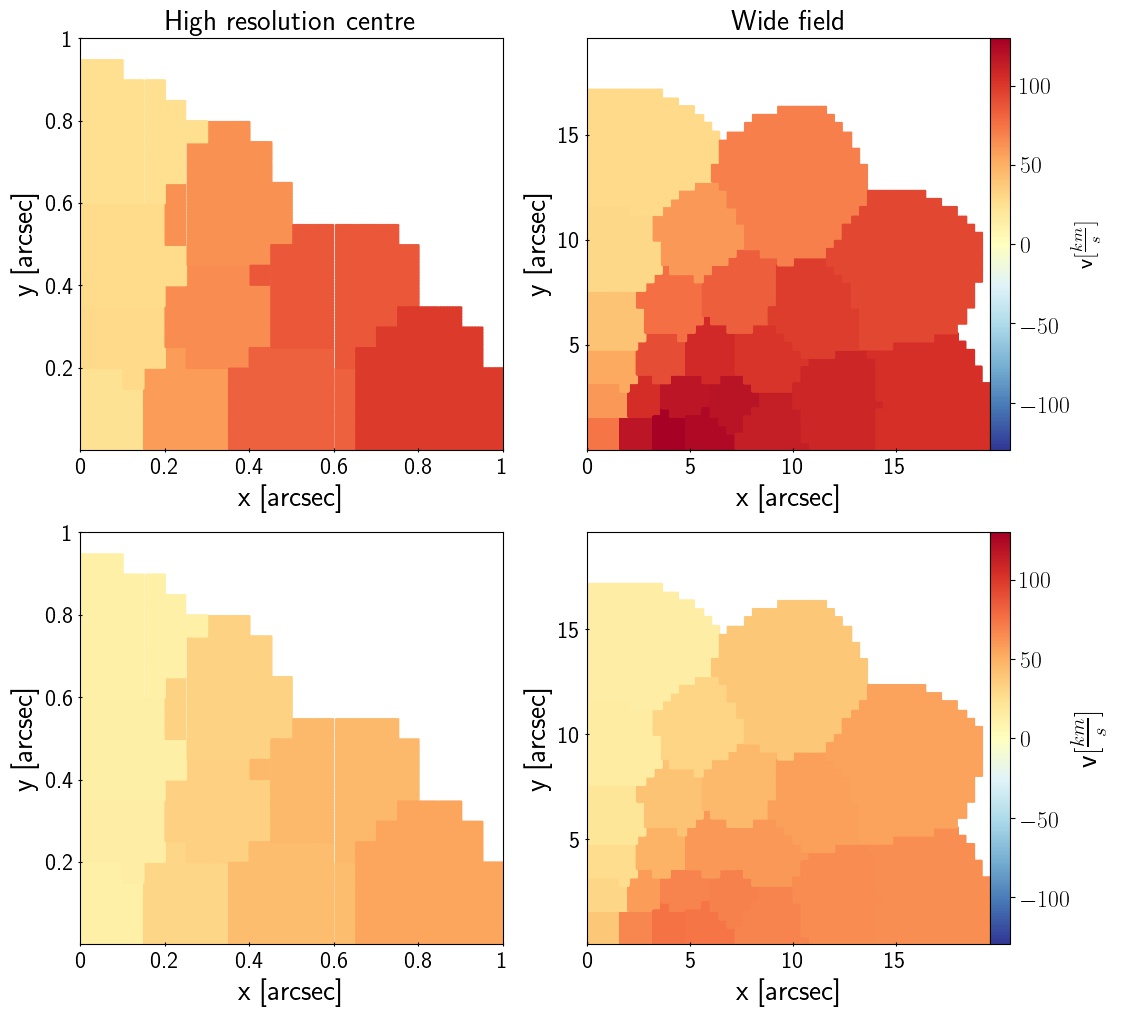}
    \caption{The (noise-free) rotation maps of two toy galaxies with angular momentum bias $\lambda=1.0$ (\textit{top panels}) and an intermediate $\lambda=0.5$ (\textit{bottom panels}). Panels on the left side show the centre with the higher-resolution data, panels on the right show the simultaneously modelled wide-field data.}
    \label{fig:rotation_maps}
\end{figure}

This dependence becomes especially apparent when comparing the dynamical modelling results of rotating and non-rotating galaxies.  We investigated this by fitting mock data sets of a sequence of toy galaxies which were identical except for the angular momentum bias $\lambda$ (equation~\ref{eqn: rotation}). One toy galaxy was generated with $\lambda=0.0$ and represents a non-rotating maximum-entropy galaxy (Galaxy A in Tab.~\ref{tab:toy_gal_table}). The two other toy galaxies were constructed to be rotating by setting  $\lambda=0.5$ (Galaxy D) and $\lambda=1.0$ (Galaxy E). The resulting (noise-free) rotation maps of the latter two toy galaxies are shown in Fig.~\ref{fig:rotation_maps}. After having added noise we analysed the three toy galaxies with an identical orbit library and estimated the flexibility of the fits using the bootstrap calculations of section~\ref{subsec:bootstrapping} above. We repeated this for five different inclinations from $i=50\degr$ to $i=90\degr$.
\begin{figure}
	\includegraphics[width=\columnwidth]{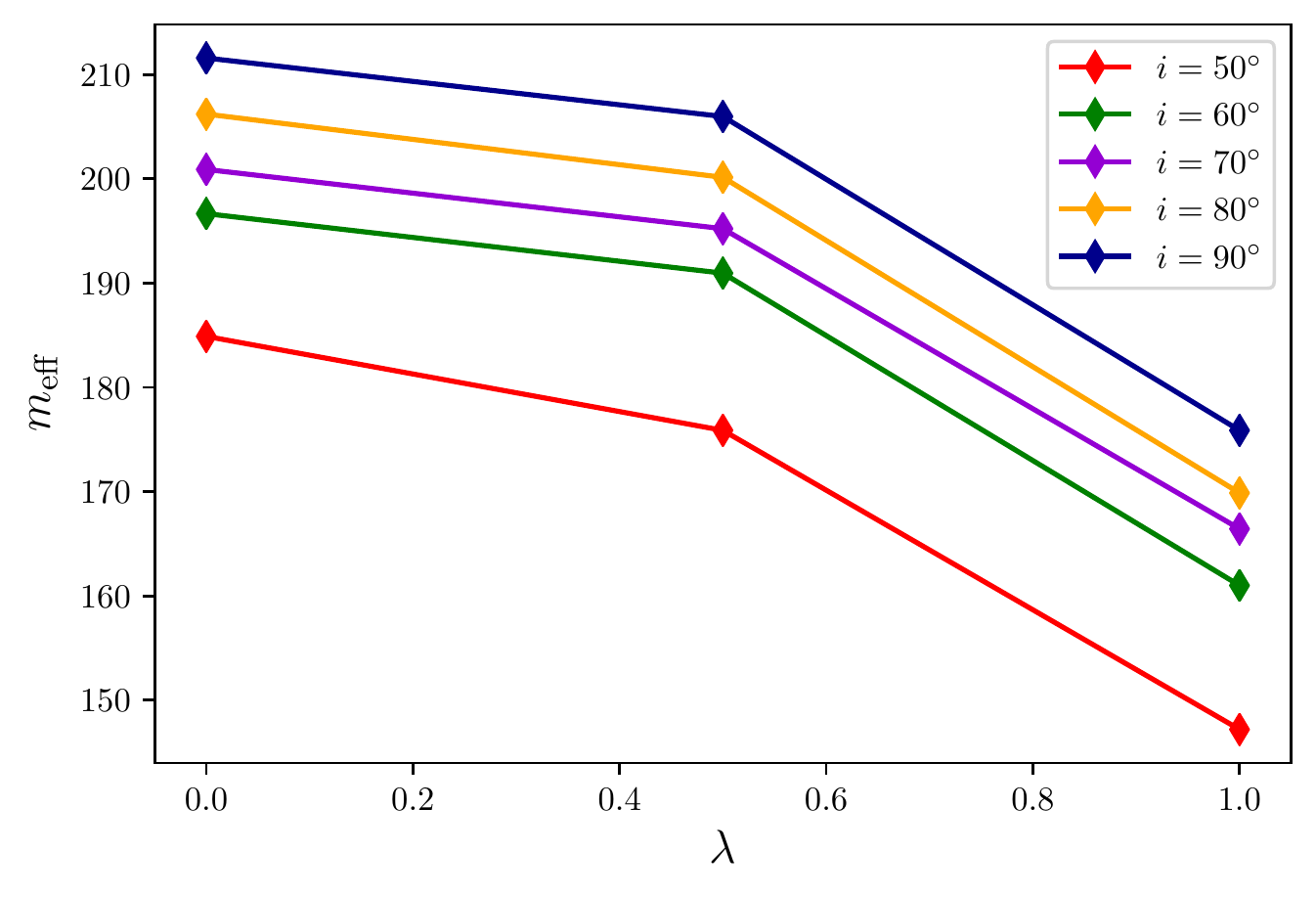}
    \caption{The number of effective parameters $m_\mathrm{eff}$ estimated for axisymmetric Schwarzschild models of rotating toy galaxies with different angular momentum bias $\lambda$. For each toy galaxy, the results are averaged over 10 mocks and the regularisation parameter was set to $\alpha = 1.67$. Even though models at the same inclination $i$ were obtained with identical orbit libraries, the number of effective parameters $m_\mathrm{eff}$ does not stay the same but decreases with the angular momentum bias. This is true at any inclination. }
    \label{fig:rotation_dependence}
\end{figure}

Fig.~\ref{fig:rotation_dependence} shows the resulting estimated number of effective parameters. Note that for a given inclination, the mock data of the toy galaxies with different $\lambda$ were fitted with exactly the same orbit library. Despite this fact, however, the model's ability to fit the data changes with $\lambda$. On the one hand, we again see how the overall flexibility of the model increases with $i$. On the other hand, however, we also see that the more net rotation is present in the underlying galaxy, the less flexible the model appears. We argue that the reason for this dependency is the fact, that with increasing rotation the light in the observed LOSVDs is more and more concentrated on either positive or negative velocities. While the amount of light on one side of the LOSVD therefore increases, it decreases on the opposite side and more and more velocity bins in each LOSVD tend to carry only little or even no light when the angular momentum bias $\lambda$ is strong. After the addition of Gaussian noise such (nearly) dark velocity bins can formally even carry a negative signal. Since we require the orbital weights to be positive, such formally negative LOSVD values can not be fitted. Thus, because with more rotation more LOSVD bins are prone to carry a negative signal due to noise, the model's flexibility as estimated by the number of effective parameters will shrink.     

This correlation of $m_\mathrm{eff}$ with the rotation implies that the flexibility of Schwarzschild models depends on the underlying phase-space distribution function of the galaxy under investigation. The same orbit library can be more responsive to one data set than to another. This is in contrast to the flexibility of linear models where the number of free parameters is independent of the underlying data generating process (as long as no constraints on the free parameters are applied). We conclude that the number of effective parameters of a given orbit model should be estimated individually for each new fit. When modelling the kinematics in the different quadrants of the same galaxy separately, for instance, even if one samples the mass parameters from the same parameter grid,  the number of effective parameters needs to be estimated separately for each fit in each quadrant. 

\section{Biases in other selection parameters}
\label{sec:mass_parameters}
The dependence of $\chi^2$ on the model flexibility is most prominent for differently inclined models, however, in principle any selection parameter can suffer from biases introduced by a variable model flexibility. We investigated this by modelling toy galaxy E with different model inclinations and mass-to-light ratios $\Upsilon$.

\begin{figure}
	\includegraphics[width=\columnwidth]{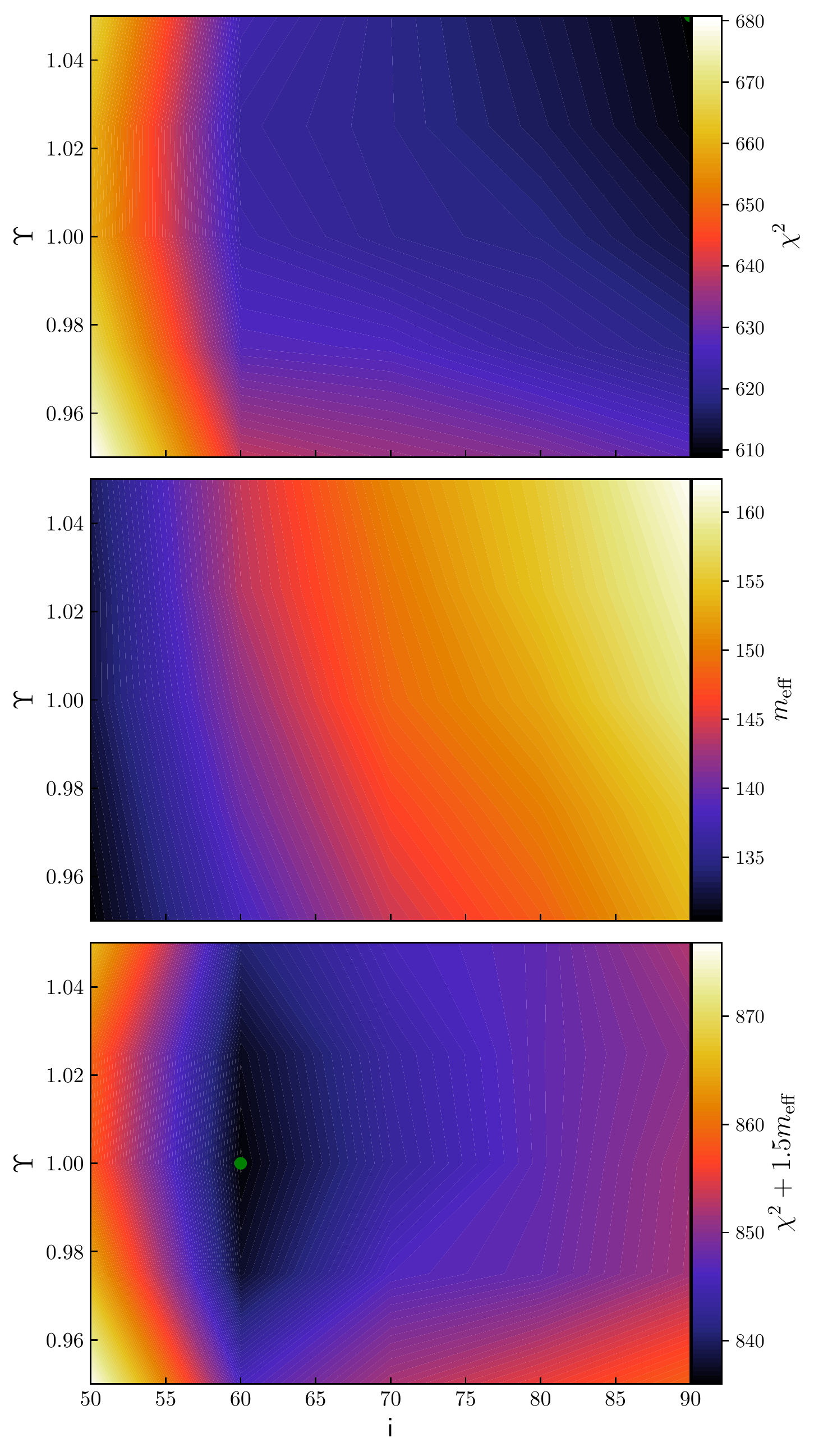}
    \caption{\textit{Top panel:} The $\chi^2$ distribution averaged over dynamical fits to 10 mock data sets of the flattened toy galaxy E with $i_{\mathrm{true}}=60\degr$ and $\Upsilon_{\mathrm{true}}=1.0$. \textit{Middle panel:} The corresponding distribution of the average number of effective parameters. \textit{Bottom panel:} The model selection distribution with the calibrated approach (Eq.~\ref{eqn: calibrated} with $w_m=1.5$). The green dot locates the parameters of the best model. A simple $\chi^2$ minimisation leads to a bias in $i$ and $\Upsilon$ because models with larger $i$ and larger $\Upsilon$ are more flexible. Model selection removes this bias and allows to clearly identify the correct model.}
    \label{fig:ml-bias}
\end{figure}

The result is displayed in the top two panels of Fig.~\ref{fig:ml-bias}, which show contours of  $\chi^2$ and $m_\mathrm{eff}$ as a function of the mass-to-light ratio $\Upsilon$ and the inclination $i$, averaged over 10 mocks. In the direction of the inclination we observe the edge-on bias established in the previous sections. However, the contours reveal an additional bias along the mass-to-light ratio axis. A simple $\chi^2$-minimization does not result in a minimum at the true mass-to-light ratio $\Upsilon_{\mathrm{true}}=1.0$ but instead is biased towards slightly larger ratios. The fact that the $\chi^2$ behaviour is mirrored by the number of effective parameters $m_\mathrm{eff}$ suggests that this increase in model flexibility causes the bias. While this bias of the mass-to-light ratio is present for all modelled inclinations, it appears to be most significant towards $i=90\degr$.  Thus, a model evaluation based on a simple $\chi^2$-analysis would not only overestimate the inclination but would also misjudge the total mass of the system.

We argue that the reason for this $\Upsilon$-$m_\mathrm{eff}$-correlation is due to the higher escape velocity in models with larger total mass. Orbits with a higher escape velocity can potentially occupy velocity bins that are unattainable at lower masses. Models at higher masses will therefore be able to fit more LOSVD bins (if the LOSVDs are sampled out to sufficiently large velocities). This means that they have an increased model flexibility when compared to their lower $\Upsilon$ counterparts.

If we take this effect into account and judge the dynamical models within the model selection framework, then the recovery of the original mass-to-light ratio and inclination becomes highly accurate. This is shown in the bottom panel of Fig.~\ref{fig:ml-bias} which shows the constraints obtained with the calibrated model selection approach (i.e. $w_m = 1.5$ Eq.~\ref{eqn: calibrated}). Both inclination and mass-to-light ratio biases have disappeared and the constraints are now tightly centred around the correct model with $60\degr$ model and $\Upsilon=1$. Similar to the inclination an intermediate calibration weight of $w_{m}\approx1.5$ proves to be most successful in recovering the correct mass-to-light ratio. The intuitive and the Akaike approach also manage to select the correct model on average, however, their contours are slightly skewed towards larger $\Upsilon$ for $w_{m}=1$ and smaller $\Upsilon$ for $w_{m}=2$.  

The orbit models shown in this and the previous section~\ref{sec:simulated} were modelled with a fixed regularization $\alpha$. However, the constraining power of the model selection framework for $\Upsilon$ and $i$ can be further improved by treating the regularization parameter as an additional selection parameter, as we will demonstrate in section~\ref{sec:regularization}.

\section{Refinement: The role of the regularization parameter}
\label{sec:regularization}
As mentioned in Sec.~\ref{subsec:parameter_estimation} the regularization parameter $\alpha$ is important to control the smoothness of the orbital weights. However, up until now we have simply chosen $\alpha$ such that $\chi^2(\alpha)$ has roughly converged. In the following we will denote such a regularization parameter that was chosen using this convergence criterion as $\alpha_{\infty}$. For the modelling of the toy galaxies in table~\ref{tab:toy_gal_table} this happens for $\alpha \gtrsim 1$, which motivated our somewhat arbitrary choice of $\alpha_{\infty}=1.67$. This means we almost only considered the most flexible models as candidate models, since $\chi^2(\alpha)$ has roughly converged at this regularization value, meaning a further increase in $\alpha$ does not significantly reduce $\chi^2$ further and the corresponding number of effective parameters has plateaued (cf. Fig.~\ref{fig:regularization}). While this choice guarantees a good fit to a given kinematic data set, it may not be suitable to restrict our test only to models with such a large $\alpha$ as they are prone to overfitting the data. And this, in turn, could potentially weaken the constraints on the selection parameters. Moreover, the optimal weight $w_{m}$ of the $m_\mathrm{eff}$ in the model evaluation (cf. eq.~\ref{eqn: calibrated}) could depend on $\alpha$, meaning that a calibration found for one data set of one galaxy may not be applicable to another data set of another galaxy. It is common practice to use Monte Carlo simulations of the galaxy under investigation to determine its optimal smoothing \citep[e.g.][]{2000AJ....119..153S,2004MNRAS.347L..31C,2005MNRAS.360.1355T,2012MNRAS.422.1571M,2020MNRAS.500.1437N}. In order to avoid smoothing-induced biases in the models, the mock data should be as realistic as possible and adapted to the particular data set at hand (resolution, signal-to-noise etc.). In addition, since the optimum value of $\alpha$ will depend on the underlying galaxy structure such simulations need to be repeated for every galaxy anew. However, a more targeted approach that optimizes $\alpha$ directly from the observed data may be advantageous as it would not rely on the choice of mock galaxy that is required for the Monte-Carlo simulations and is supposed to emulate the real galaxy.

For this purpose we again employed the concept of selection parameters. Section~\ref{subsec:selection_parameters} characterized selection parameters as those that need to be specified to single out an individual model out of the {\it family of models} $\mathcal{F}(M_\bullet,\Upsilon,i,\ldots)$. The respective orbital weights $w_i$ are its free parameters. In general, there may exist many distribution functions (or equivalently vectors $\mathbfit{w}$) that satisfy the observational constraints posed by $\chi^2(\mathbfit{w})$ for a given set of selection parameters. The reason is that $\chi^2(\mathbfit{w})$ is not necessarily {\it strictly} convex \citep[cf.][]{2020MNRAS.500.1437N}. In practice, this issue is circumvented by adopting a penalized maximum-likelihood estimation by adding a penalty function, such as the entropy term in eq.~\ref{eq:entropy} that makes the solution for the orbital weights unique. However, this means out of all the allowed and statistically viable DFs only a single, privileged DF is identified, a possible issue already pointed out by \citet{2006MNRAS.373..425M}. In the following we will show that our model selection framework facilitates an evaluation of \textit{multiple} distribution functions compatible with a single set of $M_\bullet,\Upsilon,i,\ldots$ if we extend the selection parameters to include the regularization parameter $\alpha$. 

Which distribution function out of all the compatible DFs (or $\mathbfit{w}$) is privileged by a penalized maximum-likelihood approach is determined by the specific value of the weight parameter used for the penalty function (in our case $\alpha$). In that sense the regularization parameter $\alpha$ is a prior that, like the selection parameters, constraints the parameter space available to the orbital weights. Furthermore, like models with different selection parameters, models with different $\alpha$ also have varying number of effective parameters. In fact, $\alpha$ plays the dominant role in determining the flexibility of a dynamical model, as demonstrated in section~\ref{subsec:bootstrapping}. Consequently, we may attempt to extend the concept of selection parameters to include prior parameters, such as $\alpha$, that push the orbital weights towards certain solutions. The regularization does not necessarily have to be the only such prior parameter that is used to identify a specific DF out all the DFs compatible with a set of selection parameters $(M_\bullet,\Upsilon,i,\ldots)$. For example, it is not necessary to use the phase volumes $V_{i}$ of the orbits within the entropy-term. Instead one could use a free $\omega_i$ to bias the solution towards a specific solution \citep[cf.][]{2020MNRAS.500.1437N}. In that case the $\omega_i$ are simply another set of selection parameters that could potentially be constrained by the establishment of candidate models with different $\omega_i$. However, in our case the phase volumes are not themselves independent selection parameters as they are completely determined by the other selection parameters $(M_\bullet,\Upsilon,i,\ldots)$ that were used to create an orbit library. Therefore we focus on the regularization parameter $\alpha$ that is used to control the smoothness via the entropy term in eq.~\ref{eq:entropy}. 

If $\alpha$ can be thought of as simply another selection parameter, it may be possible to constrain it using the kinematic data by constructing candidate models with various $\alpha$, analogously to any other mass or library parameter. While a recovery using only the kinematic goodness-of-fit is not sensible as it would always favor the maximum $\alpha$ models, it may be possible to constrain the regularization by evaluating the models within a model selection framework that takes into account the number of effective parameters. 
To investigate this possibility we modelled 10 mock data sets of toy galaxy A again, but this time we varied not only the two selection parameters $\Upsilon$ and i, yet also the regularization $\alpha$. Then, as described in Sec.~\ref{sec:simulated}, for each model defined by ($\alpha, \Upsilon,i$) we estimated $m_\mathrm{eff}$ and selected the best model via the general model selection approach $\chi^2+w_{m} \, m_\mathrm{eff}$.

Considering that Galaxy A was created using a maximum-entropy Schwarzschild model with $\alpha=10^{-10}$ we would expect that an optimal model selection approach should return  a model with the same $\alpha=10^{-10}$ as the best model. The reason is that the orbital weights are uniquely determined if ($\alpha, \Upsilon,i$) are given. In general, models obtained with the same $\Upsilon$ and $i$ but different $\alpha$ will have different orbital weights and so, by construction, $\alpha = 10^{-10}$ should return the weights closest to the input model. The regularization parameter itself is not a physically relevant parameter but simply a choice of prior that restricts the freedom of the orbital weights. However, as a consequence it significantly impacts the form of the resulting distribution function of a dynamical model and its respective kinematic moments. Therefore one should ideally use a regularization parameter that allows the best possible approximation of the true, underlying kinematic structure. To quantify this we characterised the difference between the underlying true orbit model and a fitted model by the average root mean square deviations (RMSD) of the first and second order velocity moments. Apart from the deviations of the individual moments we also define a total kinematic deviation:
\begin{equation}
\Delta_{\mathrm{kin}}=RMSD_{v}+RMSD_{\sigma_{r}}+RMSD_{\sigma_{t}}
\label{eq:kin_rec}
\end{equation} 
Here $v$ is the mean stellar rotation velocity of the model around its z-axis, while $\sigma_{r}$ and $\sigma_{t}$ are the radial and tangential velocity dispersions.

\begin{figure}
	\includegraphics[width=\columnwidth]{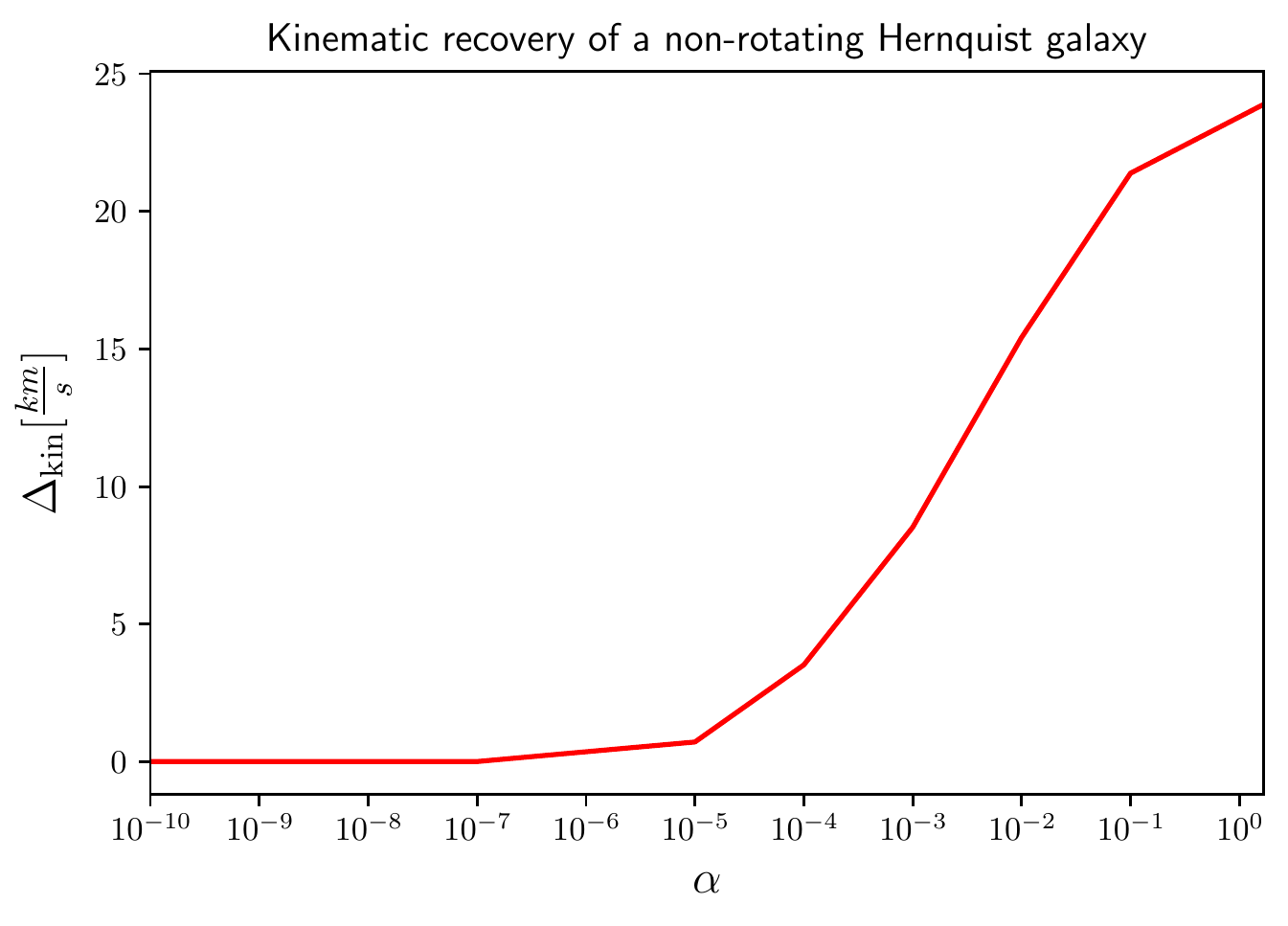}
    \caption{The total deviation in the intrinsic first and second velocity moments (Eq.~\ref{eq:kin_rec}) of fits to the non-rotating Hernquist galaxy (Galaxy A) for different values of the regularization parameter $\alpha$. The inclination and mass-to-light ratio were fixed at their true values ($\Upsilon=1.0$, $i=60\degr$). Larger $\alpha$ imply less influence of the entropy term and lead to an overfit of the noise. This increases the discrepancy between input model and fit. The lowest $\alpha$ allow an almost perfect reconstruction of the input model.}
    \label{fig:RMS_profile}
\end{figure}

\begin{figure}
	\includegraphics[width=\columnwidth]{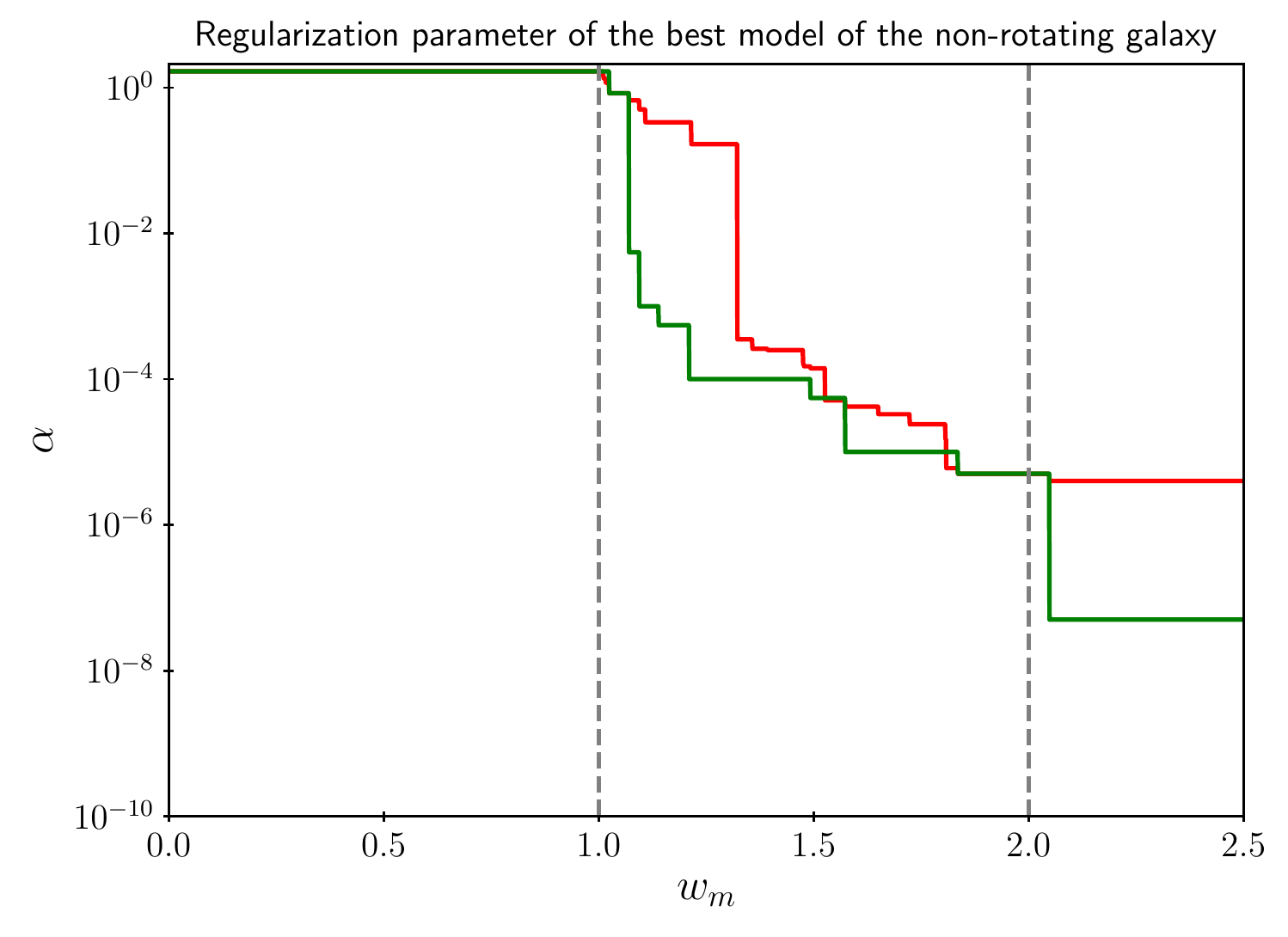}
    \caption{The mean (\textit{red}) and median (\textit{green}) regularization parameter of the best models to a non-rotating Hernquist mock galaxy (Galaxy A) as a function of the model-selection calibration parameter $w_m$. For every given $w_m$ the $\alpha$ of the best model selected from all the possible candidate models with free selection parameters $\alpha$,$\Upsilon$ and $i$ is shown. The dashed vertical lines locate the model selection using the intuitive ($w_{m}=1$) and the AIC ($w_{m}=2$) approach (cf. eq.~\ref{eq:intuitve} and \ref{eqn: Akaike}). The case $w_m=0$ corresponds to a simple $\chi^2$ minimization and returns models with large $\alpha$ as the best ones. When the number of effective parameters is weighted by $w_m=2$ (AIC), then the models identified as the best ones have much smaller $\alpha$. For each of the 10 mocks AIC only selected models with $\alpha \leq 10^{-5}$. Such an $\alpha$ implies a very close match to the input model (Fig.~\ref{fig:RMS_profile}).}
    \label{fig:alpha_recovery_median}
\end{figure}

Fig.~\ref{fig:RMS_profile} shows $\Delta_{\mathrm{kin}}$ versus $\alpha$ when the inclination $i$ and mass-to-light ratio $\Upsilon$ are fixed at the correct values. Obviously, the input model is exactly recovered for the lowest $\alpha$. Larger $\alpha$ values lead to a larger discrepancy between the fitted velocity moments and the ones in the input model. The reason is that with increasing $\alpha$ the model starts to overfit the noise incorporated in the mock data. 
In analogy to Sec.~\ref{sec:simulated}, where we used the accuracy of the inclination recovery as criterion to decide upon the  optimal weight factor $w_m$ in  $\chi^2+w_{m} \, m_\mathrm{eff}$ we can now also include $\alpha$ in the optimisation of $w_m$. Fig.~\ref{fig:alpha_recovery_median} shows the median and mean $\alpha$ values of the selected models for each mock of Galaxy A as function of the calibration weight $w_{m}$.

Unsurprisingly the simple $\chi^2$ minimization selects the model with largest sampled $\alpha$, in this case 1.67, meaning the best models overfit the noise and the recovered distribution function has a higher degree of irregularity than the smooth input model. For larger $w_{m}$ we approximate the input model better as the median $\alpha$ shifts from $1.67$ to $10^{-7}$ for $1 \la w_{m} \la 2$. In the regime of the Akaike approach, the selection framework yields $10^{-10} \la \alpha \la 10^{-5}$. In terms of the quality of the reconstruction of the internal velocity moments, any $\alpha$ in this regime leads to an almost perfect recovery of the input model (Fig.~\ref{fig:RMS_profile}). 

While the model selection framework with a variable $\alpha$ seems to work very well for the mocks of Galaxy A, in particular the Akaike approach, it may not be the most realistic setup. For one because the generating model for Galaxy A is a maximum entropy model without any rotation, but mainly because one can generally not expect to have the "true" generating model among the tested models, which in the case of Galaxy A is the model with selection parameters $i=60\degr$, $\Upsilon=1.0$, $\alpha=10^{-10}$. Furthermore the generating $\alpha=10^{-10}$ model of Galaxy A is also the model with the smallest number of effective parameters and consequently the selected model is always the generating model one as long $w_{m}$ is chosen big enough, meaning that we only establish a lower boundary for the calibration weight $w_{m}$.

For these reasons we additionally tested the approach with a variable regularization on the mocks of Galaxy D, which is a rotating toy galaxy with an angular momentum bias of $\lambda=0.5$. In this case the best regularization parameter is not known in advance and the generating model is not necessarily among the tested models. 

\begin{figure}
	\includegraphics[width=\columnwidth]{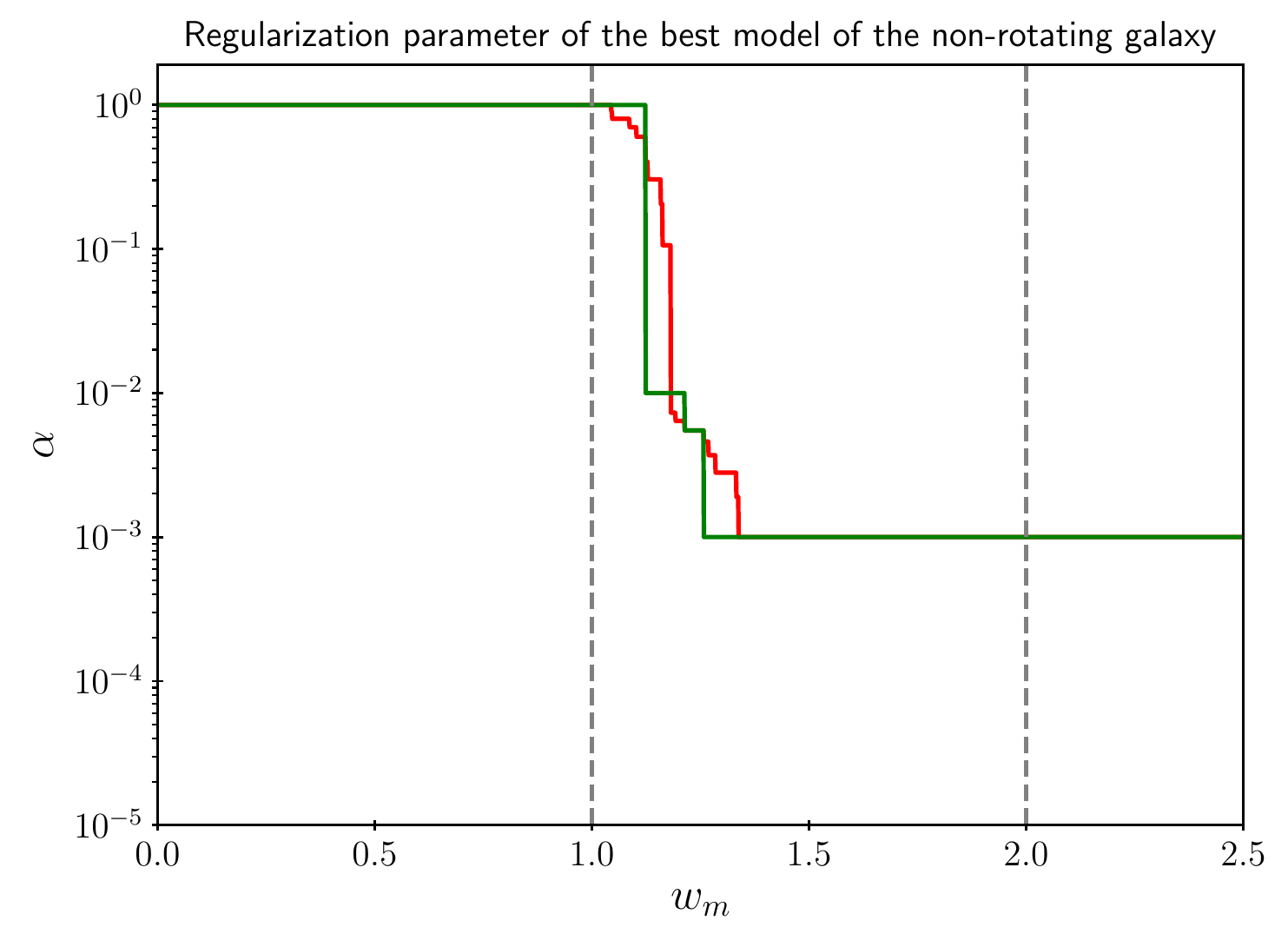}
    \caption{As Fig.~\ref{fig:alpha_recovery_median}, but for the rotating toy galaxy D. In contrast to the non-rotating galaxy, the model selection framework avoids maximum entropy models ($\alpha = 0$) here because such models would not allow an accurate reconstruction of the galaxy's net rotation induced by its angular momentum bias $\lambda=0.5$ (cf. Fig.~\ref{fig:GalaxyD_RMS_profile}).}
    \label{fig:GalD_alpha_recovery_median}
\end{figure}

Fig.~\ref{fig:GalD_alpha_recovery_median} shows the median and mean regularization parameter of the best model of Galaxy D as a function of the calibration weight $w_{m}$. We find large $\alpha$ for $w_{m} \lesssim 1$ followed by a steep falloff that turns into a stable plateau for $w_{m}\sim2$, similar to Galaxy A. However, the optimal $\alpha$ for the Akaike approach is now at an intermediate $\alpha = 10^{-3}$ and not where the models have the smallest number of effective parameters. (Only for $w_{m} \ga 4$ the approach tends towards models that simply have the smallest number of effective parameters. However, such large $w_m$ can easily be discarded because they lead to unreasonably large $\chi^{2}\gg N_{\mathrm{data}}$, i.e. they are unable to fit the mock observations of Galaxy D). 

\begin{figure}
	\includegraphics[width=\columnwidth]{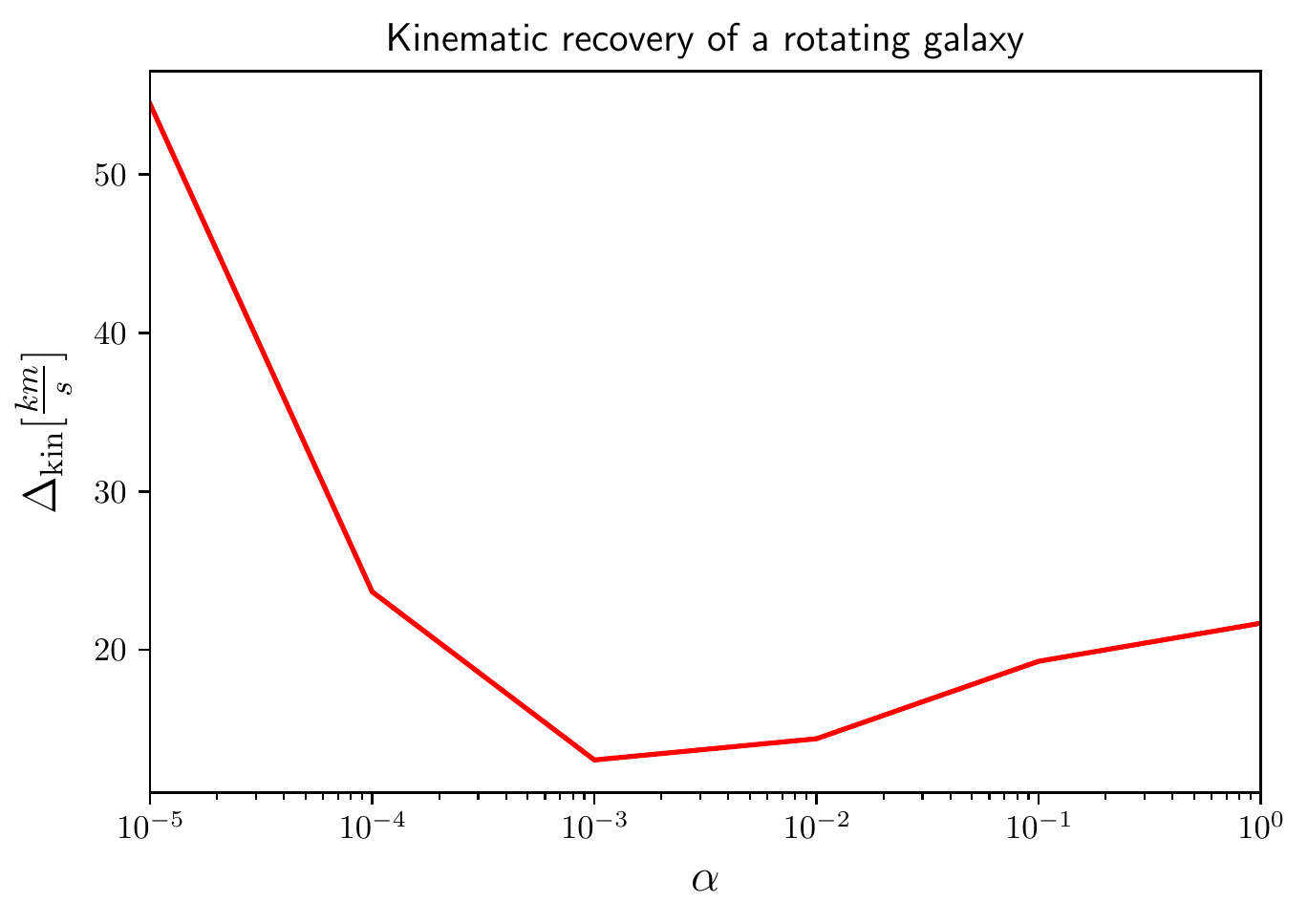}
    \caption{As Fig.~\ref{fig:RMS_profile}, but for the rotating mock galaxy D. The recovery of the intrinsic kinematic moments is best at $\alpha = 10^{-3}$. This is exactly the value, that the model selection framework identifies as best in the regime of the Akaike approach (Fig.~\ref{fig:GalD_alpha_recovery_median}). }
    \label{fig:GalaxyD_RMS_profile}
\end{figure}

The model selection framework $\chi^2+w_{m} \, m_\mathrm{eff}$ with $w_{m} \in [1.5,4]$ suggests that the optimal amount of regularization for Galaxy D is achieved with $\alpha=10^{-3}$. It becomes evident from Fig.~\ref{fig:GalaxyD_RMS_profile} that models with this regularization of $\alpha=10^{-3}$ are indeed the models that reproduce the intrinsic velocity structure of Galaxy D best. 
This demonstrates that the model selection framework $\chi^2+w_{m} \, m_\mathrm{eff}$ with $w_{m} \in [1.5,4]$ leads to an optimal recovery of the velocity structure of Galaxy D. Less regularized models, while nominally achieving a better $\chi^2$, overfit the LOSVDs and do not improve the recovered velocity moments. On the other hand more regularized models with $\alpha <10^{-3}$ are not sufficiently flexible to be able to describe the underlying non-maximum entropy distribution function.

\begin{figure}
	\includegraphics[width=\columnwidth]{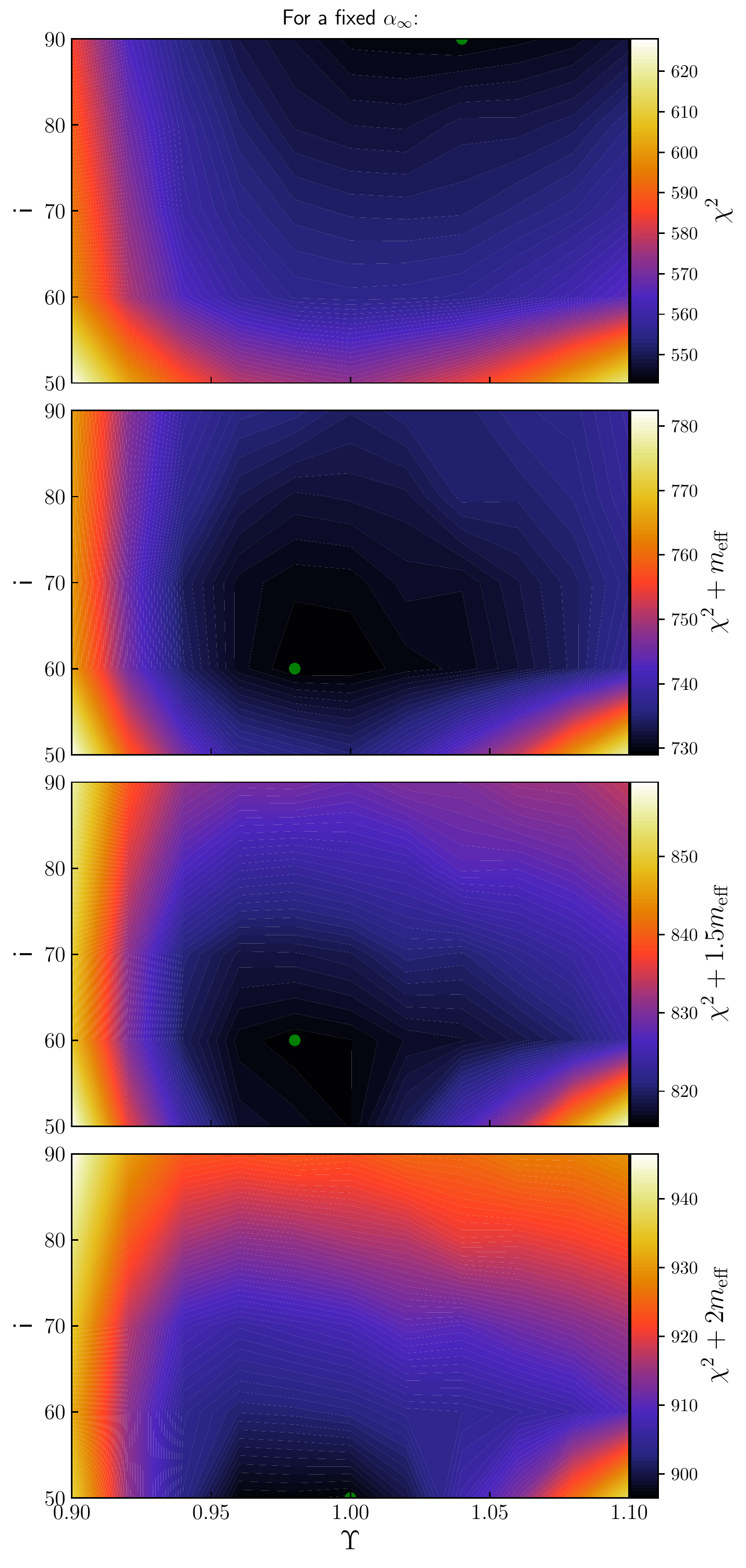}
    \caption{Constraints on the selection parameters i and $\Upsilon$ of Galaxy D for different model selection frameworks. From top to bottom: Simple $\chi^2$ minimisation, the "intuitive" approach $\chi^2+m_\mathrm{eff}$, the calibrated approach $\chi^2 + w_m \, m_\mathrm{eff}$ where $w_m$ is determined from simulations (cf. Sec.~\ref{sec:simulated}) and the Akaike approach. In all cases, the smoothing parameter was held fixed to $\alpha_{\infty}$, such that $\chi^2(\alpha)$ has approximately converged to a constant value. The green dot locates the best model in the respective model selection framework. The true values for galaxy D are $i=60\degr$ and $\Upsilon = 1.0$. As expected, the $\chi^2$ minimisation leads to a biased result, the calibrated approach is best (because the simulations to calibrate $w_m$ were made under the same assumptions for $\alpha$). When $\alpha$ is fixed, the Akaike approach also leads to biases in the selection parameters, though these biases are smaller than the ones resulting from the simple $\chi^2$ minimisation.}
    \label{fig:GalD_for_fixed_alpha}
\end{figure}

\begin{figure}
	\includegraphics[width=\columnwidth]{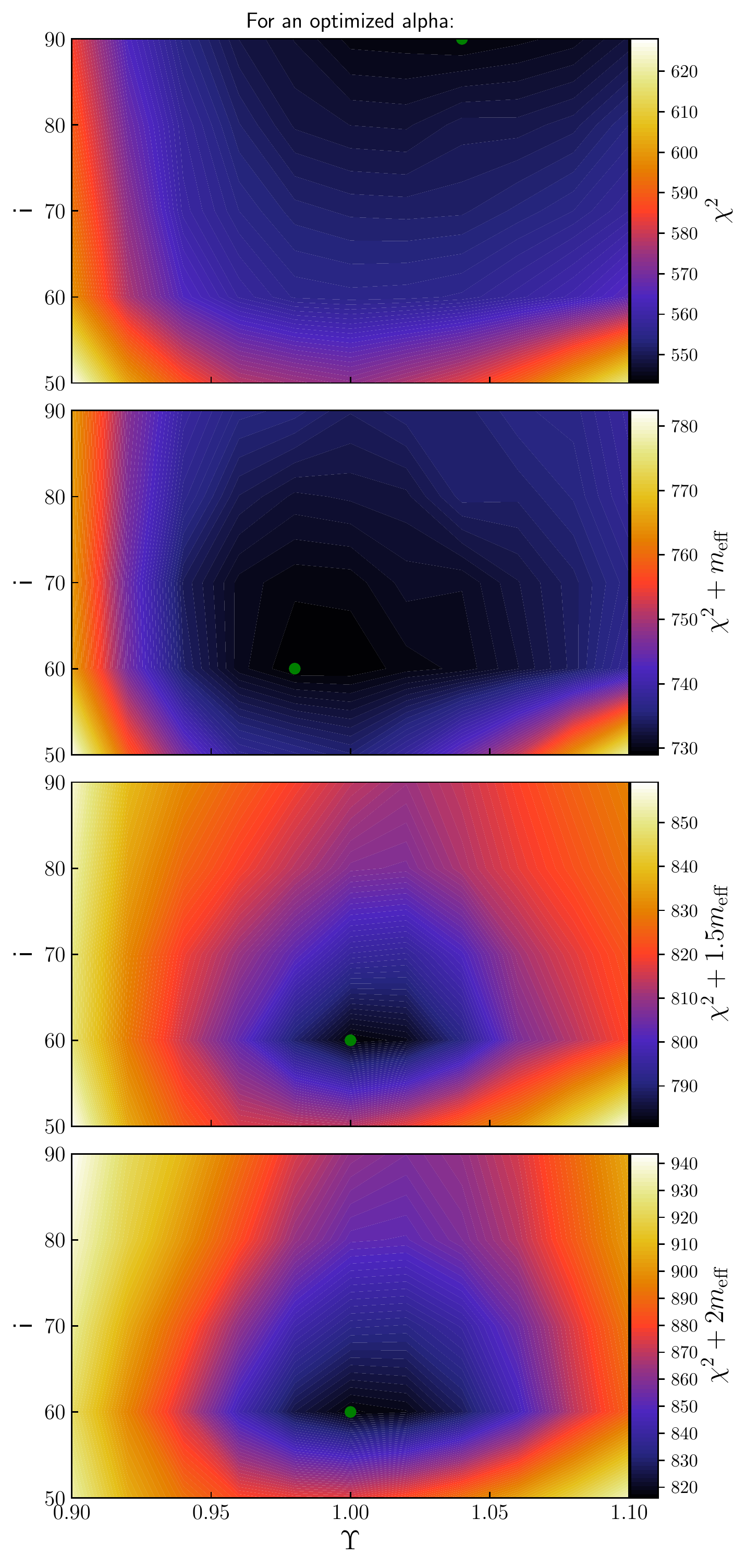}
    \caption{As Fig.~\ref{fig:GalD_for_fixed_alpha} , but the constraints on the selection parameters i and $\Upsilon$ of Galaxy D are shown for the case where $\alpha$ is treated as a free selection parameter. Compared to the case of a fixed $\alpha$ (Fig.~\ref{fig:GalD_for_fixed_alpha}) two important differences can be observed: (i) the constraints on the selection parameters have {\it improved significantly}; (ii) the results are stable, irrespective of the exact value of the weighting factor $w_m$ in the model selection. In particular, the Akaike approach leads to an unbiased model selection now. }
    \label{fig:GalD_for_variable_alpha}
\end{figure}

Treating the regularization as an additional variable selection parameter within our model selection framework, instead of arbitrarily fixing it to some value, does not merely affect the recovery of the intrinsic kinematics but also improves the constraining power of the other involved selection parameters. This is illustrated in Fig.~\ref{fig:GalD_for_fixed_alpha} and Fig.~\ref{fig:GalD_for_variable_alpha}, which show the average, recovered mass-to-light ratio and inclination of Galaxy D for different model selection frameworks for two cases: (i) the often adopted approach where the smoothing parameter is fixed to some value (Fig.~\ref{fig:GalD_for_fixed_alpha}, where $\alpha$ is fixed to 1) and (ii) the more general case where $\alpha$ is a variable model parameter (Fig.~\ref{fig:GalD_for_variable_alpha}).


\begin{figure}
	\includegraphics[width=\columnwidth]{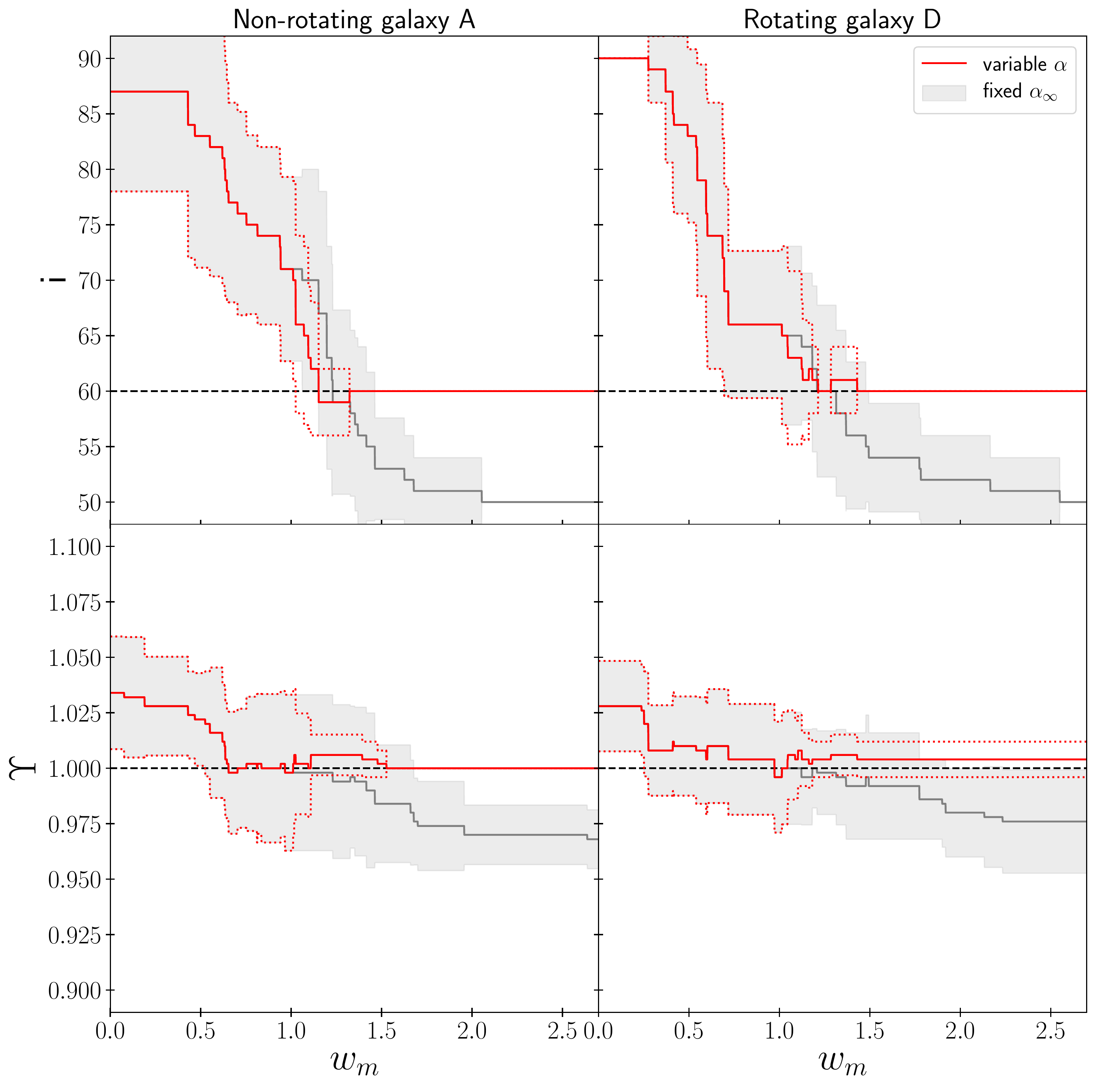}
    \caption{The recovered selection parameters of Galaxy A (\textit{left}) and Galaxy D (\textit{right}) using different model selection frameworks $\chi^2+w_{m} \, m_\mathrm{eff}$. \textit{Dashed, black line:} The correct values for $i$ and $\Upsilon$. \textit{Solid, red Line:} Recovered $\Upsilon$ and $i$ averaged over the 10 mocks of the galaxy if the regularisation parameter $\alpha$ is optimized like the other selection parameters during the fit. \textit{Dotted, red line:} The corresponding $\pm1\sigma$ error intervals. \textit{Solid, grey line:} Recovered properties when modelling with a fixed regularization $\alpha_{\infty}$. This $\alpha$ would be used if one adopts the $\chi^2(\alpha)\rightarrow const.$ criterion to chose the amount of regularization. \textit{Grey regions:} The corresponding $\pm1\sigma$ confidence regions. For models with fixed regularisation, there is a relatively small range of $w_m$ which lead to completely unbiased results for both, Galaxy A and D. Specifically, the Akaike approach is not the optimal choice for weighting the $m_\mathrm{eff}$. When the regularization is optimized, however, the model selection results become unbiased for a wide range of $w_m$ above $\approx 1.5$ and recover the correct selection parameters for both mock galaxies. Only when $w_m\gtrsim4$ the results become biased again. This suggests that model selection is optimal in the Akaike regime as long as the regularization is optimized as well.}
    \label{fig:recovery_vs_wm_GalA}
\end{figure}

As established in section~\ref{sec:mass_parameters}, a simple $\chi^2$ minimization results in an overestimation of the mass-to-light ratio of about $3$-$4\%$ and an edge-on viewing angle for both, the modelling with fixed and the one with variable $\alpha$. Only when the varying number of effective parameters of candidate models are accounted for with the model selection framework $\chi^2+w_{m} \, m_\mathrm{eff}$, one can find the correct selection parameters $\Upsilon$ and $i$ . The constraining power of the model selection framework significantly improves if the regularization is treated as a selection parameter. In addition, the results are more stable with respect to the calibration weight $w_{m}$. In fact, compared to the case with fixed regularisation, the Akaike approach yields unbiased results when the amount of regularisation is optimized during the fit. This is further illustrated in the top panels of Fig~.\ref{fig:recovery_vs_wm_GalA}, where we show the average recovered
selection parameters $i$ and $\Upsilon$ against the calibration weight $w_m$ for Galaxy A (\textit{left}) and Galaxy D (\textit{right}). With optimized regularisation, the model selection results are stable beyond $w_m \ga 1.5$, while modelling runs where we kept the regularisation to a fixed value developed a bias to less flexible models beyond $w_m > 1.5$. For the modelling with variable $\alpha$ in the Akaike regime we further note a significantly reduced scatter of both $\Upsilon$ and $i$ that appears to become smaller than our sampling steps. 

\begin{figure}
	\includegraphics[width=\columnwidth]{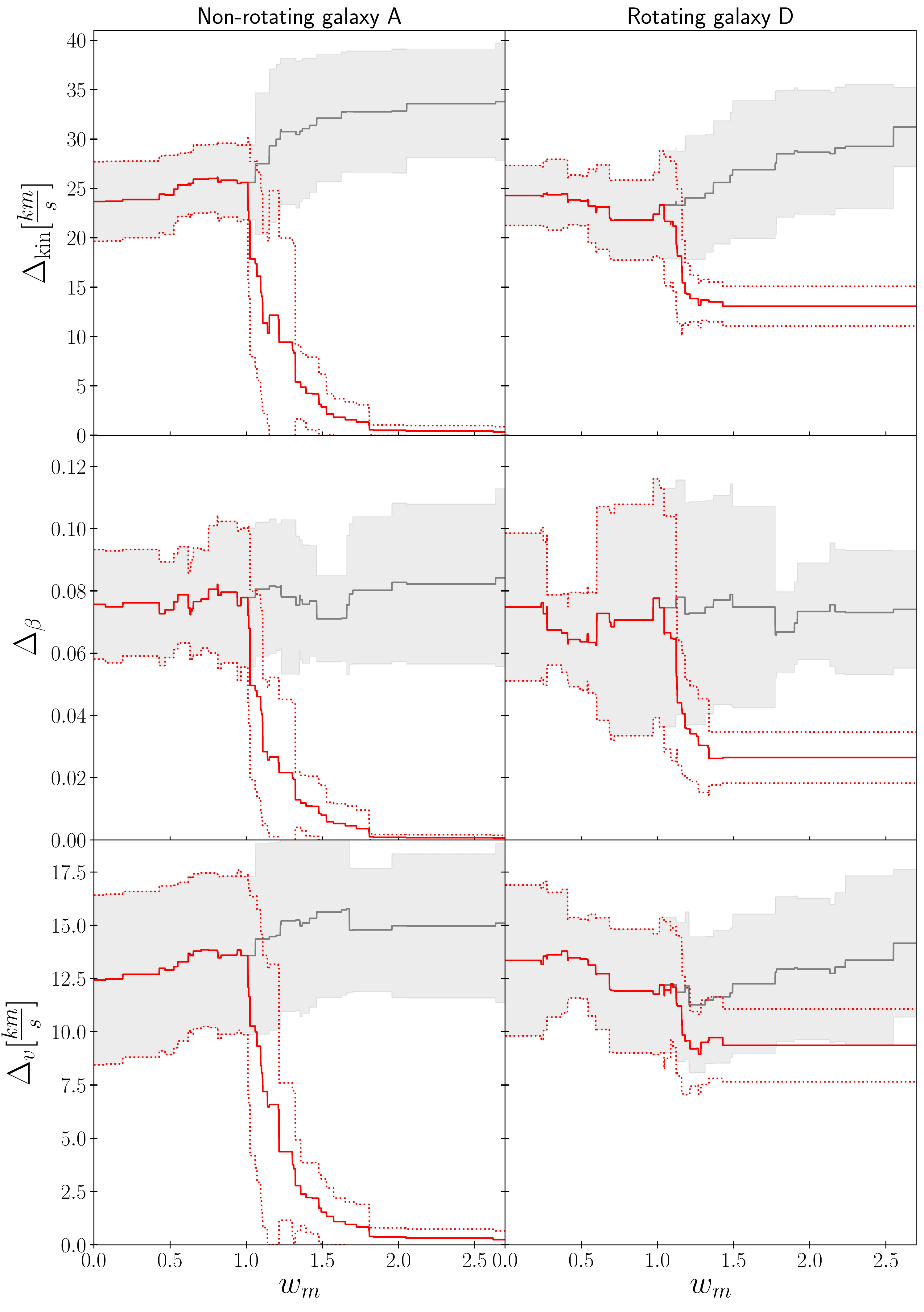}
    \caption{Similar to Fig.~\ref{fig:recovery_vs_wm_GalA}, but for the recovery of the intrinsic velocity moments of galaxies A (\textit{left}) and D (\textit{right}). \textit{Top panel:} The total kinematic Root-mean-square deviations $\Delta_{\mathrm{kin}}$ of the models selected with $\chi^2+w_{m} m_\mathrm{eff}$. \textit{Middle panel:} The corresponding RMSDs $\Delta_{\beta}$ of the anisotropy parameter $\beta$. \textit{Bottom panel:} The RMSDs for the net rotation, $\Delta_{v}$. Both the anisotropy and rotation are emulated well only if $\alpha$ is optimized simultaneously to the other selection parameters and the best model is selected using $\chi^2+w_m m_{\mathrm{eff}}$ with a $w_{m}$ in the Akaike regime. For Galaxy D the Akaike approach with $\alpha$ as a free selection parameter also selects models with intrinsic moments closer to the true kinematics, albeit not as close as it is the case for Galaxy A.
    } 
    \label{fig:kinematic_vs_wm_GalA}
\end{figure}

Given the strong improvements on the constraints for the global selection parameters, we now investigate the intrinsic velocity distributions (the orbital anisotropy) under optimized regularisation. In Fig.~\ref{fig:kinematic_vs_wm_GalA} we show the total kinematic RMS deviations $\Delta_{\mathrm{kin}}$ of the best models to Galaxy A and D with respect to the kinematics of the generating model. The Figure also includes the RMSDs of the net velocity $v$ and the anisotropy parameter $\beta=1-\sigma_{t}^2/\sigma_{r}^2$. 

For both galaxies the approach with fixed and variable regularization are initially congruent for small $w_{m} \la 0.3$ as the model selection is dominated by $\chi^2$ and consequently, even in the case of variable regularisation, models with $\alpha$ values as large as the one that we assumed for the fixed regularisation case are favoured. When $\alpha$ is fixed, an increase of $w_{m}$ beyond $w_m \ga 0.3$ counteracts the inclination and mass-to-light ratio biases, while the recovery of the intrinsic kinematics does not improve. For $\alpha \sim 1$ a "sweet-spot" $1.2 \la w_{m} \la1.5$ appears, where the correct inclination and mass-to-light ratio are obtained on average. This is in line with our calibration of the optimal $w_m$ in $\chi^2+w_{m} \, m_\mathrm{eff}$ for $\alpha=1.67$ described in Sec.~\ref{sec:simulated}. A further increase of $w_{m}$ overcompensates the mass-to-light ratio and inclination biases, suggesting that the choice of the specific model selection framework would be crucial when modelling real galaxies. 

Remarkably, the choice of the calibration weight is simplified when the candidate models are optimized with respect to the degree of regularization: the correct selection parameters are recovered for a wide range of $w_{m}$, including the Akaike $w_m = 2.0$. Moreover, the recovery of the intrinsic kinematics, i.e. the rotation and the anisotropy in the second order velocity moments, of both toy galaxies have {\it significantly improved} by treating the regularization as an additional selection parameter in the construction of the candidate models. With the right model selection, the intrinsic anisotropy can be recovered more accurately with uncertainties no larger than $\Delta \beta \sim 0.04$ in the mean. 

Simply fixing the regularization complicates rather than simplifies the selection of the best candidate model and in the worst case may even skew the results for other selection selection parameters. Therefore dynamical modelling of galaxies with an a priori fixed amount of regularization should be avoided if possible. For example, \citet{2007MNRAS.382..657T} show for their sample of Coma galaxies that whether a galaxy is radially anisotropic or tangentially anisotropic is rather independent of the assumed regularisation. However, the strength of the actual anisotropy, i.e. how much radial or how much tangential the orbit distribution is, does indeed depend on the amount of regularisation applied in the models.

\section{A real Galaxy: Dynamical Modelling of NGC 3368}
\label{sec:NGC3368}
We now turn to the application of our model selection approach to the real galaxy NGC 3368. This galaxy is part of the SINFONI black hole survey \citep[cf.][]{2016ApJ...818...47S} and was chosen by us as follow-up of a previous analysis by \citet{2010MNRAS.403..646N}. 

Due to the fact that NGC 3368 is a disk galaxy, its inclination can be estimated independently of the dynamical modelling. It is therefore a useful testbed for the inclination recovery. The observed ellipticity $\epsilon\approx 0.37$ of the outer disk implies a minimum inclination of $i_{\mathrm{min}}=51\degr$ for an axisymmetric razor-thin disk. Assuming a typical (intrinsic) disk axis ratio of $q\approx0.2$, NGC3368 is expected to be inclined at a slightly larger angle of $i=53\degr$ to project to the observed ellipticity. \citet{2010MNRAS.403..646N} also estimated the inclination of NGC3368 via the Tully-Fisher relation, which yields $i\approx48\degr$.

In section~\ref{subsec:NGC3368_data} we will provide a short overview of the imaging and spectroscopy data. More details can be found in \citet{2010MNRAS.403..646N}. Section~\ref{subsec:NGC3368_chi2} reviews the modelling results obtained from a traditional $\chi^2$ minimisation at fixed regularization. Model selection results with optimized regularisation follow in Sec.~\ref{subsec:NGC3368_var_alpha}.

\subsection{Overview - Photometric and kinematic data for NGC 3368}
\label{subsec:NGC3368_data}
NGC 3368 is a double-barred spiral galaxy with a composite bulge consisting of a larger pseudobulge and a smaller classical bulge \citep[cf.][]{2004A&A...415..941E}. The galaxy is classified as a LINER2 \citep{1997ApJS..112..315H}. The activity is so weak that we do not expect it to influence the dynamical modelling results. We assume that NGC 3368 lies at a distance of $d=10.4\mathrm{Mpc}$ \citep{2001ApJ...546..681T} where 1$\arcsec$ corresponds to $50.4 \, \mathrm{pc}$.

The SINFONI kinematics of NGC 3368 were obtained with the K-band grating in the $100 \mathrm{mas}$ resolution mode. The average adaptive-optics corrected PSF had a FWHM of $\approx0.165\arcsec$. Non-parametric LOSVDs were obtained from the CO bandheads adopting the approach of \citet{2000AJ....119.1157G}. More details on the data reduction process, AO correction, selection of template stars, and the resulting kinematic maps of the galaxies can be found in \citet{2010MNRAS.403..646N}. 

The photometry acquired for NGC 3368 is a combination of different data sets: Sloan Digital Sky Survey observations \citep[][]{2000AJ....120.1579Y} in the r-band were used to find the ellipticities of the outermost parts of the galaxy's disk. Intermediate radii of NGC 3368 are mostly covered by a dust-corrected HST NICMOS2 F160W image and a K-band (2.2$\mu m$) image of \citet{2003MNRAS.344..527K} observed with the Isaac Newton Group Red Imaging Device (INGRID). The photometry is completed by the collapsed K-band data cube images taken with the integral field spectrograph SINFONI at the Very Large Telescope (VLT). For more detailed information on the image matching, seeing, dust-correction and an isophote analysis see \citet{2010MNRAS.403..646N}.

As noted before, the photometric bulge of NGC 3368 has a composite structure. The central 2$\arcsec$ are dominated by an almost round and kinematically hot "classical" bulge. Outside this region a more flattened and  kinematically cooler structure emerges: the disky pseudobulge. Since the SINFONI FoV has a size of $3\arcsec \times 3 \arcsec$, both components are relevant for the modelling. To account for potentially different mass-to-light ratios in the respective stellar populations, we deproject the classical bulge separately from the pseudobulge and assign each deprojection its own mass-to-light ratio $\Upsilon$. To this end, 
we adopt the photometric bulge-disk decomposition of \citet{2010MNRAS.403..646N}. It assumes that the disky pseudobulge is the inner extension of the galaxy's large scale disk and both are combined into the photometric disk component. The other photometric component, respectively, consists only of the "classical" bulge. In practice, the deprojections were again obtained using the Metropolis-Algorithm of \citet{1999MNRAS.302..530M} without any shape prior. Non-axisymmetric features in the photometric data, like bars and spiral arms, are averaged over as described in \citet{2010MNRAS.403..646N}. We tested a grid of assumed inclinations ranging from $i=53\degr$ to $i=90\degr$. In the following we simply refer to the classical bulge component as the bulge, and the pseudobulge+disk as the disk component.

\subsection{$\chi^2$-modelling of NGC 3368}
\label{subsec:NGC3368_chi2}
This section illustrates the results of the "traditional" dynamical modelling approach for NGC 3368, which is based (i) on evaluating different mass models based on a pure $\chi^2$-analysis (i.e. assuming $m_\mathrm{eff} = \mathrm{const}$) and (ii) on a fixed strength of the regularisation. These results are later contrasted to the model selection framework with optimized regularisation (Sec.~\ref{subsec:NGC3368_var_alpha}).

For our final mass model we add a central SMBH to the two stellar mass components:
\begin{equation}
\rho=M_{\bullet}\cdot \frac{\delta\left(r\right)}{4\pi r^2}+\Upsilon_{\mathrm{bulge}} \cdot \nu_{\mathrm{bulge}}(i)+\Upsilon_{\mathrm{disk}} \cdot \nu_{\mathrm{disk}}(i),
\label{eqn:mass model_3368}
\end{equation}
where $\nu_{\mathrm{bulge}}(i)$ and $\nu_{\mathrm{disk}}(i)$ are the luminosity densities at assumed inclination $i$. For the $\chi^2$ analysis of this Section we considered only two inclinations: $i=53\degr$ and $i=90\degr$. A denser inclination grid is used for the model selection approach (Sec.~\ref{subsec:NGC3368_var_alpha}). Even though it seems likely that a considerable amount of the total galaxy mass is contributed by a dark matter (DM) halo, the addition of a DM component to this model of the galaxy center is not required. \citet{2018MNRAS.473.2251E} demonstrated that the addition of a DM halo to a two-component stellar model of (the inner parts of) a disk galaxy neither improves the fit significantly, nor does it change the black hole mass $M_{\bullet}$ or bulge mass-to-light ratio $\Upsilon_{\mathrm{bulge}}$ drastically. It's only that the disk mass-to-light ratio $\Upsilon_{\mathrm{disk}}$ of such a two-component model without DM is larger than the $\Upsilon_{\mathrm{disk}}$ of a mass model that does include DM (and probably larger than the actual $\Upsilon$ of the stars in the disk). This is because the disk "absorbs" the dynamical role of the ignored dark mass, which leads to a $\Upsilon_{\mathrm{disk}}$ that is biased high.

With the above mass model we constructed trial models by varying the three traditionally relevant selection parameters that determine the orbit library: the mass-to-light ratios of bulge and disk, the black hole mass. With the goal to achieve an efficient, yet nonetheless sufficiently dense sampling of these parameters we narrowed down the approximate location of the $\chi^2$ minimum with an initial round of trial models that sparsely sampled a wide range of mass parameters. Using the information gained from this initial $\chi^2$ minimization we decided for the following final sampling grid: The black hole mass is sampled linearly in the interval [$1.0\cdot 10^6M_{\sun}$, $19.0\cdot 10^6M_{\sun}$] with a stepsize of $2.0\cdot 10^6M_{\sun}$ and the bulge mass-to-light ratio is linearly sampled 15 times in the interval [$0.20$, $0.90$]. It turned out that the SINFONI FoV is too small (inner $3\arcsec$ of the galaxy) to constrain $\Upsilon_\mathrm{disk}$ . Therefore we decided to sample only $\Upsilon_\mathrm{disk} \in \left\{0.2,0.4,0.6 \right\}$ to save computation time. This indeterminacy of $\Upsilon_{\mathrm{disk}}$ was already noted in \citet{2010MNRAS.403..646N}. For the smoothing we applied the old approach to set $\alpha=1.67$ guaranteeing that $\chi^2(\alpha)$ has converged. 

Estimating the \textit{statistical} errors of the dynamical models is non-trivial, mostly because unlike in our toy galaxies, the measurement errors of real LOSVDs are strongly correlated \citep[e.g.][]{2006MNRAS.367....2H}. If these correlations are unknown or neglected then the calculated $\chi^2$ is systematically smaller than $N_\mathrm{data}-m_\mathrm{eff}$ (cf. Tabs.~\ref{tab:3368_chi2} and \ref{tab:3368_variable_alpha}). The absolute $\chi^2$ values and, consequently, also the $\chi^2+2 \, m_\mathrm{eff}$ significance intervals become inaccurate then. Ideally this issue could be solved by taking the full covariance matrix of the observations into account in the modelling, we plan to investigate this in the future. Another way to circumvent the issue would be the modelling of multiple observations of the same galaxy. Then one could calculate the RMS of the recovered selection parameters, analogously to the modelling of 10 different mocks of our toy galaxies (e.g. Sec.~\ref{sec:simulated}). In practice, this is unfeasible. However, if the investigated galaxy is perfectly axisymmetric then each quadrant provides essentially an independent observation of the same underlying galaxy structure. This simple approach of error estimation can also be adopted for galaxies that deviate from axisymmetry, in which case the resulting RMS will be increased by the systematic structural differences between the quadrants \citep[cf.][]{2010MNRAS.403..646N,2013AJ....146...45R}. 
 
\begin{figure}
	\includegraphics[width=\columnwidth]{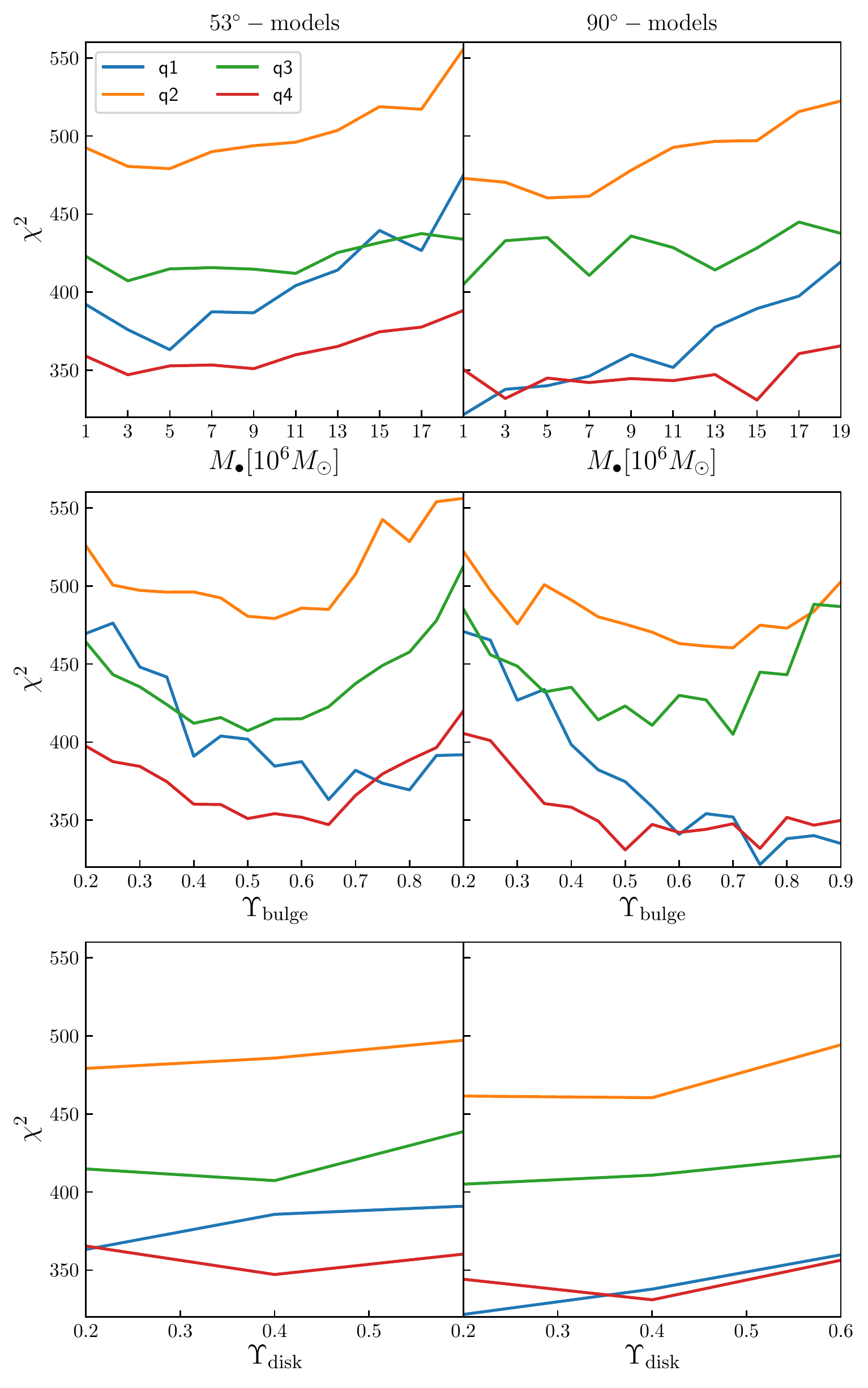}
    \caption{\textit{Left:} $\chi^2$ constraints for the mass parameters of the orbit models of NGC 3368 assuming an inclination of $i=53\degr$. \textit{Right:} The corresponding $\chi^2$ constraints assuming the galaxy is seen edge-on. We modelled the kinematic data of the four galaxy quadrants separately (colored curves; the nomenclature of the quadrants is the same as in \citet{2010MNRAS.403..646N}). For all four quadrants an edge-on model achieves a better fit than the more inclined model.}
    \label{fig:chi2_3368}
\end{figure}

Fig.~\ref{fig:chi2_3368} illustrates the resulting $\chi^2$ constraints for the mass-to-light ratios and black holes mass for both, the $53\degr$ models and the $90\degr$ models in the four galaxy quadrants. 
The models obtained under an assumed inclination of $i=53\degr$ (motivated by the disk flattening as described above) yield black hole masses $M_{\bullet} \in [3\cdot 10^6 M_{\sun},5\cdot 10^6 M_{\sun}]$. The formal average over the four quadrants is $M_{\bullet}=(4.0\pm1.0)\cdot 10^{6}M_{\sun}$. However, in some quadrants black hole masses of up to $13\cdot 10^6 M_{\sun}$ are hardly ruled out when looking at the detailed $\chi^2$ curves. The mass-to-light ratio of the bulge is well constrained with an average value of $\Upsilon_{\mathrm{bulge}}=0.59\pm0.07$. As already mentioned, the disk mass-to-light ratio is essentially unconstrained by the SINFONI kinematics. The $\chi^2$ constraints for the black hole mass at $i=90\degr$ are much noisier than at $i=53\degr$. Averaged over the four quadrants, we find $M_{\bullet}=(5.5\pm5.7)\cdot 10^{6}M_{\sun}$ (with a more than five times larger scatter than in the $i=53\degr$ case). In fact, neither very small black black hole masses $M_{\bullet}<1\cdot10^{6}M_{\sun}$ nor very large ones  $> 15\cdot10^{6}M_{\sun}$ can be ruled out at a significant confidence level. While we do not observe such a strongly increased scatter in $\Upsilon_{\mathrm{bulge}}$ we do see evidence for a slight shift towards larger mass-to-light ratios when modelling the galaxy edge-on: $\Upsilon_{\mathrm{bulge}}=0.66\pm0.10$. 

In the case of NGC 3368 the observed disk ellipticity strongly suggests that the models at  $i=53\degr$ are more realistic than the ones at $i=90\degr$, but  we do not have such prior information about a galaxy's inclination in every case, in particular not if it is a generic early-type galaxy. In that case one would be reliant on the $\chi^2$ values of the differently inclined models. Table~\ref{tab:3368_chi2} shows the selection parameters and corresponding minimum $\chi^2$ values of the best fitting model for each quadrant and inclination. In all quadrants, the edge-on models have a lower $\chi^2$ value than the models at $i=53\degr$. In other words, if NGC 3368 would be modelled without external knowledge about its inclination a pure $\chi^2$ analysis would erroneously conclude that the galaxy is seen edge-on, when, in fact, an independent inclination measurement would indicate  $i \approx 48\degr-55\degr$. 
\begin{table*}
	\centering
	\caption{Best fitting models according to a simple $\chi^2$ minimization. \textit{Column 1-2:} The modelled quadrant of NGC 3368 and the assumed inclination. \textit{Columns 3-5:} K-Band mass-to-light ratios $\Upsilon_{\mathrm{bulge}}$, $\Upsilon_{\mathrm{disk}}$ and black hole mass of the best fitting model. \textit{Column 6-7:} The $\chi^2$ and $\chi^2/N_{\mathrm{data}}$ of this model. $N_\mathrm{data}=945$.}
	\label{tab:3368_chi2}
	\begin{tabular}{ccccccc} 
    Quadrant & i & $\Upsilon_{\mathrm{bulge}}[M_{\sun}/L_{\sun}]$ & $\Upsilon_{\mathrm{disc}}[M_{\sun}/L_{\sun}]$ & $M_{\bullet}  [10^{6}M_{\sun}]$ & $\chi^2$ & $\frac{\chi^2}{N_{\mathrm{data}}}$  \\ \hline
     1 &  $53\degr$ & 0.65 & 0.20 & 5.0 & 363.230 & 0.3844 \\ 
     2 &  $53\degr$ & 0.55 & 0.20 & 5.0 & 479.117 & 0.5070 \\ 
     3 &  $53\degr$ & 0.50 & 0.40 & 3.0 & 407.302 & 0.4310 \\ 
     4 &  $53\degr$ & 0.65 & 0.40 & 3.0 & 347.183 & 0.3674 \\ 
     1 &  $90\degr$ & 0.75 & 0.20 & 1.0 & 321.628 & 0.3403 \\ 
     2 &  $90\degr$ & 0.70 & 0.40 & 5.0 & 460.396 & 0.4872 \\ 
     3 &  $90\degr$ & 0.70 & 0.20 & 1.0 & 405.026 & 0.4286 \\ 
     4 &  $90\degr$ & 0.50 & 0.40 & 15.0 & 331.003 & 0.3503 \\ 
		
	\end{tabular}
\end{table*} 
These results are not surprising. In Sec.~\ref{sec:simulated} we have seen that a simple $\chi^2$ minimisation tends to yield results that are biased towards $i=90\degr$. Moreover, in Sec.~\ref{sec:mass_parameters} we have seen that this inclination bias comes along with a noticeable mass bias (of the order of 10 per cent). Assuming that NGC3368 is seen under an inclination angle close to the expected $i=53\degr$, than our $\chi^2$ analysis of NGC3368 is fully consistent with our expectations based on the simulated toy galaxies discussed in the previous Sections.

\subsection{Model selection of NGC 3368}
\label{subsec:NGC3368_var_alpha}
In Sec.~\ref{sec:regularization} above, we have demonstrated that the most accurate reconstruction of a galaxy's mass distribution and internal structure is achieved when using a model selection approach rather than a simple $\chi^2$ minimisation. Moreover, treating the regularisation strength in a similar way as other selection parameters turned out (i) to improve the accuracy of the galaxy reconstruction and (ii) to reduce the sensitivity of the results on the weight $w_m$ for the effective parameters $m_\mathrm{eff}$. Here, we transfer this approach to NGC3368 and minimise $\chi^2 + 2.0 \, m_\mathrm{eff}$. The choice for the Akaike weight $w_m = 2.0$ directly follows from the results of Sec.~\ref{sec:regularization}. We adopt the same mass model (Eq.~\ref{eqn:mass model_3368}) as for the $\chi^2$ analysis in the previous Sec.~\ref{subsec:NGC3368_chi2}, consisting of a central black hole, a disk and a bulge. Neglecting potential effects of the deprojection degeneracy this implies that the orbit models are completely determined by the 5 selection parameters $M_{\bullet}$, $\Upsilon_{\mathrm{bulge}}$, $\Upsilon_{\mathrm{disk}}$, $i$, $\alpha$. However, in order to reduce the computation time we restricted ourselves to models with a fixed $\Upsilon_{\mathrm{disk}}=0.4$, since $\Upsilon_\mathrm{disk}$ is almost unconstrained over the SINFONI FoV (Fig.~\ref{fig:chi2_3368}). 

\begin{figure}
	\includegraphics[width=\columnwidth]{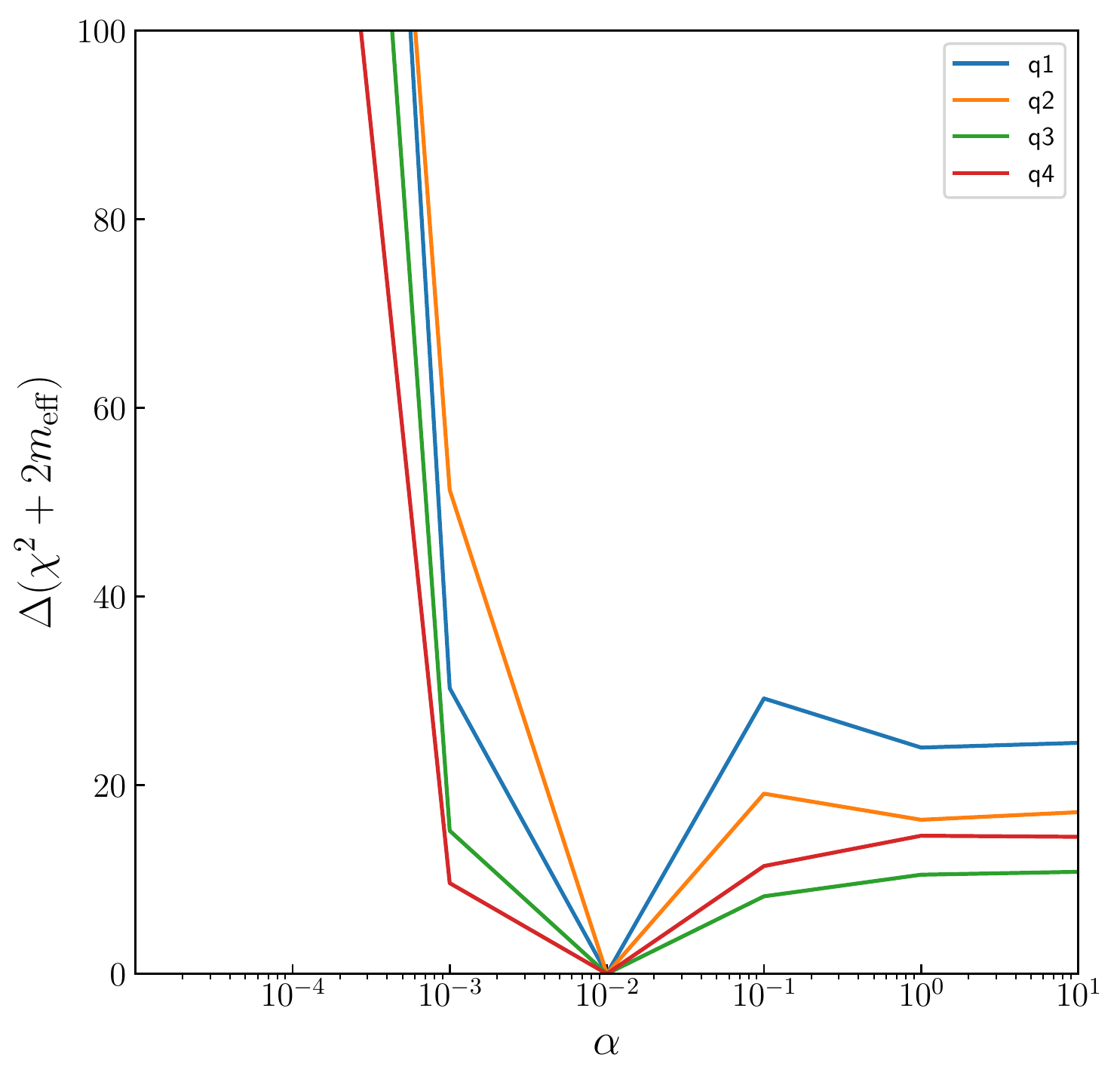}
    \caption{Constraints on the optimal smoothing for the orbit models of NGC 3368. At each $\alpha$, we show the smallest achieved $\chi^2 + 2.0 \, m_\mathrm{eff}$. Different colors represent models of different galaxy quadrants and the minimum $\chi^2 + 2.0 \, m_\mathrm{eff}$ of each quadrant has been subtracted. The optimal smoothing is well determined in each quadrant.}
    \label{fig:regularization_3368}
\end{figure}

\begin{figure}
	\includegraphics[width=\columnwidth]{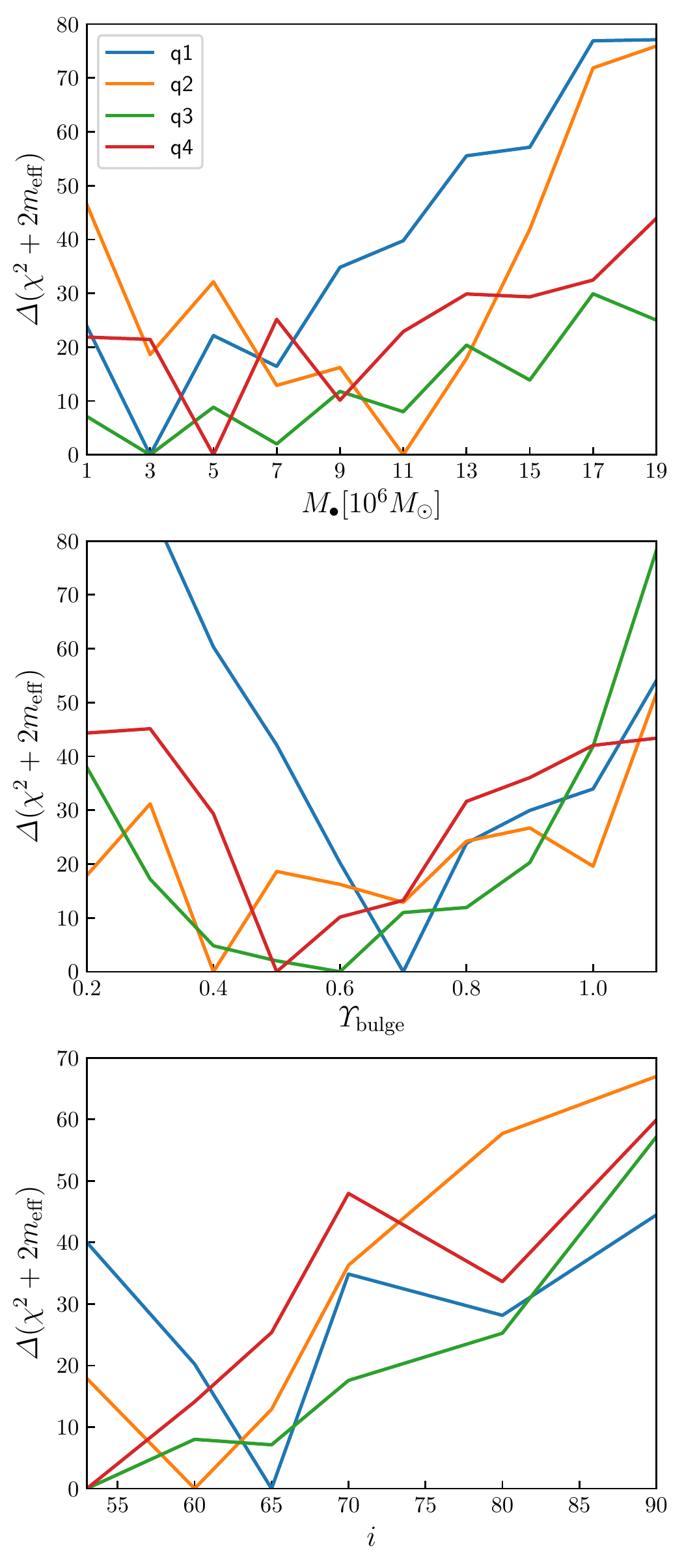}
    \caption{Similar to Fig.~\ref{fig:regularization_3368}, but here we plot the black hole mass (top), bulge mass-to-light ratio (middle) and inclination $i$ (bottom) against $\Delta (\chi^2 + 2.0 \, m_\mathrm{eff})$. }
    \label{fig:AIC_3368}
\end{figure}

Fig.~\ref{fig:regularization_3368} shows that the optimal amount of regularisation is well constrained by the data: within the Akaike model selection framework all four quadrants of NGC 3368 are best modelled with an intermediate regularization of $\alpha=10^{-2}$.  Fig.~\ref{fig:AIC_3368} shows the corresponding constraints for $\Upsilon_\mathrm{bulge}$, $M_\bullet$ and $i$ (all results are listed in Tab.~\ref{tab:3368_variable_alpha}). 
\begin{table*}
	\centering
	\caption{Selection parameters derived from the model selection approach with optimized regularisation for each quadrant of NGC3368. The number of effective parameters $m_\mathrm{eff}$ has been weighted with $w_m = 2.0$ as in the Akaike information criterion and in accordance with the simulations in Sec.~\ref{sec:regularization}.}
	\label{tab:3368_variable_alpha}
	\begin{tabular}{cccccc} 
    Quadrant & $\alpha$ & i & $\Upsilon_{\mathrm{bulge}} [M_{\sun}/L_{\sun}]$ & $M_{\bullet} [10^{6}M_{\sun}]$ & $(\chi^2+ 2.0 \, m_\mathrm{eff})_{\mathrm{min}}$ \\ \hline
     1 &  $10^{-2}$ & 65\degr & 0.7 & 3.0 & 861.02744 \\ 
     2 &  $10^{-2}$ & 60\degr & 0.4 & 11.0 & 934.158201\\ 
     3 &  $10^{-2}$ & 53\degr & 0.6 & 3.0 & 882.207510 \\ 
     4 &  $10^{-2}$ & 53\degr & 0.5 & 5.0 & 812.338033 \\ 
	\end{tabular}
\end{table*}
Averaging the modelling results of the four quadrants we find that the AIC estimation yields a black hole mass of $M_{\bullet}=(5.5\pm3.3)\cdot10^{6}M_{\sun}$, a mass-to-light ratio $\Upsilon_{\mathrm{bulge}}=(0.55\pm0.11)M_{\sun}/L_{\sun}$, and an inclination angle $i=(57.8\pm5.1)\degr$. Compared to the results of the simple $\chi^2$-minimisation in Sec.~\ref{subsec:NGC3368_chi2} we immediately see that the inclination bias has {\it disappeared}. The model selection approach with optimized regularisation yields an inclination that is consistent with the inclination angle derived independently from the disk flattening. This result was expected after the analysis of the toy galaxies above. Still, it is somewhat astonishing given how small the FoV of the SINFONI kinematic data actually is. However, this underlines the power of the model selection approach to extract all the information contained in the 945 measured LOSVD data points in each quadrant. Another difference to the simple $\chi^2$ minimisation is the fact that the model selection approach yields a 10-15 percent smaller $\Upsilon_\mathrm{bulge}$. Again, this is fully consistent with the toy galaxy results, where the recovered masses were biased high by up to $\sim10$ percent when a simple $\chi^2$ minimisation was performed, while there was no mass bias in the model selection framework. The only significant difference to the simulation results is that the scatter in 
$\Upsilon_\mathrm{bulge}$ of NGC3368 is not reduced in the model selection framework compared to the $\chi^2$ minimisation. This could be caused by a multitude of issues:
\begin{enumerate}
    \item Systematic differences between the quadrants of NGC 3368 reflecting the galaxy's intrinsic non-axisymmetry.
    \item Inaccurate errors of the observed LOSVDs.
    \item Negligence of the correlations between the measurements of the same LOSVD at different line-of-sight velocities in the modelling and bootstrap iterations.
    \item A potential non-alignment of disk and bulge.
\end{enumerate}
We used an updated version of our axisymmetric Schwarzschild code for this analysis, compared to \citet{2010MNRAS.403..646N}. Therefore, the results of the pure $\chi^2$ analysis can not be compared easily. However, we note our black hole mass is slightly smaller than the $M_{\bullet}=(7.5\pm1.5) \cdot 10^6  M_{\sun}$ quoted in that paper while our $\Upsilon_\mathrm{bulge}$ is about 25 per cent larger. These small discrepancies could also partly be caused by the different choice of mass parameter sampling and regularization. 

\begin{figure}
	\includegraphics[width=\columnwidth]{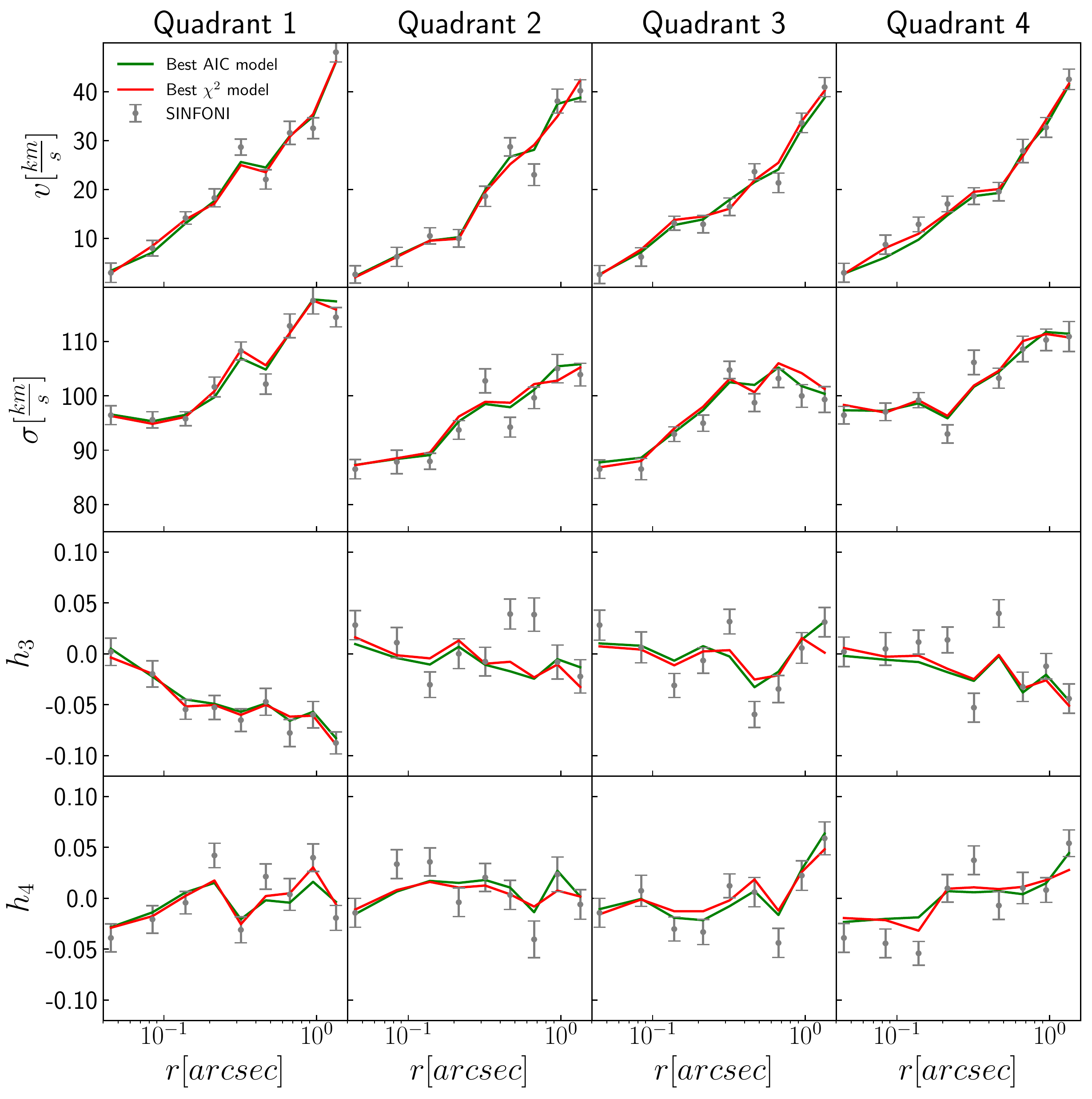}
    \caption{The Gauss-Hermite coefficients along the major axis of NGC 3368. \textit{Grey:} SINFONI observations (the error-bars are estimated neglecting possible correlations in the LOSVD data; \textit{Red:} Projected kinematics of the best $\chi^2$; \textit{Green:} Best model according to AIC.}
    \label{fig:major_axis_fit_3368}
\end{figure}

After selecting a dynamical model with AIC one should always confirm that the selected model can in fact reproduce the observed data well as AIC  only compares the models relative to another but makes no statement about their absolute quality. Fig.~\ref{fig:major_axis_fit_3368} demonstrates this for the models of NGC 3368 by showing the major-axis kinematics of the best AIC and $\chi^2$ models in relation to the observed SINFONI data. The errorbars for the SINFONI data are the $1\sigma$ errors estimated from \textit{uncorrelated} Monte-Carlo realisations of the non-parametric LOSVD data, thus, they are slightly underestimated. Both, AIC and $\chi^2$ models, fit the data well. However, as one would expect, the AIC model appears to be slightly smoother. It does not react to outliers as much as the $\chi^2$ models do. Instead of a validation by eye as done here for the major axis of NGC 3368, it may be advantageous to cross-check the goodness-of-fit of AIC models more systematically by confirming whether $\chi^2+m_{\mathrm{eff}}\sim N_{\mathrm{data}}$ holds for the selected models (cf. the discussion in the next Sec.~\ref{sec:discussion}). However, this is not a viable option as long as the correlations in the real LOSVDs are not implemented in the calculation of $\chi^2$ and $m_{\mathrm{eff}}$ as this causes both values to be underestimated.

\section{Discussion}
\label{sec:discussion}
In the following we examine the \textit{statistical} objective of AIC model selection in the context of dynamical modelling and address potential problems that may arise when estimating intrinsically degenerate properties with AIC (Sec.~\ref{AIC-degneracies}). Subsequently we review earlier studies attempting an inclination recovery using axisymmetric models (Sec.~\ref{inc-recovery-context}) and discuss the implications for the mass reconstruction when selecting models based on their $\chi^2$ alone (Sec.~\ref{mass-reconstruction}).

\subsection{Model selection and intrinsic degeneracies}
\label{AIC-degneracies}
\begin{figure}
	\includegraphics[width=\columnwidth]{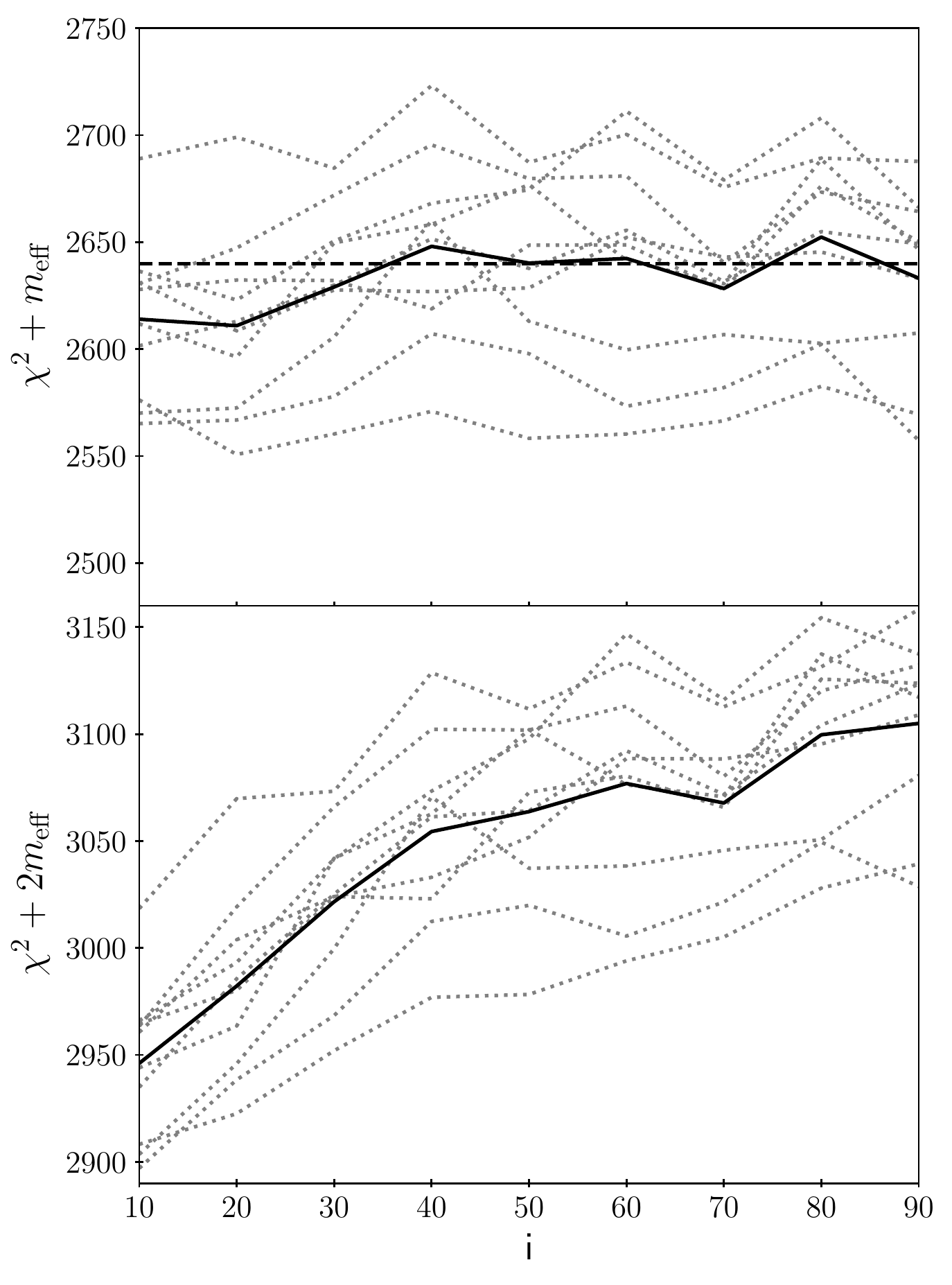}
    \caption{Model selection results for the Hernquist sphere (cf. Fig.~\ref{fig:spherical_chi2} for the selection using only $\chi^2$). \textit{Dotted lines:} The selected best models for the individual mocks. \textit{Solid Line:} The corresponding mean. \textit{Dashed line:} The number of kinematic data points. For this spherical galaxy AIC appears to favor the models with the smallest inclination which also have the smallest number of effective parameters. The $\chi^2+m_\mathrm{eff}$ curves exhibit the inclination degeneracy one would expect for a spherical galaxy: $\chi^2+m_{\mathrm{eff}} \sim N_{\mathrm{data}}$.}
    \label{fig:nbody_vs_inclination}
\end{figure}

The toy galaxy recoveries discussed in Secs.~\ref{sec:simulated} and \ref{sec:regularization} had in common that the selection parameters under investigation had a unique "true" value. In the model selection setup, this true value could be recovered in an unbiased way with high precision. In order to assess how the model selection framework operates in a degenerate case where such a true value does not exist, or where the data is insufficient to constrain a selection parameter, it is useful to get back to the example of the inclination recovery in a sphere (cf. Sec.~\ref{subsec:spherical}): in a kinematically isotropic sphere all viewing angles should be equivalent. Fig.~\ref{fig:nbody_vs_inclination} shows $\chi^2+m_\mathrm{eff}$ and $\chi^2+2.0 \, m_\mathrm{eff}$ versus the inclination for the same N-body models as in Fig.~\ref{fig:spherical_chi2} (cf. Sec.~\ref{subsec:spherical}). At each tested inclination we optimized the regularization strength $\alpha$ as in Sec.~\ref{sec:regularization}. Apart from the regularization we only varied the inclination, meaning that the orbit libraries are the exact same for all models. In contrast to the non-spherical toy galaxies that we tested in the previous sections, the AIC approach now appears to be biased, in the sense that it prefers low inclinations although all viewing angles should be equivalent (lower panel of Fig.~\ref{fig:nbody_vs_inclination}). The "intuitive" model selection framework displays the degeneracy one would generally expect for a spherical model: no inclination is preferred over the other (upper panel of Fig.~\ref{fig:nbody_vs_inclination}). We will come back to this below.

The reason why AIC assesses models of a spherical galaxy differently depending on the inclination can be understood by elaborating what the Akaike criterion is designed to achieve in a more general, statistical sense. Coming from information theory AIC estimates the information loss when modelling an underlying structure/process with a statistical model and selects the model that has the least information loss. The information loss of a statistical model is generally quantified by the Kullback–Leibler divergence (KLD) between the fitted model and the underlying (noise-free) structure it is supposed to represent. Naturally, since the latter is usually unknown, AIC is merely an estimate of the actual information loss. 

When transferring these considerations to the context of dynamical models it is crucial to clarify what the \textit{statistical} model and the underlying structure actually are in this case. The data being fitted by dynamical models are the observed kinematic data, implying that statistically speaking the underlying structure are the noise-free LOSVDs $\mathbfit{l}_{\mathrm{true}}$ and not the galaxy's distribution function or mass structure. Analogously, the statistical model is not the orbit model itself but is given by the LOSVDs $\mathbfit{l}_\mathrm{mod}$ it produces (cf. Sec.~\ref{sec:overview+problem}). The internal structure of the model, i.e. how one arrives at $\mathbfit{l}_\mathrm{mod}$, is only of secondary concern when evaluating the models.

What this means for the modelling of the isotropic N-body sphere is that the AIC is supposed to minimize the differences between the model LOSVDs and the true LOSVDs. As mentioned above, this difference is quantified by the KLD from $\mathbfit{l}_\mathrm{mod}$ to $\mathbfit{l}_{\mathrm{true}}$ but can also be illustrated more heuristically by the RMSD $\Delta \mathbfit{l}$ of the LOSVDs. 

\begin{figure}
	\includegraphics[width=\columnwidth]{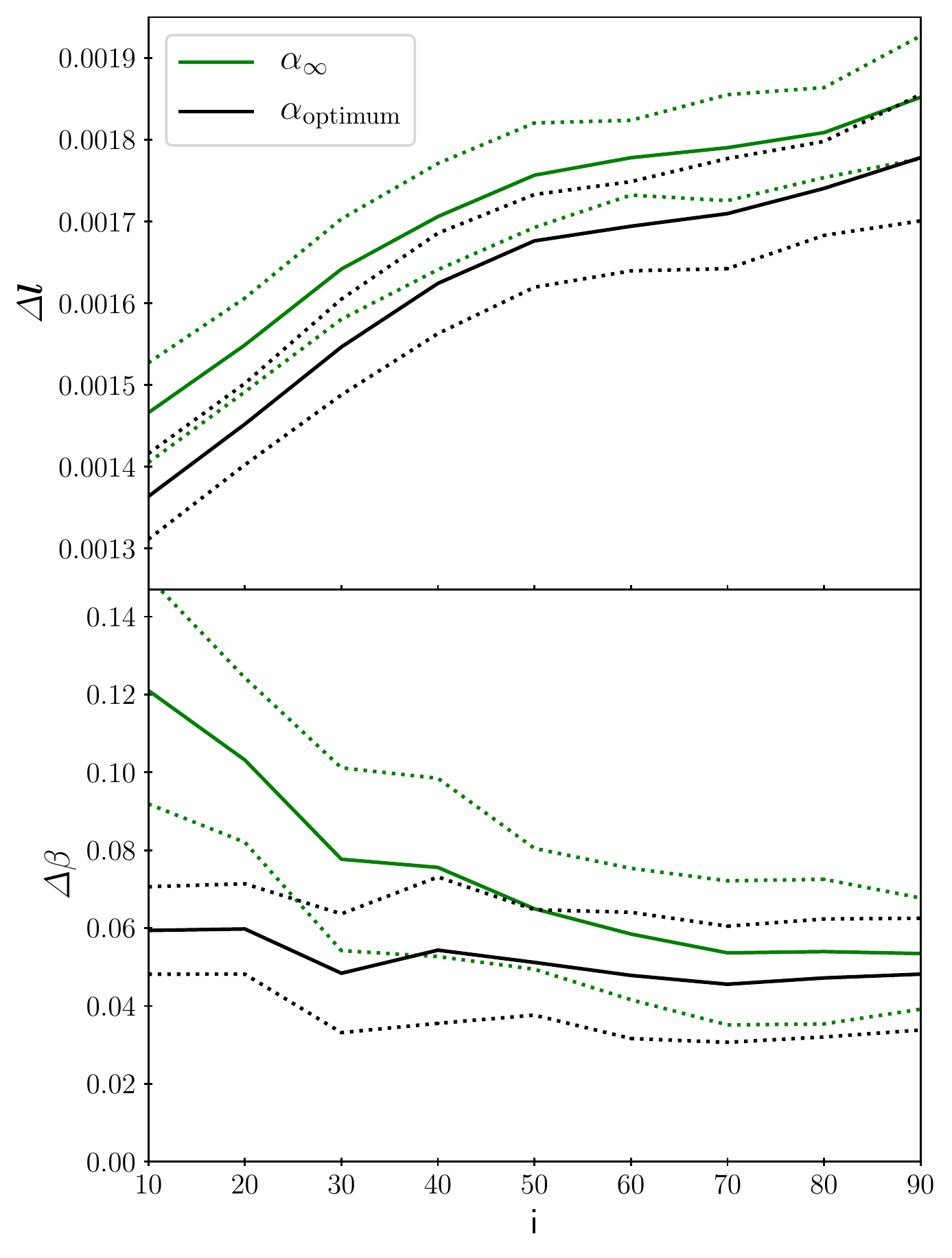}
    \caption{\textit{Top panel:} The LOSVD RMSDs of the orbit models for the Hernquist sphere. \textit{Solid lines:} The arithmetic mean of the 10 mocks. \textit{Dotted line:} The corresponding estimated $1\sigma$ errors. Models with $\alpha_{\infty}$ are shown in \textit{green} and models where the regularization was optimized using AIC are shown in \textit{black}. These latter models are the same as in Fig~\ref{fig:nbody_vs_inclination}. \textit{Bottom panel:} As the upper panel but for the RMSD of the anisotropy parameter $\beta$. Selecting models by their AIC values minimizes the information loss and leads to an optimized recovery of the underlying LOSVDs, unlike a selection based on $\chi^2$ alone which prefers edge-on models with $\alpha_{\infty}$. For the spherical galaxy the models that approximate the Hernquist LOSVDs best are at $i=10\degr$ and have $\alpha_{\mathrm{optimum}}=10^{-2}$. However, at fixed regularisation, improving the recovery of the LOSVDs does not necessarily entail an improved recovery of the internal kinematics as illustrated by the green lines.}
    \label{fig:nbody_vs_RMSD_LOSVD}
\end{figure}

As shown in the top panel of Fig.~\ref{fig:nbody_vs_RMSD_LOSVD}, $\Delta \mathbfit{l}$ behaves indeed very similar to $\chi^2+2 \, m_{\mathrm{eff}}$ (cf. Fig~\ref{fig:nbody_vs_inclination}): both have their minimum where the orbit library is viewed close to face-on.
In that sense the AIC selection achieves exactly what it is intended to do, namely to select the model that emulates the underlying LOSVDs best. That the true LOSVDs happen to be represented slightly better at one inclination than another is related to the design of our model. While the sphere itself has no preferred viewing angles, our axisymmetric model {\it does} have such a preferred axis. E.g., while we have pairs of orbits that only differ in the sign of $L_z$ we do not have the equivalent pairs of orbits that only differ in the sign of other angular-momentum components. We expect that an orbit sampling using a 5d starting space \citep{2020MNRAS.500.1437N} will lead to a different behaviour with respect to the assumed viewing angles in a sphere.

Irrespective of the question which particular inclination happens to yield the {\it best LOSVDs} we can ask whether a selection parameter (here: the inclination) is constrained by the data or not. As already mentioned above, the top panel of Fig.~\ref{fig:nbody_vs_inclination} shows that $\chi^2 + m_\mathrm{eff} \sim N_\mathrm{data}$ at all inclinations. This is exactly the expected behaviour when the data does not constrain the exact value of $i$: For a viable statistical model we expect  $E(\chi^2) = N_\mathrm{data}-m_\mathrm{eff}$ (Sec.~\ref{sec:ansaetze}) and if this can be achieved for all $i$, then the data does not constrain the inclination. Among all these viable models the model selection picks up the one with the smallest $m_\mathrm{eff}$. This is a generic result, because when $\chi^2 + m_\mathrm{eff} \sim N_\mathrm{data}$, then 
\begin{equation}
\chi^2 + 2 \, m_\mathrm{eff} \sim N_\mathrm{data} + m_\mathrm{eff}. 
\end{equation}
In other words, when a selection parameter is not constrained by the data and $\chi^2 + m_\mathrm{eff} \sim N_\mathrm{data}$ over some extended interval, then the model selection may be biased in the sense that it will pick up that particular model inside the interval which has the lowest $m_\mathrm{eff}$. 

In terms of the selection parameter recovery such a bias is undesirable. However, as we have seen above, it still comes along with a reduction of the information loss in the {\it statistical} model. One can therefore also ask how it affects other intrinsic properties of the orbit model. Unsurprisingly, it does not necessarily entail an improvement in these intrinsic properties either as they are also not subject to the AIC optimization. However, in practice, it often does improve the situation. It should be noted that in general, a model's intrinsic properties (like the orbital anisotropy) are only meaningful when a priori knowledge about a galaxy's typical structure is added. In principle, any $\mathbfit{l}_{\mathrm{mod}}$ could be a good {\it statistical} model without the need of secondary properties such as inclination, orbital weights etc. (cf. Sec.~\ref{sec:overview+problem}). An improvement in such secondary properties can only be achieved if a property correlates with an improvement of the model LOSVDs, or in other words: if the property can be constrained by ideal (noise-free) data. As an example, the bottom panel of Fig.~\ref{fig:nbody_vs_RMSD_LOSVD} shows the RMSDs of $\beta$ analogous to the $\Delta \mathbfit{l}$. Unlike the model LOSVDs the anisotropy recovery does not necessarily improve with the AIC model selection. For example, when $\alpha$ is not optimized the anisotropy recovery worsens considerably by choosing the more face-on models. However, as seen in Sec.~\ref{sec:regularization}, a substantial improvement in the orbital recovery can be made by optimizing the regularization, and in that case $\Delta \beta$ is approximately independent of the inclination, as expected. In practice, therefore, intrinsic model properties can be improved via the model selection even in a situation where parts of the model are not well constrained. 

We suppose that an approach similar to our model selection can also be transferred to other modelling frameworks. For example, in a Bayesian setting it may be worthwhile to marginalize over the entire high-dimensional space of orbital weights \citep[e.g.][]{2006MNRAS.373..425M,2014MNRAS.437.2230M} and nuisance parameters \citep[e.g.][]{2018MNRAS.473.2288B} given some suitable prior. In such a Bayesian setting the remaining parameters which are not marginalized out, play a similar role as the selection parameters do in our model selection framework. The best set of (selection) parameters is then determined according to the resulting marginalized likelihoods of different trial models and will depend on the choice of the prior. Our results seem to make a good case that a prior choice that aims to minimize the KLD of the LOSVDs -- such that the marginalized likelihoods rank models equivalently to the Akaike model selection \citep[KLD-prior, e.g.][]{burnham_anderson_burnham_2002} -- is very powerful. It has the additional benefit of being easy to implement even for complex models.

In summary, the AIC selection will always choose the model with the least amount of effective parameters out of all models that achieved a good fit. Often this can be very advantageous, especially when determining the regularization parameter $\alpha$ from the data (Sec.~\ref{sec:regularization}). However, in the case of degeneracies, when 
the information contained in the LOSVD data is enough to constrain a selection parameter only up to some interval, then the selection {\it within} this degenerate interval can be biased with respect to the true values. 
The fact that we never encountered any related bias in our flattenend toy galaxies for either $i$, $\Upsilon$, or $\alpha$ suggests that all these selection parameters are {\it well constrained} by the kinematic observations of the kind we tested: {\it fully resolved LOSVDs} over a field of view typical for modern integral-field spectrographs. On top of that, secondary properties such as the anisotropy $\beta$ are likewise well determined by this kind of data. This is consistent with the recent findings of \citet{2020MNRAS.500.1437N} in the triaxial case and shows how powerful orbit models are if all the information contained in modern data is exploited. 

\subsection{Inclination recovery of axisymmetric galaxies}
\label{inc-recovery-context}
We are not aware of any systematic attempts yet to study the inclination recovery of axisymmetric galaxies using orbit models. Early works on M32, using a $\chi^2$ minimization, yielded promisingly good constraints ($i=70\degr\pm 5\degr$, \citealt[]{2002MNRAS.335..517V}). However, the galaxy's intrinsic axis ratio $q=0.68\pm0.03$ at this inclination is just slightly different from the $q'=0.73$ expected for $i=90\degr$ and the errors were estimated using the $3\sigma$ intervals of a $\Delta \chi^2$ distribution with only three degrees of freedom (one for each varied selection parameter). Hence, the significance of the assigned inclination uncertainty is hard to judge. A subsequent detailed modelling of the early-type galaxy NGC 2974 by \citet[]{2005MNRAS.357.1113K} yielded a similarly well constrained inclination $i=65\degr\pm2.5\degr$ according to a $\chi^2$-minimization. Despite the small statistical uncertainty, however, \citet{2005MNRAS.357.1113K} concluded that the expected kinematic differences between models at different inclinations are so small that systematics in the data and/or in the models hamper a robust inclination recovery. Modelling tests with a two-integral toy model designed to represent NGC 2974 also supported the conclusion that a determination of the inclination is likely unfeasible. These studies did not exploit all the information contained in fully resolved LOSVDs, but were based on Gauss-Hermite expansions. Noteworthy, these early studies did not exhibit an apparent edge-on bias that dominates the $\chi^2$ surfaces. In comparison, our modelling of the toy galaxies and NGC 3368 were visibly impacted by the greater flexibility of more inclined models. Besides the galaxies/toy models presented in this paper, a dominant edge-on bias was present in the axisymmetric modelling results of NGC 4151 by \citet[]{2007ApJ...670..105O}. They compared models at the most likely inclination of the galaxy's large-scale disk ($i=23\degr$) and at $i=90\degr$. The edge-on models achieved a better fit. \citet{2000AJ....119.1157G} modelled NGC 3379 with inclinations ranging from $i=29\degr$ to $90\degr$ and found that the edge-on models are strongly preferred. \citet[]{2007MNRAS.381.1672T} modelled mock data derived from a suite of N-body binary merger simulations with differently inclined models yet the edge-on model consistently fitted the mocks best. \citet{2007MNRAS.382..657T} modelled a sample of galaxies in the Coma cluster assuming three different inclination angles per galaxy. Most of the galaxies were found to be edge-on, but since the sample was selected towards significantly flattened galaxies a potential inclination bias was difficult to quantify. More recently \citet{2020ApJ...891....4L} modelled NGC 1453 using different assumed inclinations with the triaxial Schwarzschild implementation of \citet{2008MNRAS.385..647V} in the axisymmetric limit and found that the kinematics are fitted better the closer the inclination is to an edge-on configuration. 

The fact that some of the early studies did not experience a noticeable edge-on bias could be caused by discreteness effects or other systematics being more significant than the bias induced by the  model flexibility. E.g., in early applications of the Schwarzschild models less orbits were used than are typically used now. Alternatively, the inclination bias could also depend on the galaxy structure. For example, if a galaxy is highly flattened its intrinsic structure might involve large intrinsic azimuthal velocities (like in a rotating disk). Edge-on models might have trouble to reach such large projected net velocities due to their rounder intrinsic shapes unless the $\cos(i)$ factor in the velocity projection overcompensates the lower intrinsic rotation (at fixed mass). In most cases this would be no problem for edge-on orbit libraries as they can usually follow the rotation signal by re-balancing the weights of retrograde to prograde orbits or by increasing the mass. However, if the edge-on model is already "at its limits", for example if all light contribution already comes from either retrograde or prograde orbits, then it may become significant. We were able to construct such extreme toy galaxies for which the edge-on bias (at fixed mass) was weaker or even absent because the edge-on models were simply not able to adequately fit the kinematic data and therefore the edge-on bias was not a dominant feature in the $\chi^2$ surfaces. However, we regard this explanation unlikely, since these extreme toy galaxies came along with a positive (global) correlation between the projected rotation velocity $v$ and the Gauss-Hermite parameter $h_3$, in contrast to the observed anticorrelation between $v$ and $h_3$ in real galaxies (\citealt{1994MNRAS.269..785B}; a positive correlation has only been observed vary rarely and even then only in small areas of a galaxy, e.g. \citealt{2016A&A...591A.143G}). 

Beyond the inclination in axisymmetric Schwarzschild implementations the viewing angles and associated shapes of triaxial systems also often prove difficult to recover correctly unless the underlying galaxy exhibits very distinct kinematic features \citep[e.g.][]{2008MNRAS.385..647V}. By modelling simulations using triaxial Schwarzschild models \citet{2019MNRAS.486.4753J} find that the flattening of the intermediate and minor axis are on average slightly overestimate, thus, biasing the models to more spherical shapes. This bias may be responsible for the large fraction of nearly spherical systems in a sample of 149 early-type galaxies modelled triaxially by \citet{2020MNRAS.491.1690J}. 

Unlike Schwarzschild models isotropic/anisotropic Jeans-models often provide strong inclination constraints and kinematic maps obtained from Jeans models are noticeably different when viewed under different angles \citep[cf.][]{2006MNRAS.366.1126C,2008MNRAS.390...71C}. However, these constraints may be overly optimistic as Jeans models only model the first and second velocity moments and are not as general as Schwarzschild models as they rely on assumptions about the velocity ellipsoid and anisotropy of the galaxy. The resulting Jeans models are not guaranteed to correspond to non-negative phase-space distribution functions and only a restricted subspace of all \textit{possible} non-negative distribution function compatible with a given potential is sampled by Jeans models. These restrictions may be advantageous in breaking the aforementioned inclination degeneracy if the assumptions hold approximately valid for a given galaxy, but if not then Jeans modelling could generate artificially constrained inclination results.

While there do exist alternative methods to constrain the inclination that are not based on dynamical modelling they are usually only applicable to disk galaxies, meaning that the inclination and consequently the intrinsic flattening of elliptical galaxies is often undetermined. However, since viewing angles are necessary parameters in the construction of any Schwarzschild model it is often simply assumed that elliptical galaxies are viewed edge-on or one adopts the inclination derived from Jeans models for the construction of Schwarzschild models \citep[e.g.][]{2006MNRAS.366.1126C,2019A&A...625A..62T}. 
A successful recovery of the inclination with dynamical modelling implies that the intrinsic 3d stellar mass distribution that corresponds to the tested viewing angles results in gravitational potentials (and thus the orbit models) that differ significantly enough to be detectable in the model selection framework. This stellar mass/luminosity distribution is obtained by deprojecting the observed surface brightness distribution for a given inclination. However, the deprojection of axisymmetric systems is ambiguous as it is only unique in the special case where the galaxy is viewed edge-on. For all other viewing angles the luminosity density is underdetermined as their exists a range of deprojections compatible with a given surface brightness distribution and inclination \citep[cf.][]{1987IAUS..127..397R,1990ApJ...362...52R}. As demonstrated by \citet{1996MNRAS.279..993G} one can construct many physically reasonable boxy and disky density distributions that projected to the same surface brightness distribution for a given inclination. This means there may exist many physically reasonable deprojections that could be used for the construction of our orbit libraries at a given inclination $i<90\degr$. Even though we obtained the (non-parametric) deprojections used in this paper by applying an Metropolis-Algorithm \citep[cf.][]{1999MNRAS.302..530M} which is able to explore the full range of physically allowed deprojections, we only picked a single "privileged" deprojection per assumed inclination, as is common in modelling applications to real galaxies. Given our results are suggesting that it is possible to distinguish between different inclinations and the differently flattened stellar distributions associated with them, it may also be possible to further dynamically discriminate between the many boxy/disky deprojections compatible with a given surface brightness and inclination. This possibility of breaking the (photometric) deprojection degeneracy using kinematics was explored before by \citet{1999MNRAS.302..530M} who has shown for \textit{two-integral} models that disks which are undetectable in the photometry can leave easily detectable features in the corresponding kinematics. We plan to investigate the deprojection degeneracy in the future using our model selection framework. 

\subsection{Mass reconstruction}
\label{mass-reconstruction}
Similar to the inclination other selection parameters can be biased, in particular mass parameters. For example we found the mass-to-light ratios of the toy galaxies A and D to be biased high by about $3$-$4\%$ when evaluated with a $\chi^2$ minimization because their intrinsic model flexibility is positively correlated with $\Upsilon$. In addition to this explicit dependence on the $m_{\mathrm{eff}}$ we also expect that mass parameters are implicitly affected if the galaxy is modelled assuming the wrong viewing angles. \citet{2007MNRAS.381.1672T} discuss via the tensor virial theorem how this depends on the intrinsic shape of a galaxy. E.g. the masses of oblate objects seen face-on are typically underestimated, while the masses of prolate objects seen end-on will be overestimated. In the triaxial modelling of the Milky Way's nuclear star \citet{2017MNRAS.466.4040F} find that the mass-to-light ratio is positively correlated with the triaxial shape parameter p (or the equivalent viewing angle) while the black hole mass only varies slightly. However, the $\chi^2$ surfaces for both mass parameters broaden as said shape parameter is increased. In the axisymmetric case the dynamical modelling of NGC 4151 by \citet[]{2007ApJ...670..105O} indicates a strong correlation of both, black hole mass and mass-to-light ratio, with the assumed inclination. Our modelling of NGC 3368 only very tentatively exhibits such a correlation between the two mass parameters $M_{\bullet}$ and $\Upsilon$ with the assumed inclination (Fig.\ref{fig:chi2_3368}), however, we do notice an increased scatter when modelling the galaxy edge-on.  
We generally expect that mass parameters which are less constrained by the available kinematic data (often these are the parameters of a dark matter halo or the mass of a central SMBH) are more sensitive to varying model flexibilities and, thus, are more likely prone to a bias. In multicomponent mass models an additional complication arises from the cross-talk between the various components \citep[e.g.][]{2018MNRAS.473.2251E}. Whether a specific component could be over- or underestimated is therefore not entirely clear. A more focused and systematic investigation with a large galaxy sample or with multi-component toy models could give better insight. We plan to address these questions and also the triaxial case in future papers.

\section{Summary and Conclusions}
\label{sec:conclusions}
When modelling the kinematic data of a galaxy to determine the mass of its SMBH, the stellar initial mass function or the structure of its dark matter halo, a huge number of trial dynamical models with different assumed mass distributions have to be fitted to the data. Then, based on the observations and the quality of the fit, one needs to decide which of these models represents the true structure of the galaxy best. 
We have motivated that the commonly used approach to judge the models solely by their goodness-of-fit is often not well defined. Moreover, we have shown that it can lead to substantial biases in estimated galaxy properties. The reason is that the process of identifying the best fit involves the comparison of {\it different models} with intrinsically different model flexibilities. In the case of axisymmetric modelling this point is most apparent when trying to recover a galaxy's inclination, where it causes a dominant bias towards edge-on models. If not corrected for, this effect inhibits the possibility of constraining the inclination via dynamical modelling. However, the issue of varying model flexibility is not contained to the inclination alone, we  demonstrated that it also introduces an overestimation of galaxy masses.

Quantifying the flexibility of Schwarzschild models is non-trivial due to various complications like non-linearity or priors that restrict the parameter space accessible to the orbital weights $w_{i}$. We introduced the concept of the number of effective free parameters and presented two calculation methods which rely on bootstrap iterations. Although computationally expensive, this bootstrap approach is robust, flexible and can be applied to more general modelling techniques. Once the model flexibility (i.e. the number of effective parameters $m_\mathrm{eff}$) is known, model selection techniques can be applied to choose the best model out of a given set of fits.

We tested three different model selection frameworks, the "intuitive" one, an information-based one using the Akaike Information Criterion (AIC) and a more generalised approach with an adjustable parameter $w_{m}$. The only difference between the approaches is how the model flexibility is weighted in the model selection process. The "intuitive" approach considers models equivalent that have the same reduced $\chi^2$ ($\chi^2 + m_\mathrm{eff} \rightarrow \mathrm{min}$), the AIC weighs the effective parameters twice as large as in the intuitive approach ($\chi^2 + 2 \, m_\mathrm{eff} \rightarrow \mathrm{min}$) and yields the least flexible model out of all models with the same reduced $\chi^2$. In the generalised approach ($\chi^2 + w_m \, m_\mathrm{eff} \rightarrow \mathrm{min}$) the weight of the model flexibility is parameterised by $w_m$. It includes the other two approaches as special cases.

We applied these model selection schemes to realistic mock data sets of a number of axisymmetric toy galaxies with the goal to recover their inclination and mass-to-light ratio $\Upsilon$. We confirmed that an evaluation based solely on $\chi^2$ {\it always} favors the edge-on orbit models and is biased towards higher mass-to-light ratios. A model assessment based on $\chi^2$ alone will limit the potential constraining power of dynamical models, meaning that better kinematic data will not lead to a corresponding improved accuracy in the estimated galaxy properties. Model evaluation within a model selection framework can correct these issues, enabling the recovery of the correct galaxy inclination and mass with very small uncertainties.

In a second step we extended the model selection approach to also encompass the strength of the entropy regularisation that is applied in the models. The amount of regularization is a crucial choice as it can negatively affect the recovery of the intrinsic dynamics and the stability of the calibration weight $w_{m}$. This led to a further significant improvement of the results. Based on the simulated toy galaxies we found that
\begin{itemize}
\item the model selection not only returns always the correct masses and inclinations but also returns the model that matches the toy galaxy most closely in terms of its orbital dynamics
\item while the best results of the model selection with fixed smoothing required a $w_m \approx 1.5$ depending on the galaxy under study, the best results with optimized smoothing were {\it robustly} obtained always for the {\it same} $w_m =2$ (AIC) 
\item the constraining power of the data improved, i.e. the confidence regions of the derived galaxy masses, inclinations and orbital anisotropies tightened significantly
\end{itemize}

This suggests that, in order to achieve optimal results, one should construct models by varying the degree of regularization among other parameters and evaluate models with different selection parameters using the information-based Akaike criterion. 

With modern integral-field spectrographs on 10m-class telescopes it is possible to measure the LOSVDs of galaxies non-parametrically at hundreds of positions spread over the galaxy on the sky \citep[e.g.][]{2019ApJ...887..195M}. Our orbit models are designed to deal with the full amount of information that is contained in these non-parametric LOSVD fields. With the simulated toy galaxies that we used to mimick such observations in a realistic fashion we could show that orbit models allow to reconstruct galaxy inclinations, masses and anisotropies with an uncertainty no larger than 1-2 percent in mass and $\Delta \beta \sim 0.04$ in the anisotropy. As long as the regularization is optimized the viewing angle and consequently the intrinsic flattening $q$ of the toy galaxies are well constrained by an AIC framework with an uncertainty smaller than the sampling size $\Delta i=10\degr$ of the trial models used here. This demonstrates the power of orbit superposition models and the prospects of the model selection ansatz. 

The edge-on bias in the $\chi^2$-minimization is also occurring for the dynamical modelling of real galaxies, as demonstrated by the modelling of the spiral galaxy NGC 3368. Even though the ellipticity  of the galaxy's large-scale disk suggests a moderate inclination of $i=53\degr$, edge-on models fit kinematic data obtained with SINFONI systematically better than dynamical models constructed with $i=53\degr$. When the varying flexibilities are included, however, the recovered inclination is in agreement with the observed ellipticity. Specifically, when applying the full model selection with optimized regularisation, we find a black hole mass $M_{\bullet}=(5.5\pm3.3)\cdot10^{6}M_{\sun}$, a bulge mass-to-light ratio $\Upsilon_{\mathrm{bulge}}=(0.55\pm0.11)M_{\sun}/L_{\sun}$ and an inclination angle $i=(57.8\pm5.1)\degr$ in agreement with the independent inclination estimation. The fact that we could recover the galaxy's inclination from just the SINFONI kinematics (which cover only the inner $\sim1$ arcsec of the galaxy, roughly 5-10 times the sphere of influence of the central SMBH) again underlines the prospect of the model selection technique. The above Monte-Carlo simulations suggest that the accuracy of the involved selection parameters can be significantly improved by modelling more extended, high-resolution kinematic data. In addition, we suspect that a modelling that also takes into account more accurate, correlated error patterns will further improve the dynamical modeling. Thus, the obvious next step will be to incorporate the full error correlation matrix of the observed LOSVDs within the modelling procedure.

The model selection approach is versatile and and may well be used to optimize other intrinsic library parameters which impact the model flexibilty like, e.g., the number of orbits. We plan to investigate this in a companion paper. Since the objective of the AIC selection is to minimize the KLD of the \textit{statistical} model (i.e. in our case the LOSVDs) it does not rely on the internal structure of the underlying model. Therefore the approach may be adopted for other dynamical modelling techniques as well. It may even be possible that the extension of the model selection ansatz to determine the optimal smoothing based on data can be applied in other non-parametric methods as well, e.g. in the recovery of non-parameteric LOSVDs (Thomas et al. in preparation), non-parametric deprojections \citep[e.g.][]{2020MNRAS.496.3076D} and non-parametric source reconstructions in strong gravitational lensing etc.

\section*{Acknowledgements}
We are grateful to Dr. Roberto Saglia for his active support and for many valuable discussions throughout the project. We also thank the anonymous referee for helpful comments. This research was supported by the Excellence Cluster ORIGINS which is funded by the Deutsche Forschungsgemeinschaft (DFG, German Research Foundation) under Germany's Excellence Strategy - EXC-2094-390783311. Simulations have been carried out on the computing facilities of the Computational Center for Particle and Astrophysics (C2PAP) and at the COBRA HPC system at the Max Planck Computing and Data Facility (MPCDF), Germany.

\section*{Data availability statement}
The data underlying this article will be shared on reasonable request to the corresponding author.

\bibliographystyle{mnras}
\bibliography{literature}

\begin{thebibliography}{}
\makeatletter
\relax
\def\mn@urlcharsother{\let\do\@makeother \do\$\do\&\do\#\do\^\do\_\do\%\do\~}
\def\mn@doi{\begingroup\mn@urlcharsother \@ifnextchar [ {\mn@doi@}
  {\mn@doi@[]}}
\def\mn@doi@[#1]#2{\def\@tempa{#1}\ifx\@tempa\@empty \href
  {http://dx.doi.org/#2} {doi:#2}\else \href {http://dx.doi.org/#2} {#1}\fi
  \endgroup}
\def\mn@eprint#1#2{\mn@eprint@#1:#2::\@nil}
\def\mn@eprint@arXiv#1{\href {http://arxiv.org/abs/#1} {{\tt arXiv:#1}}}
\def\mn@eprint@dblp#1{\href {http://dblp.uni-trier.de/rec/bibtex/#1.xml}
  {dblp:#1}}
\def\mn@eprint@#1:#2:#3:#4\@nil{\def\@tempa {#1}\def\@tempb {#2}\def\@tempc
  {#3}\ifx \@tempc \@empty \let \@tempc \@tempb \let \@tempb \@tempa \fi \ifx
  \@tempb \@empty \def\@tempb {arXiv}\fi \@ifundefined
  {mn@eprint@\@tempb}{\@tempb:\@tempc}{\expandafter \expandafter \csname
  mn@eprint@\@tempb\endcsname \expandafter{\@tempc}}}

\bibitem[\protect\citeauthoryear{{Andrae}, {Schulze-Hartung}  \&
  {Melchior}}{{Andrae} et~al.}{2010}]{2010arXiv1012.3754A}
{Andrae} R.,  {Schulze-Hartung} T.,   {Melchior} P.,  2010, arXiv e-prints,
  \href {https://ui.adsabs.harvard.edu/abs/2010arXiv1012.3754A} {p.
  arXiv:1012.3754}

\bibitem[\protect\citeauthoryear{{Baes}, {Dejonghe}  \& {Buyle}}{{Baes}
  et~al.}{2005}]{2005A&A...432..411B}
{Baes} M.,  {Dejonghe} H.,   {Buyle} P.,  2005, \mn@doi [\aap]
  {10.1051/0004-6361:20041907}, \href
  {https://ui.adsabs.harvard.edu/abs/2005A&A...432..411B} {432, 411}

\bibitem[\protect\citeauthoryear{{Bender}, {Saglia}  \& {Gerhard}}{{Bender}
  et~al.}{1994}]{1994MNRAS.269..785B}
{Bender} R.,  {Saglia} R.~P.,   {Gerhard} O.~E.,  1994, \mn@doi [\mnras]
  {10.1093/mnras/269.3.785}, \href
  {https://ui.adsabs.harvard.edu/abs/1994MNRAS.269..785B} {269, 785}

\bibitem[\protect\citeauthoryear{Binney \& Tremaine}{Binney \&
  Tremaine}{2008}]{Binney2008}
Binney J.,  Tremaine S.,  2008, Galactic Dynamics - Second Edition.
Princeton Univ. Press, Princeton

\bibitem[\protect\citeauthoryear{{Bovy}, {Kawata}  \& {Hunt}}{{Bovy}
  et~al.}{2018}]{2018MNRAS.473.2288B}
{Bovy} J.,  {Kawata} D.,   {Hunt} J. A.~S.,  2018, \mn@doi [\mnras]
  {10.1093/mnras/stx2402}, \href
  {https://ui.adsabs.harvard.edu/abs/2018MNRAS.473.2288B} {473, 2288}

\bibitem[\protect\citeauthoryear{Burnham, Anderson  \& Burnham}{Burnham
  et~al.}{2002}]{burnham_anderson_burnham_2002}
Burnham K.~P.,  Anderson D.~R.,   Burnham K.~P.,  2002, Model selection and
  multi-model inference: A practical information-theoretic approach.
Springer, New York

\bibitem[\protect\citeauthoryear{{Cappellari}}{{Cappellari}}{2008}]{2008MNRAS.390...71C}
{Cappellari} M.,  2008, \mn@doi [\mnras] {10.1111/j.1365-2966.2008.13754.x},
  \href {https://ui.adsabs.harvard.edu/abs/2008MNRAS.390...71C} {390, 71}

\bibitem[\protect\citeauthoryear{{Cappellari} \& {Copin}}{{Cappellari} \&
  {Copin}}{2003}]{2003MNRAS.342..345C}
{Cappellari} M.,  {Copin} Y.,  2003, \mn@doi [\mnras]
  {10.1046/j.1365-8711.2003.06541.x}, \href
  {https://ui.adsabs.harvard.edu/abs/2003MNRAS.342..345C} {342, 345}

\bibitem[\protect\citeauthoryear{{Cappellari}, {Verolme}, {van der Marel},
  {Verdoes Kleijn}, {Illingworth}, {Franx}, {Carollo}  \& {de
  Zeeuw}}{{Cappellari} et~al.}{2002}]{2002ApJ...578..787C}
{Cappellari} M.,  {Verolme} E.~K.,  {van der Marel} R.~P.,  {Verdoes Kleijn}
  G.~A.,  {Illingworth} G.~D.,  {Franx} M.,  {Carollo} C.~M.,   {de Zeeuw}
  P.~T.,  2002, \mn@doi [\apj] {10.1086/342653}, \href
  {https://ui.adsabs.harvard.edu/abs/2002ApJ...578..787C} {578, 787}

\bibitem[\protect\citeauthoryear{{Cappellari} et~al.,}{{Cappellari}
  et~al.}{2006}]{2006MNRAS.366.1126C}
{Cappellari} M.,  et~al., 2006, \mn@doi [\mnras]
  {10.1111/j.1365-2966.2005.09981.x}, \href
  {https://ui.adsabs.harvard.edu/abs/2006MNRAS.366.1126C} {366, 1126}

\bibitem[\protect\citeauthoryear{{Cretton} \& {Emsellem}}{{Cretton} \&
  {Emsellem}}{2004}]{2004MNRAS.347L..31C}
{Cretton} N.,  {Emsellem} E.,  2004, \mn@doi [\mnras]
  {10.1111/j.1365-2966.2004.07374.x}, \href
  {https://ui.adsabs.harvard.edu/abs/2004MNRAS.347L..31C} {347, L31}

\bibitem[\protect\citeauthoryear{{Cretton}, {de Zeeuw}, {van der Marel}  \&
  {Rix}}{{Cretton} et~al.}{1999}]{1999ApJS..124..383C}
{Cretton} N.,  {de Zeeuw} P.~T.,  {van der Marel} R.~P.,   {Rix} H.-W.,  1999,
  \mn@doi [\apjs] {10.1086/313264}, \href
  {https://ui.adsabs.harvard.edu/abs/1999ApJS..124..383C} {124, 383}

\bibitem[\protect\citeauthoryear{{Emsellem} et~al.,}{{Emsellem}
  et~al.}{2011}]{2011MNRAS.414..888E}
{Emsellem} E.,  et~al., 2011, \mn@doi [\mnras]
  {10.1111/j.1365-2966.2011.18496.x}, \href
  {https://ui.adsabs.harvard.edu/abs/2011MNRAS.414..888E} {414, 888}

\bibitem[\protect\citeauthoryear{{Erwin}}{{Erwin}}{2004}]{2004A&A...415..941E}
{Erwin} P.,  2004, \mn@doi [\aap] {10.1051/0004-6361:20034408}, \href
  {https://ui.adsabs.harvard.edu/abs/2004A&A...415..941E} {415, 941}

\bibitem[\protect\citeauthoryear{{Erwin}, {Thomas}, {Saglia}, {Fabricius},
  {Rusli}, {Seitz}  \& {Bender}}{{Erwin} et~al.}{2018}]{2018MNRAS.473.2251E}
{Erwin} P.,  {Thomas} J.,  {Saglia} R.~P.,  {Fabricius} M.,  {Rusli} S.~P.,
  {Seitz} S.,   {Bender} R.,  2018, \mn@doi [\mnras] {10.1093/mnras/stx2499},
  \href {https://ui.adsabs.harvard.edu/abs/2018MNRAS.473.2251E} {473, 2251}

\bibitem[\protect\citeauthoryear{{Feldmeier-Krause}, {Zhu}, {Neumayer}, {van de
  Ven}, {de Zeeuw}  \& {Sch{\"o}del}}{{Feldmeier-Krause}
  et~al.}{2017}]{2017MNRAS.466.4040F}
{Feldmeier-Krause} A.,  {Zhu} L.,  {Neumayer} N.,  {van de Ven} G.,  {de Zeeuw}
  P.~T.,   {Sch{\"o}del} R.,  2017, \mn@doi [\mnras] {10.1093/mnras/stw3377},
  \href {https://ui.adsabs.harvard.edu/abs/2017MNRAS.466.4040F} {466, 4040}

\bibitem[\protect\citeauthoryear{{Gebhardt} et~al.,}{{Gebhardt}
  et~al.}{2000}]{2000AJ....119.1157G}
{Gebhardt} K.,  et~al., 2000, \mn@doi [\aj] {10.1086/301240}, \href
  {https://ui.adsabs.harvard.edu/abs/2000AJ....119.1157G} {119, 1157}

\bibitem[\protect\citeauthoryear{{Gebhardt} et~al.,}{{Gebhardt}
  et~al.}{2003}]{2003ApJ...583...92G}
{Gebhardt} K.,  et~al., 2003, \mn@doi [\apj] {10.1086/345081}, \href
  {https://ui.adsabs.harvard.edu/abs/2003ApJ...583...92G} {583, 92}

\bibitem[\protect\citeauthoryear{{Gerhard}}{{Gerhard}}{1993}]{1993MNRAS.265..213G}
{Gerhard} O.~E.,  1993, \mn@doi [\mnras] {10.1093/mnras/265.1.213}, \href
  {https://ui.adsabs.harvard.edu/abs/1993MNRAS.265..213G} {265, 213}

\bibitem[\protect\citeauthoryear{{Gerhard} \& {Binney}}{{Gerhard} \&
  {Binney}}{1996}]{1996MNRAS.279..993G}
{Gerhard} O.~E.,  {Binney} J.~J.,  1996, \mn@doi [\mnras]
  {10.1093/mnras/279.3.993}, \href
  {https://ui.adsabs.harvard.edu/abs/1996MNRAS.279..993G} {279, 993}

\bibitem[\protect\citeauthoryear{{Gu{\'e}rou}, {Emsellem}, {Krajnovi{\'c}},
  {McDermid}, {Contini}  \& {Weilbacher}}{{Gu{\'e}rou}
  et~al.}{2016}]{2016A&A...591A.143G}
{Gu{\'e}rou} A.,  {Emsellem} E.,  {Krajnovi{\'c}} D.,  {McDermid} R.~M.,
  {Contini} T.,   {Weilbacher} P.~M.,  2016, \mn@doi [\aap]
  {10.1051/0004-6361/201628743}, \href
  {https://ui.adsabs.harvard.edu/abs/2016A&A...591A.143G} {591, A143}

\bibitem[\protect\citeauthoryear{{H{\"a}fner}, {Evans}, {Dehnen}  \&
  {Binney}}{{H{\"a}fner} et~al.}{2000}]{2000MNRAS.314..433H}
{H{\"a}fner} R.,  {Evans} N.~W.,  {Dehnen} W.,   {Binney} J.,  2000, \mn@doi
  [\mnras] {10.1046/j.1365-8711.2000.03242.x}, \href
  {https://ui.adsabs.harvard.edu/abs/2000MNRAS.314..433H} {314, 433}

\bibitem[\protect\citeauthoryear{Hastie, Tibshirani  \& Friedman}{Hastie
  et~al.}{2013}]{2013}
Hastie T.,  Tibshirani R.,   Friedman J.,  2013, The Elements of Statistical
  Learning - Data Mining, Inference, and Prediction.
Springer Science \& Business Media, Berlin Heidelberg

\bibitem[\protect\citeauthoryear{{Hernquist}}{{Hernquist}}{1990}]{1990ApJ...356..359H}
{Hernquist} L.,  1990, \mn@doi [\apj] {10.1086/168845}, \href
  {https://ui.adsabs.harvard.edu/abs/1990ApJ...356..359H} {356, 359}

\bibitem[\protect\citeauthoryear{{Ho}, {Filippenko}  \& {Sargent}}{{Ho}
  et~al.}{1997}]{1997ApJS..112..315H}
{Ho} L.~C.,  {Filippenko} A.~V.,   {Sargent} W. L.~W.,  1997, \mn@doi [\apjs]
  {10.1086/313041}, \href
  {https://ui.adsabs.harvard.edu/abs/1997ApJS..112..315H} {112, 315}

\bibitem[\protect\citeauthoryear{{Houghton}, {Magorrian}, {Sarzi}, {Thatte},
  {Davies}  \& {Krajnovi{\'c}}}{{Houghton} et~al.}{2006}]{2006MNRAS.367....2H}
{Houghton} R.~C.~W.,  {Magorrian} J.,  {Sarzi} M.,  {Thatte} N.,  {Davies}
  R.~L.,   {Krajnovi{\'c}} D.,  2006, \mn@doi [\mnras]
  {10.1111/j.1365-2966.2005.09713.x}, \href
  {https://ui.adsabs.harvard.edu/abs/2006MNRAS.367....2H} {367, 2}

\bibitem[\protect\citeauthoryear{{Jin}, {Zhu}, {Long}, {Mao}, {Xu}, {Li}  \&
  {van de Ven}}{{Jin} et~al.}{2019}]{2019MNRAS.486.4753J}
{Jin} Y.,  {Zhu} L.,  {Long} R.~J.,  {Mao} S.,  {Xu} D.,  {Li} H.,   {van de
  Ven} G.,  2019, \mn@doi [\mnras] {10.1093/mnras/stz1170}, \href
  {https://ui.adsabs.harvard.edu/abs/2019MNRAS.486.4753J} {486, 4753}

\bibitem[\protect\citeauthoryear{{Jin}, {Zhu}, {Long}, {Mao}, {Wang}  \& {van
  de Ven}}{{Jin} et~al.}{2020}]{2020MNRAS.491.1690J}
{Jin} Y.,  {Zhu} L.,  {Long} R.~J.,  {Mao} S.,  {Wang} L.,   {van de Ven} G.,
  2020, \mn@doi [\mnras] {10.1093/mnras/stz3072}, \href
  {https://ui.adsabs.harvard.edu/abs/2020MNRAS.491.1690J} {491, 1690}

\bibitem[\protect\citeauthoryear{{Knapen}, {de Jong}, {Stedman}  \&
  {Bramich}}{{Knapen} et~al.}{2003}]{2003MNRAS.344..527K}
{Knapen} J.~H.,  {de Jong} R.~S.,  {Stedman} S.,   {Bramich} D.~M.,  2003,
  \mn@doi [\mnras] {10.1046/j.1365-8711.2003.06846.x}, \href
  {https://ui.adsabs.harvard.edu/abs/2003MNRAS.344..527K} {344, 527}

\bibitem[\protect\citeauthoryear{{Kowalczyk}, {{\L}okas}  \&
  {Valluri}}{{Kowalczyk} et~al.}{2017}]{2017MNRAS.470.3959K}
{Kowalczyk} K.,  {{\L}okas} E.~L.,   {Valluri} M.,  2017, \mn@doi [\mnras]
  {10.1093/mnras/stx1520}, \href
  {https://ui.adsabs.harvard.edu/abs/2017MNRAS.470.3959K} {470, 3959}

\bibitem[\protect\citeauthoryear{{Krajnovi{\'c}}, {Cappellari}, {Emsellem},
  {McDermid}  \& {de Zeeuw}}{{Krajnovi{\'c}}
  et~al.}{2005}]{2005MNRAS.357.1113K}
{Krajnovi{\'c}} D.,  {Cappellari} M.,  {Emsellem} E.,  {McDermid} R.~M.,   {de
  Zeeuw} P.~T.,  2005, \mn@doi [\mnras] {10.1111/j.1365-2966.2005.08715.x},
  \href {https://ui.adsabs.harvard.edu/abs/2005MNRAS.357.1113K} {357, 1113}

\bibitem[\protect\citeauthoryear{{Krajnovi{\'c}}, {McDermid}, {Cappellari}  \&
  {Davies}}{{Krajnovi{\'c}} et~al.}{2009}]{2009MNRAS.399.1839K}
{Krajnovi{\'c}} D.,  {McDermid} R.~M.,  {Cappellari} M.,   {Davies} R.~L.,
  2009, \mn@doi [\mnras] {10.1111/j.1365-2966.2009.15415.x}, \href
  {https://ui.adsabs.harvard.edu/abs/2009MNRAS.399.1839K} {399, 1839}

\bibitem[\protect\citeauthoryear{{Leung} et~al.,}{{Leung}
  et~al.}{2018}]{2018MNRAS.477..254L}
{Leung} G. Y.~C.,  et~al., 2018, \mn@doi [\mnras] {10.1093/mnras/sty288}, \href
  {https://ui.adsabs.harvard.edu/abs/2018MNRAS.477..254L} {477, 254}

\bibitem[\protect\citeauthoryear{{Levison} \& {Richstone}}{{Levison} \&
  {Richstone}}{1985}]{1985ApJ...295..349L}
{Levison} H.~F.,  {Richstone} D.~O.,  1985, \mn@doi [\apj] {10.1086/163379},
  \href {https://ui.adsabs.harvard.edu/abs/1985ApJ...295..349L} {295, 349}

\bibitem[\protect\citeauthoryear{{Liepold}, {Quenneville}, {Ma}, {Walsh},
  {McConnell}, {Greene}  \& {Blakeslee}}{{Liepold}
  et~al.}{2020}]{2020ApJ...891....4L}
{Liepold} C.~M.,  {Quenneville} M.~E.,  {Ma} C.-P.,  {Walsh} J.~L.,
  {McConnell} N.~J.,  {Greene} J.~E.,   {Blakeslee} J.~P.,  2020, \mn@doi
  [\apj] {10.3847/1538-4357/ab6f71}, \href
  {https://ui.adsabs.harvard.edu/abs/2020ApJ...891....4L} {891, 4}

\bibitem[\protect\citeauthoryear{{Magorrian}}{{Magorrian}}{1999}]{1999MNRAS.302..530M}
{Magorrian} J.,  1999, \mn@doi [\mnras] {10.1046/j.1365-8711.1999.02135.x},
  \href {https://ui.adsabs.harvard.edu/abs/1999MNRAS.302..530M} {302, 530}

\bibitem[\protect\citeauthoryear{{Magorrian}}{{Magorrian}}{2006}]{2006MNRAS.373..425M}
{Magorrian} J.,  2006, \mn@doi [\mnras] {10.1111/j.1365-2966.2006.11054.x},
  \href {https://ui.adsabs.harvard.edu/abs/2006MNRAS.373..425M} {373, 425}

\bibitem[\protect\citeauthoryear{{Magorrian}}{{Magorrian}}{2014}]{2014MNRAS.437.2230M}
{Magorrian} J.,  2014, \mn@doi [\mnras] {10.1093/mnras/stt2031}, \href
  {https://ui.adsabs.harvard.edu/abs/2014MNRAS.437.2230M} {437, 2230}

\bibitem[\protect\citeauthoryear{{Mehrgan}, {Thomas}, {Saglia}, {Mazzalay},
  {Erwin}, {Bender}, {Kluge}  \& {Fabricius}}{{Mehrgan}
  et~al.}{2019}]{2019ApJ...887..195M}
{Mehrgan} K.,  {Thomas} J.,  {Saglia} R.,  {Mazzalay} X.,  {Erwin} P.,
  {Bender} R.,  {Kluge} M.,   {Fabricius} M.,  2019, \mn@doi [\apj]
  {10.3847/1538-4357/ab5856}, \href
  {https://ui.adsabs.harvard.edu/abs/2019ApJ...887..195M} {887, 195}

\bibitem[\protect\citeauthoryear{{Morganti} \& {Gerhard}}{{Morganti} \&
  {Gerhard}}{2012}]{2012MNRAS.422.1571M}
{Morganti} L.,  {Gerhard} O.,  2012, \mn@doi [\mnras]
  {10.1111/j.1365-2966.2012.20733.x}, \href
  {https://ui.adsabs.harvard.edu/abs/2012MNRAS.422.1571M} {422, 1571}

\bibitem[\protect\citeauthoryear{{Navarro}, {Frenk}  \& {White}}{{Navarro}
  et~al.}{1996}]{1996ApJ...462..563N}
{Navarro} J.~F.,  {Frenk} C.~S.,   {White} S. D.~M.,  1996, \mn@doi [\apj]
  {10.1086/177173}, \href
  {https://ui.adsabs.harvard.edu/abs/1996ApJ...462..563N} {462, 563}

\bibitem[\protect\citeauthoryear{{Neureiter} et~al.,}{{Neureiter}
  et~al.}{2021}]{2020MNRAS.500.1437N}
{Neureiter} B.,  et~al., 2021, \mn@doi [\mnras] {10.1093/mnras/staa3014}, \href
  {https://ui.adsabs.harvard.edu/abs/2021MNRAS.500.1437N} {500, 1437}

\bibitem[\protect\citeauthoryear{{Nowak}, {Thomas}, {Erwin}, {Saglia}, {Bender}
   \& {Davies}}{{Nowak} et~al.}{2010}]{2010MNRAS.403..646N}
{Nowak} N.,  {Thomas} J.,  {Erwin} P.,  {Saglia} R.~P.,  {Bender} R.,
  {Davies} R.~I.,  2010, \mn@doi [\mnras] {10.1111/j.1365-2966.2009.16167.x},
  \href {https://ui.adsabs.harvard.edu/abs/2010MNRAS.403..646N} {403, 646}

\bibitem[\protect\citeauthoryear{{Onken} et~al.,}{{Onken}
  et~al.}{2007}]{2007ApJ...670..105O}
{Onken} C.~A.,  et~al., 2007, \mn@doi [\apj] {10.1086/522220}, \href
  {https://ui.adsabs.harvard.edu/abs/2007ApJ...670..105O} {670, 105}

\bibitem[\protect\citeauthoryear{{Press}, {Teukolsky}, {Vetterling}  \&
  {Flannery}}{{Press} et~al.}{1992}]{1992nrfa.book.....P}
{Press} W.~H.,  {Teukolsky} S.~A.,  {Vetterling} W.~T.,   {Flannery} B.~P.,
  1992, {Numerical recipes in FORTRAN. The art of scientific computing}

\bibitem[\protect\citeauthoryear{{Richstone} \& {Tremaine}}{{Richstone} \&
  {Tremaine}}{1988}]{1988ApJ...327...82R}
{Richstone} D.~O.,  {Tremaine} S.,  1988, \mn@doi [\apj] {10.1086/166171},
  \href {https://ui.adsabs.harvard.edu/abs/1988ApJ...327...82R} {327, 82}

\bibitem[\protect\citeauthoryear{{Rix} \& {White}}{{Rix} \&
  {White}}{1990}]{1990ApJ...362...52R}
{Rix} H.-W.,  {White} S. D.~M.,  1990, \mn@doi [\apj] {10.1086/169242}, \href
  {https://ui.adsabs.harvard.edu/abs/1990ApJ...362...52R} {362, 52}

\bibitem[\protect\citeauthoryear{{Rix}, {de Zeeuw}, {Cretton}, {van der Marel}
  \& {Carollo}}{{Rix} et~al.}{1997}]{1997ApJ...488..702R}
{Rix} H.-W.,  {de Zeeuw} P.~T.,  {Cretton} N.,  {van der Marel} R.~P.,
  {Carollo} C.~M.,  1997, \mn@doi [\apj] {10.1086/304733}, \href
  {https://ui.adsabs.harvard.edu/abs/1997ApJ...488..702R} {488, 702}

\bibitem[\protect\citeauthoryear{{Rusli} et~al.,}{{Rusli}
  et~al.}{2013}]{2013AJ....146...45R}
{Rusli} S.~P.,  et~al., 2013, \mn@doi [\aj] {10.1088/0004-6256/146/3/45}, \href
  {https://ui.adsabs.harvard.edu/abs/2013AJ....146...45R} {146, 45}

\bibitem[\protect\citeauthoryear{{Rybicki}}{{Rybicki}}{1987}]{1987IAUS..127..397R}
{Rybicki} G.~B.,  1987, in {de Zeeuw} P.~T.,  ed.,  IAU Symposium Vol. 127,
  Structure and Dynamics of Elliptical Galaxies. p.~397,
  \mn@doi{10.1007/978-94-009-3971-4_41}

\bibitem[\protect\citeauthoryear{{Saglia}, {Kronawitter}, {Gerhard}  \&
  {Bender}}{{Saglia} et~al.}{2000}]{2000AJ....119..153S}
{Saglia} R.~P.,  {Kronawitter} A.,  {Gerhard} O.,   {Bender} R.,  2000, \mn@doi
  [\aj] {10.1086/301153}, \href
  {https://ui.adsabs.harvard.edu/abs/2000AJ....119..153S} {119, 153}

\bibitem[\protect\citeauthoryear{{Saglia} et~al.,}{{Saglia}
  et~al.}{2016}]{2016ApJ...818...47S}
{Saglia} R.~P.,  et~al., 2016, \mn@doi [\apj] {10.3847/0004-637X/818/1/47},
  \href {https://ui.adsabs.harvard.edu/abs/2016ApJ...818...47S} {818, 47}

\bibitem[\protect\citeauthoryear{{Schwarzschild}}{{Schwarzschild}}{1979}]{1979ApJ...232..236S}
{Schwarzschild} M.,  1979, \mn@doi [\apj] {10.1086/157282}, \href
  {https://ui.adsabs.harvard.edu/abs/1979ApJ...232..236S} {232, 236}

\bibitem[\protect\citeauthoryear{{Siopis} \& {Kandrup}}{{Siopis} \&
  {Kandrup}}{2000}]{2000MNRAS.319...43S}
{Siopis} C.,  {Kandrup} H.~E.,  2000, \mn@doi [\mnras]
  {10.1046/j.1365-8711.2000.03740.x}, \href
  {https://ui.adsabs.harvard.edu/abs/2000MNRAS.319...43S} {319, 43}

\bibitem[\protect\citeauthoryear{{Thater}, {Krajnovi{\'c}}, {Cappellari},
  {Davis}, {de Zeeuw}, {McDermid}  \& {Sarzi}}{{Thater}
  et~al.}{2019}]{2019A&A...625A..62T}
{Thater} S.,  {Krajnovi{\'c}} D.,  {Cappellari} M.,  {Davis} T.~A.,  {de Zeeuw}
  P.~T.,  {McDermid} R.~M.,   {Sarzi} M.,  2019, \mn@doi [\aap]
  {10.1051/0004-6361/201834808}, \href
  {https://ui.adsabs.harvard.edu/abs/2019A&A...625A..62T} {625, A62}

\bibitem[\protect\citeauthoryear{{Thomas}, {Saglia}, {Bender}, {Thomas},
  {Gebhardt}, {Magorrian}  \& {Richstone}}{{Thomas}
  et~al.}{2004}]{2004MNRAS.353..391T}
{Thomas} J.,  {Saglia} R.~P.,  {Bender} R.,  {Thomas} D.,  {Gebhardt} K.,
  {Magorrian} J.,   {Richstone} D.,  2004, \mn@doi [\mnras]
  {10.1111/j.1365-2966.2004.08072.x}, \href
  {https://ui.adsabs.harvard.edu/abs/2004MNRAS.353..391T} {353, 391}

\bibitem[\protect\citeauthoryear{{Thomas}, {Saglia}, {Bender}, {Thomas},
  {Gebhardt}, {Magorrian}, {Corsini}  \& {Wegner}}{{Thomas}
  et~al.}{2005}]{2005MNRAS.360.1355T}
{Thomas} J.,  {Saglia} R.~P.,  {Bender} R.,  {Thomas} D.,  {Gebhardt} K.,
  {Magorrian} J.,  {Corsini} E.~M.,   {Wegner} G.,  2005, \mn@doi [\mnras]
  {10.1111/j.1365-2966.2005.09139.x}, \href
  {https://ui.adsabs.harvard.edu/abs/2005MNRAS.360.1355T} {360, 1355}

\bibitem[\protect\citeauthoryear{{Thomas}, {Jesseit}, {Naab}, {Saglia},
  {Burkert}  \& {Bender}}{{Thomas} et~al.}{2007a}]{2007MNRAS.381.1672T}
{Thomas} J.,  {Jesseit} R.,  {Naab} T.,  {Saglia} R.~P.,  {Burkert} A.,
  {Bender} R.,  2007a, \mn@doi [\mnras] {10.1111/j.1365-2966.2007.12343.x},
  \href {https://ui.adsabs.harvard.edu/abs/2007MNRAS.381.1672T} {381, 1672}

\bibitem[\protect\citeauthoryear{{Thomas}, {Saglia}, {Bender}, {Thomas},
  {Gebhardt}, {Magorrian}, {Corsini}  \& {Wegner}}{{Thomas}
  et~al.}{2007b}]{2007MNRAS.382..657T}
{Thomas} J.,  {Saglia} R.~P.,  {Bender} R.,  {Thomas} D.,  {Gebhardt} K.,
  {Magorrian} J.,  {Corsini} E.~M.,   {Wegner} G.,  2007b, \mn@doi [\mnras]
  {10.1111/j.1365-2966.2007.12434.x}, \href
  {https://ui.adsabs.harvard.edu/abs/2007MNRAS.382..657T} {382, 657}

\bibitem[\protect\citeauthoryear{{Thomas} et~al.,}{{Thomas}
  et~al.}{2009a}]{2009MNRAS.393..641T}
{Thomas} J.,  et~al., 2009a, \mn@doi [\mnras]
  {10.1111/j.1365-2966.2008.14238.x}, \href
  {https://ui.adsabs.harvard.edu/abs/2009MNRAS.393..641T} {393, 641}

\bibitem[\protect\citeauthoryear{{Thomas}, {Saglia}, {Bender}, {Thomas},
  {Gebhardt}, {Magorrian}, {Corsini}  \& {Wegner}}{{Thomas}
  et~al.}{2009b}]{2009ApJ...691..770T}
{Thomas} J.,  {Saglia} R.~P.,  {Bender} R.,  {Thomas} D.,  {Gebhardt} K.,
  {Magorrian} J.,  {Corsini} E.~M.,   {Wegner} G.,  2009b, \mn@doi [\apj]
  {10.1088/0004-637X/691/1/770}, \href
  {https://ui.adsabs.harvard.edu/abs/2009ApJ...691..770T} {691, 770}

\bibitem[\protect\citeauthoryear{{Thomas} et~al.,}{{Thomas}
  et~al.}{2011}]{2011MNRAS.415..545T}
{Thomas} J.,  et~al., 2011, \mn@doi [\mnras]
  {10.1111/j.1365-2966.2011.18725.x}, \href
  {https://ui.adsabs.harvard.edu/abs/2011MNRAS.415..545T} {415, 545}

\bibitem[\protect\citeauthoryear{{Thomas}, {Saglia}, {Bender}, {Erwin}  \&
  {Fabricius}}{{Thomas} et~al.}{2014}]{2014ApJ...782...39T}
{Thomas} J.,  {Saglia} R.~P.,  {Bender} R.,  {Erwin} P.,   {Fabricius} M.,
  2014, \mn@doi [\apj] {10.1088/0004-637X/782/1/39}, \href
  {https://ui.adsabs.harvard.edu/abs/2014ApJ...782...39T} {782, 39}

\bibitem[\protect\citeauthoryear{{Tonry}, {Dressler}, {Blakeslee}, {Ajhar},
  {Fletcher}, {Luppino}, {Metzger}  \& {Moore}}{{Tonry}
  et~al.}{2001}]{2001ApJ...546..681T}
{Tonry} J.~L.,  {Dressler} A.,  {Blakeslee} J.~P.,  {Ajhar} E.~A.,  {Fletcher}
  A.~B.,  {Luppino} G.~A.,  {Metzger} M.~R.,   {Moore} C.~B.,  2001, \mn@doi
  [\apj] {10.1086/318301}, \href
  {https://ui.adsabs.harvard.edu/abs/2001ApJ...546..681T} {546, 681}

\bibitem[\protect\citeauthoryear{{Valluri}, {Merritt}  \& {Emsellem}}{{Valluri}
  et~al.}{2004}]{2004ApJ...602...66V}
{Valluri} M.,  {Merritt} D.,   {Emsellem} E.,  2004, \mn@doi [\apj]
  {10.1086/380896}, \href
  {https://ui.adsabs.harvard.edu/abs/2004ApJ...602...66V} {602, 66}

\bibitem[\protect\citeauthoryear{{Vasiliev} \& {Athanassoula}}{{Vasiliev} \&
  {Athanassoula}}{2015}]{2015MNRAS.450.2842V}
{Vasiliev} E.,  {Athanassoula} E.,  2015, \mn@doi [\mnras]
  {10.1093/mnras/stv805}, \href
  {https://ui.adsabs.harvard.edu/abs/2015MNRAS.450.2842V} {450, 2842}

\bibitem[\protect\citeauthoryear{{Vasiliev} \& {Valluri}}{{Vasiliev} \&
  {Valluri}}{2020}]{2020ApJ...889...39V}
{Vasiliev} E.,  {Valluri} M.,  2020, \mn@doi [\apj] {10.3847/1538-4357/ab5fe0},
  \href {https://ui.adsabs.harvard.edu/abs/2020ApJ...889...39V} {889, 39}

\bibitem[\protect\citeauthoryear{{Verolme} et~al.,}{{Verolme}
  et~al.}{2002}]{2002MNRAS.335..517V}
{Verolme} E.~K.,  et~al., 2002, \mn@doi [\mnras]
  {10.1046/j.1365-8711.2002.05664.x}, \href
  {https://ui.adsabs.harvard.edu/abs/2002MNRAS.335..517V} {335, 517}

\bibitem[\protect\citeauthoryear{Ye}{Ye}{1998}]{doi:10.1080/01621459.1998.10474094}
Ye J.,  1998, \mn@doi [Journal of the American Statistical Association]
  {10.1080/01621459.1998.10474094}, 93, 120

\bibitem[\protect\citeauthoryear{{York} et~al.,}{{York}
  et~al.}{2000}]{2000AJ....120.1579Y}
{York} D.~G.,  et~al., 2000, \mn@doi [\aj] {10.1086/301513}, \href
  {https://ui.adsabs.harvard.edu/abs/2000AJ....120.1579Y} {120, 1579}

\bibitem[\protect\citeauthoryear{{de Nicola}, {Saglia}, {Thomas}, {Dehnen}  \&
  {Bender}}{{de Nicola} et~al.}{2020}]{2020MNRAS.496.3076D}
{de Nicola} S.,  {Saglia} R.~P.,  {Thomas} J.,  {Dehnen} W.,   {Bender} R.,
  2020, \mn@doi [\mnras] {10.1093/mnras/staa1703}, \href
  {https://ui.adsabs.harvard.edu/abs/2020MNRAS.496.3076D} {496, 3076}

\bibitem[\protect\citeauthoryear{{van de Ven}, {de Zeeuw}  \& {van den
  Bosch}}{{van de Ven} et~al.}{2008}]{2008MNRAS.385..614V}
{van de Ven} G.,  {de Zeeuw} P.~T.,   {van den Bosch} R.~C.~E.,  2008, \mn@doi
  [\mnras] {10.1111/j.1365-2966.2008.12873.x}, \href
  {https://ui.adsabs.harvard.edu/abs/2008MNRAS.385..614V} {385, 614}

\bibitem[\protect\citeauthoryear{{van den Bosch}, {van de Ven}, {Verolme},
  {Cappellari}  \& {de Zeeuw}}{{van den Bosch}
  et~al.}{2008}]{2008MNRAS.385..647V}
{van den Bosch} R.~C.~E.,  {van de Ven} G.,  {Verolme} E.~K.,  {Cappellari} M.,
    {de Zeeuw} P.~T.,  2008, \mn@doi [\mnras]
  {10.1111/j.1365-2966.2008.12874.x}, \href
  {https://ui.adsabs.harvard.edu/abs/2008MNRAS.385..647V} {385, 647}

\bibitem[\protect\citeauthoryear{{van der Marel} \& {Franx}}{{van der Marel} \&
  {Franx}}{1993}]{1993ApJ...407..525V}
{van der Marel} R.~P.,  {Franx} M.,  1993, \mn@doi [\apj] {10.1086/172534},
  \href {https://ui.adsabs.harvard.edu/abs/1993ApJ...407..525V} {407, 525}

\bibitem[\protect\citeauthoryear{{van der Marel}, {Cretton}, {de Zeeuw}  \&
  {Rix}}{{van der Marel} et~al.}{1998}]{1998ApJ...493..613V}
{van der Marel} R.~P.,  {Cretton} N.,  {de Zeeuw} P.~T.,   {Rix} H.-W.,  1998,
  \mn@doi [\apj] {10.1086/305147}, \href
  {https://ui.adsabs.harvard.edu/abs/1998ApJ...493..613V} {493, 613}

\makeatother
\end{thebibliography}

\appendix

\section{Basics - Constructing Schwarzschild models}
\label{appendix:modelling}
To derive the orbital structure of a galaxy or the mass of its central supermassive black hole, stars and dark matter halo via dynamical modelling, one starts from a set of photometric and kinematic observations. The photometric observations typically consist of measurements of a galaxy's surface brightness either from an isophote analysis or directly from an image. Even though the photometric information is intrinsically 2d in nature, we can always rearrange the finite number of measurements into a 1d vector $\mathbfit{sb}_{\mathrm{obs},i}$  with $i=1,\ldots,N_\mathrm{phot}$ data points and their corresponding errors $\Delta \mathbfit{sb}_{\mathrm{obs},i}$. The kinematic observations consist of a large number of measurements related to the line-of-sight velocity distributions (LOSVDs) $\mathbfit{l}_{\mathrm{obs},ij} \pm \Delta \mathbfit{l}_{\mathrm{obs},ij}$  at $i=1,\ldots,N_\mathrm{kin}$ positions spread over the galaxy on the sky (e.g. in $N_\mathrm{kin}$ Voronoi Bins). If the LOSVDs are measured non-parametrically, which is the state of the art today \citep[e.g.][]{2019ApJ...887..195M}, then in each bin on the sky we have $j=1,\ldots,N_\mathrm{vel}$ data points that sample the respective LOSVD over $N_\mathrm{vel}$ different line-of-sight velocites. For modern integral-field spectrographs, the $\mathbfit{l}_{\mathrm{obs},ij}$ form a data cube and are 3d in nature. Still, we can again rearrange the finite (though large) number of data points into a 1d vector $\mathbfit{l}_{\mathrm{obs},k}$   or short $\mathbfit{l}_\mathrm{obs}$. Several modeling steps are required to derive the mass distribution or internal structure of a galaxy from an orbit superposition model based on these observations $\mathbfit{l}_\mathrm{obs}$ and $\mathbfit{sb}_\mathrm{obs}$.

First, a trial mass distribution has to be assumed. The corresponding gravitational potential is calculated via the Poisson equation. Then a set of orbits with representative initial conditions, called the orbit library, is integrated in this potential and the properties of all orbits are stored. It is necessary to sample the available phase-space sufficiently dense with these orbits, thus their initial conditions must be chosen with care. For a comprehensive description of our orbit sampling see \citet{2004MNRAS.353..391T}. The orbits of the library are then superimposed by giving each of them an adjustable, non-negative orbital weight $w_{i}$, akin to an occupation number that represents the number of stars tracking the orbit. The non-negativity constraint is imposed on these orbital weights to guarantee the resulting phase-space distribution function of the orbit model is positive everywhere. The adjustment of the orbital weights is done such that the properties of the superposition emulate the photometric and kinematic observations of the galaxy as good as possible in the assumed trial potential. Here it turns out that the LOSVDs  $\mathbfit{l}_{\mathrm{mod},j}$ of the orbit superposition model are linear combinations of the individual contributions of each orbit. In other words, if the contribution of orbit $i$ to the kinematic measurement $j$ is $\mathbfss{l}_{\mathrm{orb},i}^{j}$, then the predicted kinematics $\mathbfit{l}_{\mathrm{mod},j}$  of the whole orbit superposition model reads
\begin{equation}
    \mathbfit{l}_{\mathrm{mod},j}=\sum_{i}^{N_\mathrm{orbit}} \mathbfit{w}_{i}\mathbfss{l}_{\mathrm{orb},i}^{j}
	\label{eq:model LOSVD}
\end{equation}
where the sum goes over the number of library orbits $N_\mathrm{orbit}$. In compact matrix notation we can write $\mathbfit{l}_\mathrm{mod} = \mathbfss{L}_\mathrm{orb} \cdot \mathbfit{w}$, where $\mathbfss{L}_\mathrm{orb}$ is the matrix with elements $\mathbfss{l}^j_{\mathrm{orb},i}$ and $\mathbfit{w}$ is the vector with the orbital weights. 

The 3d mass distribution $\rho$ required in the first step is usually unknown. A comprehensive trial mass distribution should include the most important galaxy components. Commonly, the density is composed as:
\begin{equation}
    \rho=\Upsilon \cdot \nu + M_{\bullet} \cdot \frac{\delta(r)}{4 \pi r^{2}} + \rho_{\mathrm{DM}}
	\label{eq:mass model}
\end{equation}
where the first term is the contribution of the stellar component determined by the mass-to-light ratio $\Upsilon$ and the 3d luminosity density $\nu$. The second term is a central supermassive black hole with mass $M_{\bullet}$ and the third term encompasses the contribution of the dark matter halo, which can in itself be further parametrized, for example by a Navarro-Frank-White profile \citep{1996ApJ...462..563N}. The 3d luminosity density $\nu$ is typically not a free parameter of this mass model. Instead it is calculated by deprojecting the 2D surface brightness distribution of the investigated galaxy, implying that $\nu$ and consequently $\rho$ depend on the galaxy's assumed inclination. Similarly as for the model LOSVDs it turns out that its intrinsic 3d luminosity density $\mathbfit{d}_\mathrm{mod}$ is simply the linear combination of the individual orbital contributions. In compact matrix notation we can write $\mathbfit{d}_\mathrm{mod} = \mathbfss{D}_\mathrm{orb} \cdot \mathbfit{w}$, analogously to the model's LOSVDs.

In detail, the few implementations of the Schwarzschild method that have been developed differ considerably. Some, like ours, exploit the information contained in the entire (though binned) LOSVDs of the galaxy and orbit model, while others rely on Gauss-Hermite expansions up to some finite Hermite order $n$ (often $n<8$). Likewise, some implementations fit the observed surface brightness and/or the deprojected density, while others -- like ours -- enforce full consistency of the models through equality constraints for the deprojected luminosity density $\mathbfit{d}_\mathrm{data}$. Specifically, for our implementation, we use a $\chi^2$ minimisation to derive the best-fit orbital weights from the $\mathbfit{l}_\mathrm{obs}$ where the orbital weights must fulfill the equality constrain $\mathbfit{d}_\mathrm{data} = \mathbfit{d}_\mathrm{mod}= \mathbfss{D}_\mathrm{orb} \cdot \mathbfit{w}$ to ensure self-consistency. 

When modelling a galaxy, it is often one of the main goals to determine its unknown mass distribution while the orbital weights are not of primary interest. Therefore one typically postulates a number of trial mass distributions and creates an orbit superposition for each of them. Then, the final task is to pick up the "best" model out of this set of trial mass distributions. Usually, this is also done via a $\chi^2$ comparison, which means that the model with the smallest $\chi^2$ is considered to be the best representation of the galaxy.

\section{Examples for kinematic Maps}
\label{appendix:kinmaps}

\begin{figure}
	\includegraphics[width=\columnwidth]{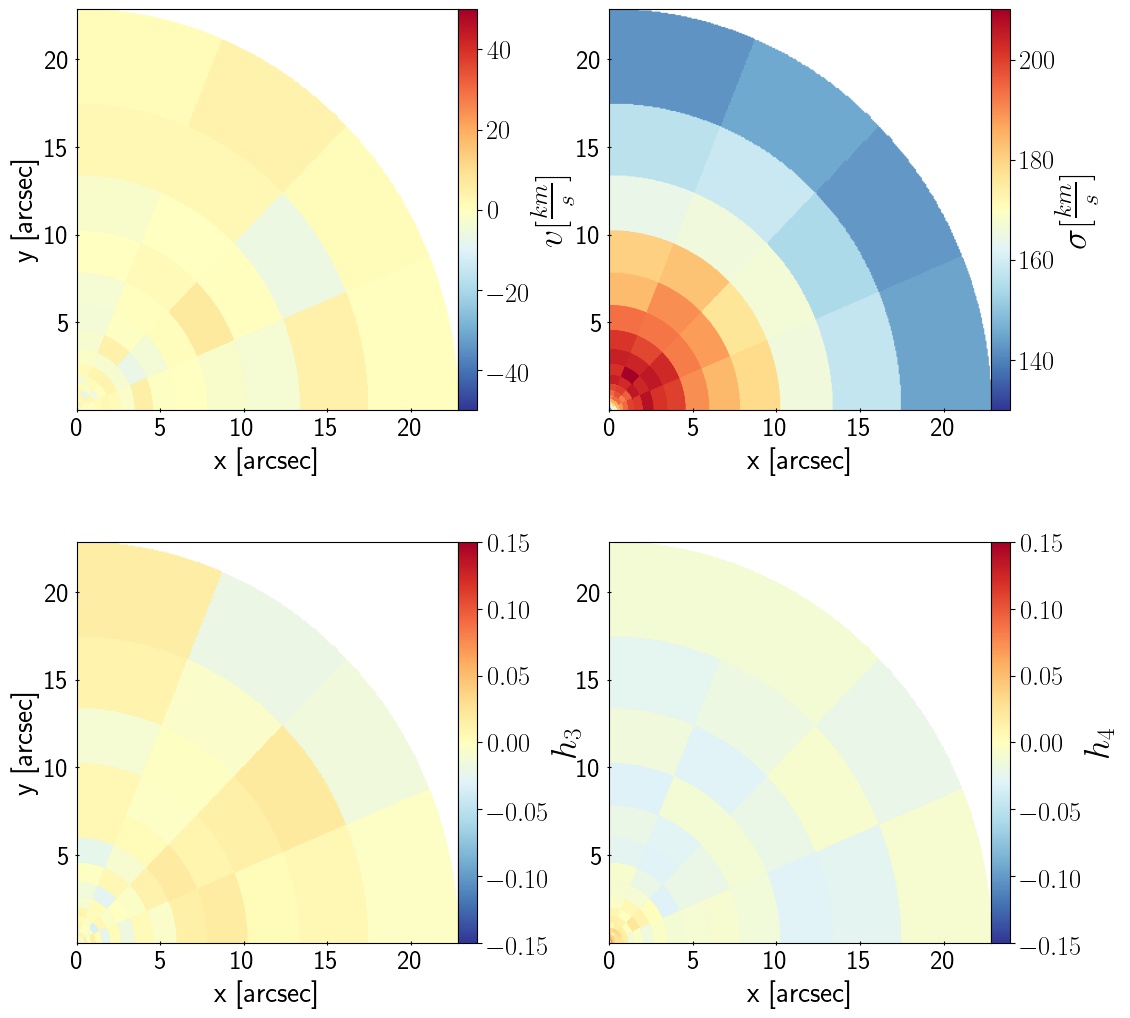}
    \caption{The mock Gauss-Hermite maps of a quadrant of the spherical Hernquist N-body. Responsible for the deviations from spherical symmetry is Gaussian distributed Monte-Carlo noise that has been added to the N-body data.}
    \label{fig:spherical_kinmap}
\end{figure}

\begin{figure}
	\includegraphics[width=\columnwidth]{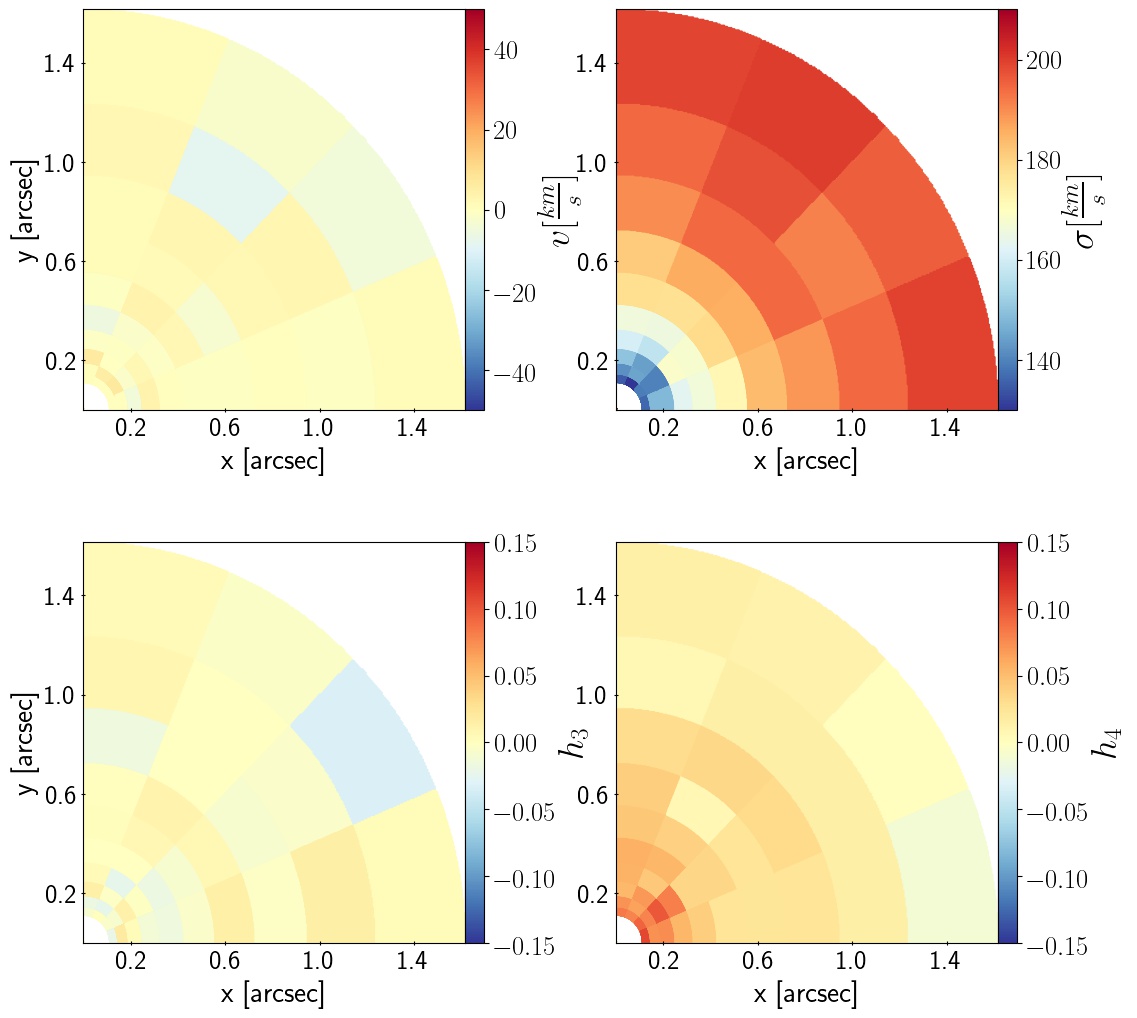}
    \caption{As Fig.~\ref{fig:spherical_kinmap} but only the central 2\arcsec are shown. Towards the centre the velocity dispersion drops while $h_4$ increases (as expected for an isotropic Hernquist galaxy, cf. \citealt{2005A&A...432..411B}).}
    \label{fig:spherical_central_kinmap}
\end{figure}

\begin{figure}
	\includegraphics[width=\columnwidth]{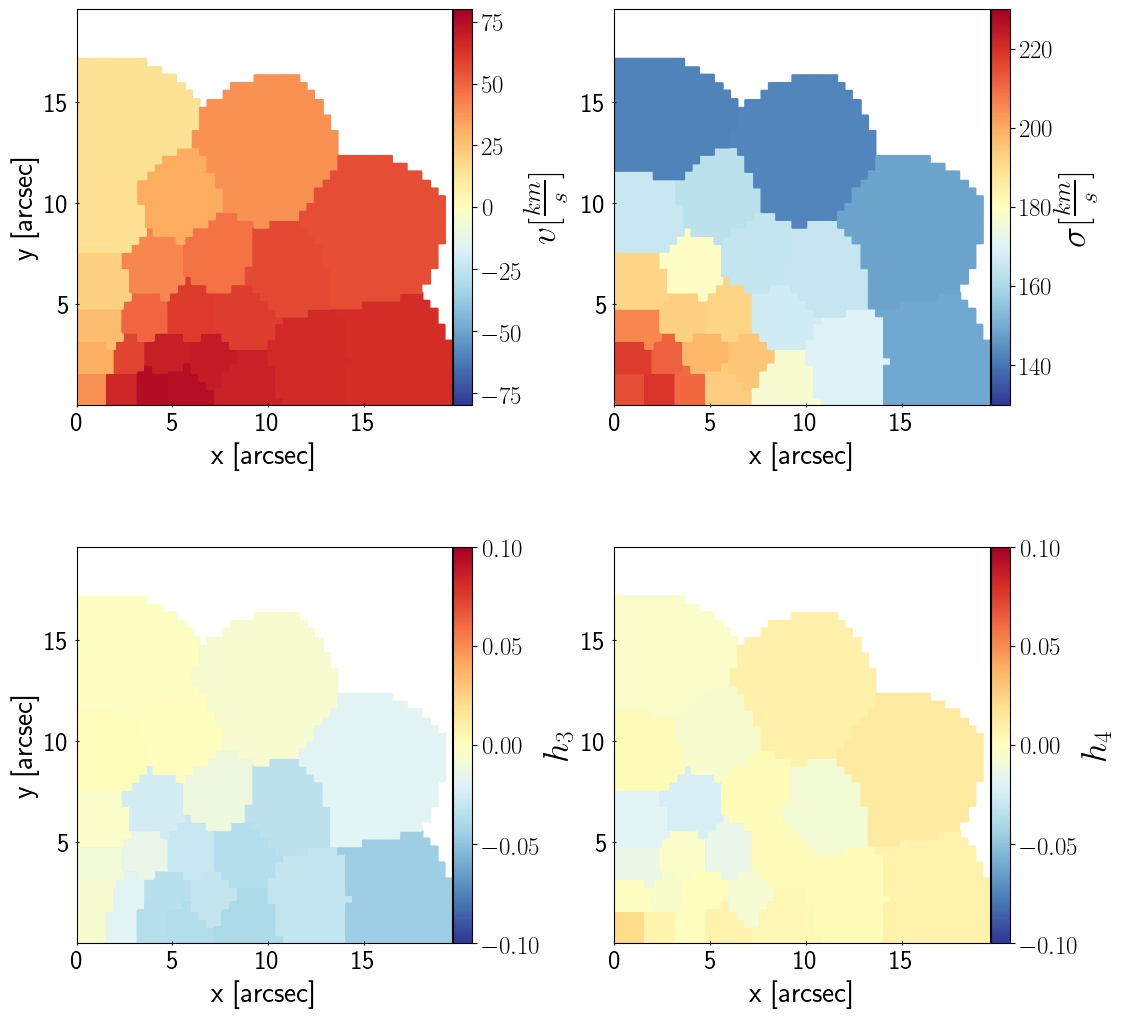}
    \caption{The kinematic map of the early-type toy galaxy D \textit{before} the addition of Monte-Carlo noise. The spatial grid shown is typical for wide-field kinematic data that extend beyond the galaxy's effective radius. In addition to the spatial bins shown here we simultaneously modeled bins with higher resolution in the centre of the galaxy (cf. Fig.~\ref{fig:rotation_maps}).}
    \label{fig:GalaxyD_muse}
\end{figure}

Shown are the Gauss-Hermite coefficients up to $h_4$ of the LOSVDs of the spherical N-body (Fig.~\ref{fig:spherical_kinmap}-~\ref{fig:spherical_central_kinmap}) and of toy galaxy D (Fig.~\ref{fig:GalaxyD_muse}) which was created using Schwarzschild models with an angular momentum bias $\lambda=0.5$.

\bsp
\label{lastpage}
\end{document}